\documentclass[
  journal=pasa,
  manuscript=research-paper, 
  year=202X,
  volume=YY,
]{cup-journal}

\usepackage{natbib,twoopt}
\usepackage{xcolor}
\usepackage[breaklinks=true, colorlinks=true, linkcolor=blue, citecolor=blue]{hyperref} 
\bibpunct{(}{)}{;}{a}{}{,}             
\makeatletter
\newcommandtwoopt{\citeads}[3][][]{\href{http://adsabs.harvard.edu/abs/#3}%
    {\def\hyper@linkstart##1##2{}%
     \let\hyper@linkend\@empty\citealp[#1][#2]{#3}}}
  \newcommandtwoopt{\citepads}[3][][]{\href{http://adsabs.harvard.edu/abs/#3}%
    {\def\hyper@linkstart##1##2{}%
     \let\hyper@linkend\@empty\citep[#1][#2]{#3}}}
  \newcommandtwoopt{\citetads}[3][][]{\href{http://adsabs.harvard.edu/abs/#3}%
    {\def\hyper@linkstart##1##2{}%
     \let\hyper@linkend\@empty\citet[#1][#2]{#3}}}
  \newcommandtwoopt{\citeyearads}[3][][]%
    {\href{http://adsabs.harvard.edu/abs/#3}
    {\def\hyper@linkstart##1##2{}%
     \let\hyper@linkend\@empty\citeyear[#1][#2]{#3}}}
\makeatother
\usepackage{graphicx}
\usepackage{soul}
\usepackage{booktabs}
\usepackage{multirow}
\usepackage{longtable}
\usepackage{comment}
\usepackage{amsmath}
\usepackage{siunitx}
\usepackage{upgreek}
\usepackage{txfonts}
\usepackage{subfig}
\usepackage{sidecap}
\sidecaptionvpos{figure}{c}

\newcommand{\ldet}{$\mathcal{L}_\mathrm{DET}$}
\newcommand{\lext}{$\mathcal{L}_\mathrm{EXT}$}
\title{The life of central radio galaxies in clusters: AGN-ICM studies of eRASS1 clusters in the ASKAP fields}

\author{Angie Veronica}
\author{Thomas H. Reiprich}
\author{Florian Pacaud}
\affiliation{Argelander-Institut f\"ur Astronomie, Universit\"at Bonn, Auf dem H\"ugel 71, 53121 Bonn, Germany}
\email[A. Veronica]{averonica@astro.uni-bonn.de}

\author{Marcus Br\"uggen}
\affiliation{Hamburger Sternwarte, Universit\"at Hamburg, Gojenbergsweg 112, 21029 Hamburg, Germany}

\author{B\"arbel Koribalski}
\affiliation{Australia Telescope National Facility, CSIRO, Space and Astronomy, P.O. Box 76, Epping, NSW 1710, Australia}
\alsoaffiliation{School of Science, Western Sydney University, Locked Bag 1797, Penrith, NSW 2751, Australia}

\author{Thomas Pasini}
\affiliation{INAF - Istituto di Radioastronomia, via P. Gobetti 101, 40129, Bologna, Italy}

\author{Tessa Vernstrom}
\affiliation{ICRAR, The University of Western Australia, 35 Stirling Hwy, 6009 Crawley, Australia}
\alsoaffiliation{Australia Telescope National Facility, CSIRO, Space and Astronomy, PO Box 1130, Bentley, WA 6151, Australia}

\author{Stefan W. Duchesne}
\affiliation{Australia Telescope National Facility, CSIRO, Space and Astronomy, PO Box 1130, Bentley, WA 6151, Australia}

\author{Kathrin B\"ockmann}
\affiliation{Hamburger Sternwarte, Universit\"at Hamburg, Gojenbergsweg 112, 21029 Hamburg, Germany}

\author{Jeremy S. Sanders}
\author{Y. Emre Bahar}
\author{Fabian Balzer}
\affiliation{Max-Planck-Institut f\"ur extraterrestrische Physik, Gie{\ss}enbachstra{\ss}e 1, 85748 Garching, Germany}

\author{Lachlan J. Barnes}
\affiliation{School of Mathematical and Physical Sciences, 12 Wally's Walk, Macquarie University, NSW 2109, Australia}

\author{Esra Bulbul}
\affiliation{Max-Planck-Institut f\"ur extraterrestrische Physik, Gie{\ss}enbachstra{\ss}e 1, 85748 Garching, Germany}

\author{Nicolas Clerc}
\affiliation{IRAP, Universit\'e de Toulouse, CNRS, UPS, CNES, 31028 Toulouse, France}

\author{Jessica E. M. Craig}
\affiliation{Lennard-Jones Laboratories, Keele University, ST5 5BG, United Kingdom}

\author{Johan Comparat}
\affiliation{Max-Planck-Institut f\"ur extraterrestrische Physik, Gie{\ss}enbachstra{\ss}e 1, 85748 Garching, Germany}

\author{Simon Dannhauer}
\affiliation{Argelander-Institut f\"ur Astronomie, Universit\"at Bonn, Auf dem H\"ugel 71, 53121 Bonn, Germany}
\alsoaffiliation{I. Physik. Institut, University of Cologne, Z\"ulpicher Str. 77, D-50937 Cologne, Germany}

\author{Jakob Dietl}
\affiliation{Argelander-Institut f\"ur Astronomie, Universit\"at Bonn, Auf dem H\"ugel 71, 53121 Bonn, Germany}

\author{Klaus Dolag}
\affiliation{Universit\"ats-Sternwarte, Fakult\"at f\"ur Physik, Ludwig-Maximilians-Universit\"at M\"unchen, Scheinerstr. 1, D-81679 M\"unchen, Germany}
\alsoaffiliation{Max Planck Institute for Astrophysics, Karl-Schwarzschild-Str. 1, D-85741 Garching, Germany}

\author{Vittorio Ghirardini}
\affiliation{Max-Planck-Institut f\"ur extraterrestrische Physik, Gie{\ss}enbachstra{\ss}e 1, 85748 Garching, Germany}
\alsoaffiliation{INAF, Osservatorio di Astrofisica e Scienza dello Spazio, via Piero Gobetti 93/3, 40129 Bologna, Italy}

\author{Sebastian Grandis}
\affiliation{Universit\"ats-Sternwarte, Fakult\"at f\"ur Physik, Ludwig-Maximilians-Universit\"at M\"unchen, Scheinerstr. 1, D-81679 M\"unchen, Germany}
\alsoaffiliation{Universit\"at Innsbruck, Institut f\"ur Astro- und Teilchenphysik, Technikerstr. 25/8, 6020 Innsbruck, Austria}

\author{Duy Hoang}
\affiliation{Th\"uringer Landessternwarte, Sternwarte 5, 07778 Tautenburg, Germany}
\alsoaffiliation{Hamburger Sternwarte, Universit\"at Hamburg, Gojenbergsweg 112, 21029 Hamburg, Germany}

\author{Andrew M. Hopkins}
\affiliation{School of Mathematical and Physical Sciences, 12 Wally's Walk, Macquarie University, NSW 2109, Australia}

\author{Zsofi Igo}
\affiliation{Max-Planck-Institut f\"ur extraterrestrische Physik, Gie{\ss}enbachstra{\ss}e 1, 85748 Garching, Germany}
\alsoaffiliation{Exzellenzcluster ORIGINS, Boltzmannstr. 2, 85748, Garching, Germany}

\author{Matthias Kluge}
\affiliation{Max-Planck-Institut f\"ur extraterrestrische Physik, Gie{\ss}enbachstra{\ss}e 1, 85748 Garching, Germany}

\author{Ang Liu}
\affiliation{Institute for Frontiers in Astronomy and Astrophysics, Beijing Normal University, Beijing 102206, China}
\alsoaffiliation{Max-Planck-Institut f\"ur extraterrestrische Physik, Gie{\ss}enbachstra{\ss}e 1, 85748 Garching, Germany}

\author{Konstantinos Migkas}
\affiliation{Leiden Observatory, Leiden University, PO Box 9513, NL-2300 RA Leiden, The Netherlands}
\alsoaffiliation{SRON Netherlands Institute for Space Research, Niels Bohrweg 4, NL-2333 CA Leiden, The Netherlands}

\author{Vanessa A. Moss}
\affiliation{Australia Telescope National Facility, CSIRO, Space and Astronomy, P.O. Box 76, Epping, NSW 1710, Australia}

\author{Miriam E. Ramos-Ceja}
\affiliation{Max-Planck-Institut f\"ur extraterrestrische Physik, Gie{\ss}enbachstra{\ss}e 1, 85748 Garching, Germany}

\author{Chris Riseley}
\affiliation{Astronomisches Institut der Ruhr-Universit\"at Bochum (AIRUB), Universit\"atsstra{\ss}e 150, 44801 Bochum, Germany}
\alsoaffiliation{Dipartimento di Fisica e Astronomia, Università degli Studi di Bologna, Bologna, Italy}
\alsoaffiliation{INAF - Istituto di Radioastronomia, via P. Gobetti 101, 40129, Bologna, Italy}

\author{Lawrence Rudnick}
\affiliation{Minnesota Institute for Astrophysics, University of Minnesota, Minneapolis, MN, USA}

\author{Mara Salvato}
\affiliation{Max-Planck-Institut f\"ur extraterrestrische Physik, Gie{\ss}enbachstra{\ss}e 1, 85748 Garching, Germany}

\author{Stanislav Shabala}
\affiliation{School of Natural Sciences, University of Tasmania, Private Bag 37, Hobart 7001, Australia}

\author{Riccardo Seppi}
\affiliation{Max-Planck-Institut f\"ur extraterrestrische Physik, Gie{\ss}enbachstra{\ss}e 1, 85748 Garching, Germany}

\author{Jacco van Loon}
\affiliation{Lennard-Jones Laboratories, Keele University, ST5 5BG, United Kingdom}

\author{Tayyaba Zafar}
\affiliation{School of Mathematical and Physical Sciences, 12 Wally's Walk, Macquarie University, NSW 2109, Australia}

\author{Xiaoyuan Zhang}
\affiliation{Max-Planck-Institut f\"ur extraterrestrische Physik, Gie{\ss}enbachstra{\ss}e 1, 85748 Garching, Germany}

\received {06 Jun 2025}
\revised  {22 Oct 2025}
\accepted {25 Nov 2025}
\published{dd Mmm YYYY}

\keywords{X-rays: galaxies: clusters; radio continuum: galaxies; galaxies: clusters: intracluster medium; galaxies: active; galaxies: clusters: general}

\begin{document}

\begin{abstract}
The mechanical feedback from the central active galactic nuclei (AGNs) can be crucial for balancing the radiative cooling of the intracluster medium (ICM) at the cluster centre.
We aim to understand the relationship between the power of AGN feedback and the cooling of gas in the centres of galaxy clusters by correlating the radio properties of the brightest cluster galaxies (BCGs) with the X-ray properties of their host clusters.
We used the catalogues from the first SRG/eROSITA All-Sky Survey (eRASS1) along with radio observations from the Australian SKA Pathfinder (ASKAP). In total, we identified 134 radio sources associated with BCGs of the 151 eRASS1 clusters located in the PS1, PS2, and SWAG-X ASKAP fields. Non-detections were treated as upper limits. We correlated the radio properties of the BCGs (radio luminosity, largest linear size/LLS, and BCG offset from the cluster centre) with the integrated X-ray luminosity of the host clusters.
We utilised the concentration parameter, $c_{R_{500}}$, to categorise the clusters into cool cores (CCs) and non-cool cores (NCCs). By combining $c_{R_{500}}$ with the BCG offset, we assessed the dynamical states of the clusters in our sample. Furthermore, we analysed the correlation between radio mechanical power and X-ray luminosity within the CC subsample.
We observe a potential positive trend between LLS and BCG offset, which may hint at an environmental influence on the morphology of central radio sources. We find a weak trend suggesting that more luminous central radio galaxies are found in clusters with higher X-ray luminosity. Additionally, there is a positive but highly scattered relationship between the mechanical luminosity of AGN jets and the X-ray cooling luminosity within the CC subsample. This finding is supported by bootstrap resampling and flux-flux analyses. The correlation observed in our CC subsample indicates that AGN feedback is ineffective in high-luminosity (high-mass) clusters. At a cooling luminosity of $L_{\mathrm{X},~r<R_\mathrm{cool}}\approx 5.50\times10^{43}\,\mathrm{erg\,s^{-1}}$, on average, AGN feedback appears to contribute only about $13\%-22\%$ of the energy needed to offset the radiative losses in the ICM.
\end{abstract}

\section{Introduction}\label{sect:intro}
The intracluster medium (ICM) radiates its thermal energy primarily through X-ray emission. In dense cluster cores, the ICM cools tremendously, leading to a short cooling time of typically $0.01-0.1$ of Hubble time \citep{Hlavacek_2022}. In the absence of a heating mechanism, the ICM would continuously cool, losing pressure support, and subsequently flow inward toward the cluster centre with a mass deposition rate of up to $1\,000~M_\odot\,\mathrm{yr^{-1}}$. This is known as the cooling flow model \citep{Fabian_1994}.
\par
The accumulation of cool gas should fuel star formation and the active galactic nucleus (AGN) of the central galaxy. However, optical and infrared observations reveal that star formation rates were only of the order of a few per cent of the cooling gas \citep{McNamara_1989, ODea_2008}. Cold atomic and molecular gas mass was also found to be lower than expected \citep{Edge_2001}. Moreover, X-ray observations of the cooling flow clusters revealed a lack of emission lines from the Fe\,L complex \citep{Boehringer_2002}, and measured a deficit of cold gas as compared to what the cooling flow model predicted \citep{Peterson_2001, Tamura_2001, Sanders_2008}. This discrepancy suggests that the gas cooling rate is overestimated, or that there exists an energy source that prevents the cooling flow or significantly suppresses the mass deposition rate. Several heating mechanisms have been suggested, such as thermal conduction from the hot outer ICM layer \citep[e.g.,][]{Voigt_2002, Zakamska_2003, Voigt_2004}, heating by cosmic rays \citep{Guo_2008}, and supernova explosions \citep{McNamara_2004}. However, these processes alone are not sufficient to counter the radiative losses \citep[e.g.,][]{Soker_2003, McNamara_2004}.
\par
AGN feedback has emerged as the most promising mechanism for heating the ICM, preventing the catastrophic cooling \citep[for reviews see, e.g.,][]{McNamara_2007, McNamara_2012, Fabian_2012, Hlavacek_2022}. This self-regulated process begins when the cold material is accreted to the supermassive black hole (SMBH) harboured by the central brightest cluster galaxies (BCGs), triggering AGN activity. The relativistic AGN jets inflate low-density, rising bubbles. In their wakes, they redistribute energy through weak shocks \citep[e.g.,][]{Randall_2011}, thermal conduction and sound waves \citep[e.g.,][]{Fabian_2005}, and turbulence \citep[e.g.,][]{Dennis_2005}.
\par
The signatures of this mechanism are evident in the wealth of \textit{Chandra} and \textit{XMM-Newton} observations of massive galaxy clusters as X-ray cavities that often coincide with radio lobes \citep[e.g.,][]{Fabian_2000, David_2001}. Statistical studies of X-ray cavities \citep{Birzan_2004, Dunn_2005, Rafferty_2008} have shown that AGN feedback provides enough energy to regulate star formation and counteract the cooling of hot halos. This finding is further supported by numerical simulations \citep[e.g.,][]{Croton_2006, Bower_2006}. Moreover, this mechanism reshapes the thermodynamical properties of the ICM, as the rising bubbles increase cluster entropy and the outflows transport metals from the core to the outskirts, altering the chemical composition of the ICM and affecting its radiative properties \citep{Kirkpatrick_2015}.
\par
In this work, we aim to gain more insight into the cooling and heating processes regulated by AGN feedback. We exploit the group and cluster catalogue of the first SRG/eROSITA All-Sky Survey \citep[eRASS1,][]{Bulbul_2024} to construct a sample of clusters located in three different fields observed by the Australian SKA Pathfinder \citep[ASKAP,][]{Hotan_2021}. We examine the relation between the radio properties of the AGN associated with the BCGs derived from the ASKAP data \citep[][Moss et al. in prep.]{Norris_2021, Boeckmann_2023} and the compiled eRASS1 X-ray properties along with the morphological parameters of the host clusters \citep{Kluge_2024, Sanders_2025}.
\par
The structure of this paper is as follows: In Section~\ref{sect:data}, we explain how the sample is constructed, and we describe the X-ray and radio properties of the sample. In Section~\ref{sect:results}, we present the results. We compare our results with previous works and discuss their implications, including the balance between central AGN feedback and ICM radiative losses in our sample in Section~\ref{sec:discussion}. We summarise our findings in Section~\ref{sect:conclusions}. The assumed cosmology in this work is a flat $\Lambda$CDM cosmology from the eRASS1 cosmological constraints \citep{erass1cosmo}, with parameters $H_0=67.77\,\mathrm{km\,s^{-1}\,Mpc^{-1}}$ and $\Omega_\mathrm{m}=0.29$. Throughout this paper, all logarithms are expressed in base 10 unless stated otherwise.

\section{Sample}\label{sect:data}
\subsection{eRASS1 group and cluster catalogues}
The extended ROentgen Survey with an Imaging Telescope Array (eROSITA) on board the Spectrum-Roentgen-Gamma (SRG) mission \citep{Sunyaev_2021} was launched on 13 July 2019 \citep{Predehl_2021} to the second Lagrangian point (L2) of the Earth-Sun system. After completion of the commissioning, calibration, and performance verification phase, eROSITA began its first all-sky survey (eRASS1) on 12 December 2019, which continued until 11 June 2020, for a total survey duration of 184 days. The complete eRASS1 X-ray catalogues of the Western Galactic hemisphere are released in \cite{Merloni_2024}. The detection of the sources is performed using the eROSITA source detection chain of the eROSITA Science Analysis Software System \citep[\texttt{eSASS},][]{Brunner_2022} for the eROSITA-DE Data Release 1 (\texttt{eSASS4DR1}).\footnote{\url{https://erosita.mpe.mpg.de/dr1/eSASS4DR1/}} In total $1\,277\,477$ sources down to detection likelihood (\texttt{DET\_LIKE\_0} in the catalogue) of five (\ldet$\geq5$) are detected in the most sensitive band of eROSITA, $0.2-2.3\,\mathrm{keV}$, of which only $\sim2\%$ are extended sources.
\par
This work utilised the eRASS1 galaxy group and cluster primary catalogue \citep[][hereafter, eRASS1 primary cluster catalogue]{Bulbul_2024}, which is derived from the Main eRASS1 catalogue. To improve the purity of the catalogue, an extent likelihood (\texttt{EXT\_LIKE}) threshold is applied (\lext$\geq3$). Further cleaning is conducted through optical identification using datasets from the DESI Legacy Survey \citep{Dey_2019} DR9 and DR10, processed using the \texttt{eROMaPPer} pipeline \citep{Chitham_2020, Kluge_2024}. This pipeline is based on the matched-filter red-sequence algorithm, \texttt{redMaPPer} \citep{Rykoff_2014, Rykoff_2016}. The optical properties of the eRASS1 groups and clusters, including redshift, richness, optical centre, and BCG position, are published in \cite{Kluge_2024}. In the $13\,116\,\mathrm{deg}^2$ survey area, there are $12\,247$ confirmed groups and clusters with a reported sample purity of $86\%$ and a completeness of $13.3\%$ for a flux limit of $4\times10^{-14}\,\mathrm{erg\,s^{-1}\,cm^{-2}}$. The completeness increases to $\sim\!36\%$ and $70\%$ for flux limits of $1\times10^{-13}\,\mathrm{erg\,s^{-1}\,cm^{-2}}$ and $3\times10^{-13}\,\mathrm{erg\,s^{-1}\,cm^{-2}}$, respectively \citep[see Figure~8 of][]{Bulbul_2024}. The redshifts of the sample range between 0.003 and 1.32 with a median of 0.31 \citep[see Figure~6 of][]{Bulbul_2024}.

\subsection{ASKAP fields: PS1, PS2, SWAG-X}\label{sect:askap}
The Australian SKA Pathfinder \citep[ASKAP,][]{Johnston_2008, McConnell_2016, Hotan_2021} is a radio telescope located at Inyarrimanha Ilgari Bundara, the CSIRO Murchison Radio-astronomy Observatory, on Wajarri Yamaji Country in remote Western Australia. It consists of an array of 36 12-m antennas, spread out over a region of 6 km in diameter. Each antenna is equipped with a phased array feed (PAF) that can be used to form 36 dual-polarization primary beams, giving the telescope an instantaneous wide field of view of about 30 deg$^2$ at $900\,\mathrm{MHz}$ and enabling rapid survey capability. The ASKAP telescope operates within a frequency range of $700-1\,800\,\mathrm{MHz}$ and has an instantaneous bandwidth of $288\,\mathrm{MHz}$. The processing of the ASKAP data was done using the ASKAP data-processing pipeline \citep[\texttt{ASKAPsoft},][]{Guzman_2019, Whiting_2020}. The last step in the \texttt{ASKAPsoft} pipeline is the automatic source-finding and cataloguing performed by the \texttt{Selavy} source finder \citep{Whiting_2012}.
The data are mosaicked into a weighted average using the ASKAP \texttt{linmos} task, and the source finding task is executed on this image (see \citealt{Norris_2021} and \citealt{Hopkins_2025} for more details of the ASKAP/EMU source cataloguing and associated image analysis). Each team may apply more advanced source finders tailored to their specific project goals.
\par
In this work, we used data collected by the ASKAP radio telescope covering three different fields. The Pilot Survey of the Evolutionary Map of the Universe with the ASKAP telescope \citep[ASKAP/EMU PS,][]{Norris_2021} was observed at a frequency of 944\,MHz. The primary aim of this pilot survey was to evaluate the readiness for the full EMU survey. The Pilot Survey Phase I (PS1) observations covered an area of $270\,\mathrm{deg}^2$, coinciding with the region covered by the Dark Energy Survey \citep{DES}. Observations were carried out between 15 July 2019 and 24 November 2019, totalling 100 hours of observation. The survey achieved a depth of $25-30\,\upmu\mathrm{Jy/beam}$ RMS, with a spatial resolution of approximately $11-18''$. From the PS1 field, about $220\,000$ sources were catalogued by the \texttt{Selavy} source finder, with approximately $180\,000$ classified as simple, single-component sources, while the remainder were identified as complex sources \citep{Norris_2021}.
\par
The Pilot Survey Phase II (PS2) aimed to evaluate the technical performance of the ASKAP/EMU pipelines, namely \texttt{ASKAPsoft} and \texttt{EMUCAT} (responsible for generating the EMU value-added catalogue), to determine the most effective approach for the main survey. The PS2 covered both a Galactic and extragalactic region. The latter, which is used in this work, is located near the celestial equator, spanning a Right Ascension (R.A.) range of about $41^\circ$ to $29^\circ$ and a Declination (Dec.) range of $-12^\circ$ to $3^\circ$. Out of $\sim180\,\mathrm{deg}^2$ area, only about $38\,\mathrm{deg}^2$ lie within the western Galactic hemisphere, where eROSITA observations are available to us. The PS2 observations were convolved to a common resolution of $18''$. Since no ASKAP source catalogue is available for this field, we did not conduct any cross-matching and directly performed manual measurements of the radio sources associated with the BCGs.
\par
The Survey With ASKAP of GAMA-09 + X-ray (SWAG-X, Moss et al. in prep.) was launched as a complementary survey to the eROSITA Final Equatorial-Depth Survey \citep[eFEDS, e.g.,][]{Liu_2022, Pasini_2022} and the 9 hours Galaxy And Mass Assembly spectroscopic observations \citep[GAMA-09,][]{Gama09} conducted at the Anglo-Australian Telescope (AAT). In the first SWAG-X data release,\footnote{\url{https://data.csiro.au/collection/csiro:44441}} 12 ASKAP fields were made publicly available, including six tiles of 8 hours integration in each 888\,MHz and 1296\,MHz band. In this work, we used the observations at 888\,MHz. All available observations in this band were convolved to a common resolution of $17.3''$ by $14.2''$ and combined into a weighted averaged dataset using \texttt{SWarp} \citep{swarp}.
After exclusion of the areas of lower sensitivity at the edge of the map due to the primary beam attenuation, we are left with a coverage area of approximately $19\times12\,\mathrm{deg}^2$. The generation of the source list for the SWAG-X field is produced by using the Python Blob Detector and Source Finder \citep[\texttt{PyBDSF},][]{pybdsf},\footnote{\url{ https://github.com/lofar-astron/PyBDSF}} which models the extended sources well. The parameters used in the \texttt{PyBDSF} run follow those described in \cite{Duchesne_2024RACS}. About $110\,000$ sources above $5\sigma$ are detected.

\begin{table}
\centering
\caption{Radio survey specifications of the ASKAP fields.}
\resizebox{\columnwidth}{!}
{\begin{tabular}{c | c c c}
\hline
\hline
Properties & PS1$^\star$ & PS2 & SWAG-X$^\ddagger$ \\
\hline\\[-2ex]
Area of survey [deg$^2$]& 270 & 180 & 228 \\
Common beam size [$''$] & $18.0 \times 18.0$ & $18.0 \times 18.0$ & $17.3 \times 14.2$ \\
Central frequency [MHz] & 944 & 944 & 888 \\
RMS sensitivity [$\upmu\mathrm{Jy/beam}$] & $25-35$ & $53-74^\dagger$ & $45-58^\dagger$\\
Total integration time & $10\times10\,\mathrm{h}$ & $6\times10\,\mathrm{h}$ & $6\times8\,\mathrm{h}$ \\
\hline
\multicolumn{4}{l}{$^\star$\cite{Norris_2021}, $^\ddagger$Moss et al.~(in prep.)}\\
\multicolumn{4}{l}{$^\dagger$\cite{Duchesne_2024}}\\
\hline
\hline
\end{tabular}}
\label{tab:askap}
\end{table}

\subsection{Sample construction and properties}\label{sect:sample}
\begin{figure*}[h!]
\centering
\includegraphics[width=\textwidth,trim=2.0cm 0.5cm 0.25cm 1.5cm, clip]{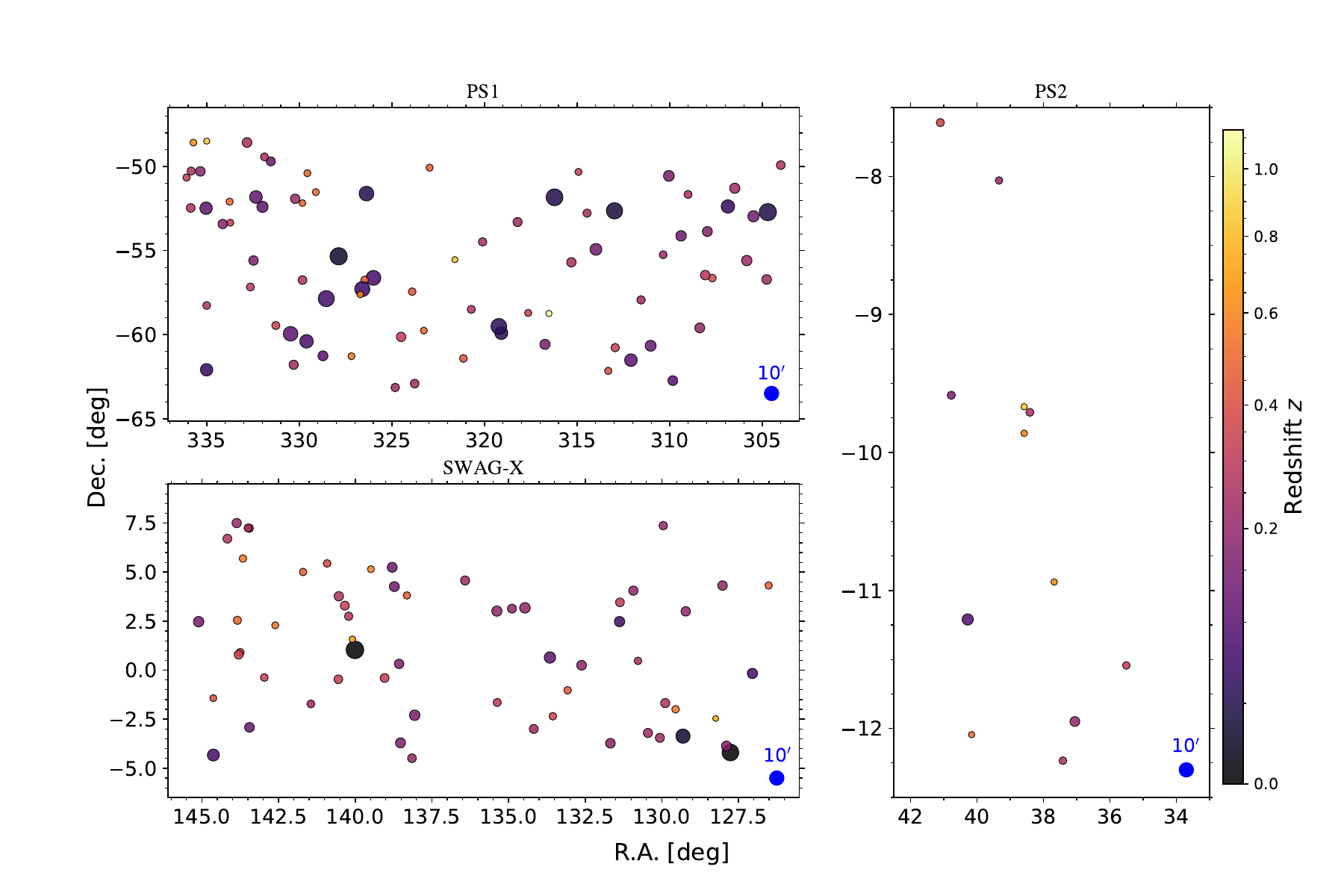}
\caption{Spatial distribution of the eRASS1 clusters with \lext\, $\ge 6$ is shown for the PS1 (\textit{top left}), PS2 (\textit{right}), and SWAG-X (\textit{bottom left}) ASKAP fields. The data points are colour-coded according to their redshift, and their sizes are equal to their $R_{500}$ in arcminutes (see the $10\,\mathrm{arcmin}$ blue circle at the bottom right corner of each plot for a scale).}
\label{fig:allskydistr}
\end{figure*}
We identified 260 eRASS1 clusters detected across the three ASKAP fields in the German eROSITA sky. Applying a \lext\, cutoff of six, we can enhance the purity of the sample to 94\% \citep[see Table~1 in][]{Bulbul_2024}. This results in a final sample of 151 clusters, comprising 80 clusters in the PS1 field, 12 in the PS2 field, and 59 in the SWAG-X field, corresponding to detection rates of $0.30$, $0.32$, and $0.26\,\mathrm{cluster\,deg^{-2}}$, respectively. The variation in detection rates between fields is likely due to different survey depths. Longer exposure times (e.g., PS2 field) lead to a higher number of detected clusters per unit area. In contrast, shallower observations typically yield lower detection rates (see Table~\ref{tab:xray_prop}).
\par
The redshifts \citep[$z$; \texttt{BEST\_Z} in the catalogue, which is the best available cluster redshift, see][]{Kluge_2024} of the entire sample (hereafter, we refer to it as the eRASS1/ASKAP cluster sample) range from 0.033 to 1.126, with a median value of 0.238. The spatial distribution of the eRASS1/ASKAP cluster sample, colour-coded by redshifts, is displayed in Figure~\ref{fig:allskydistr}. The size of the data points is scaled according to the $R_{500}$ values of the clusters in arcminutes. For the X-ray centers of the clusters, we used those that are the results of the eRASS1 image fittings by the MultiBand Projector in 2D tool \citep[\texttt{MBProj2D},][]{Sanders_2018},\footnote{\url{https://github.com/ jeremysanders/mbproj2d}} listed in \texttt{RA\_XFIT} and \texttt{DEC\_XFIT} columns in the eRASS1 primary cluster catalogue. The BCG positions are taken from the eRASS1 optical cluster catalogue, listed in \texttt{RA\_BCG} and \texttt{DEC\_BCG} columns.

\subsubsection{X-ray properties}
\begin{figure*}[!ht]
\centering
\includegraphics[width=0.49\textwidth]{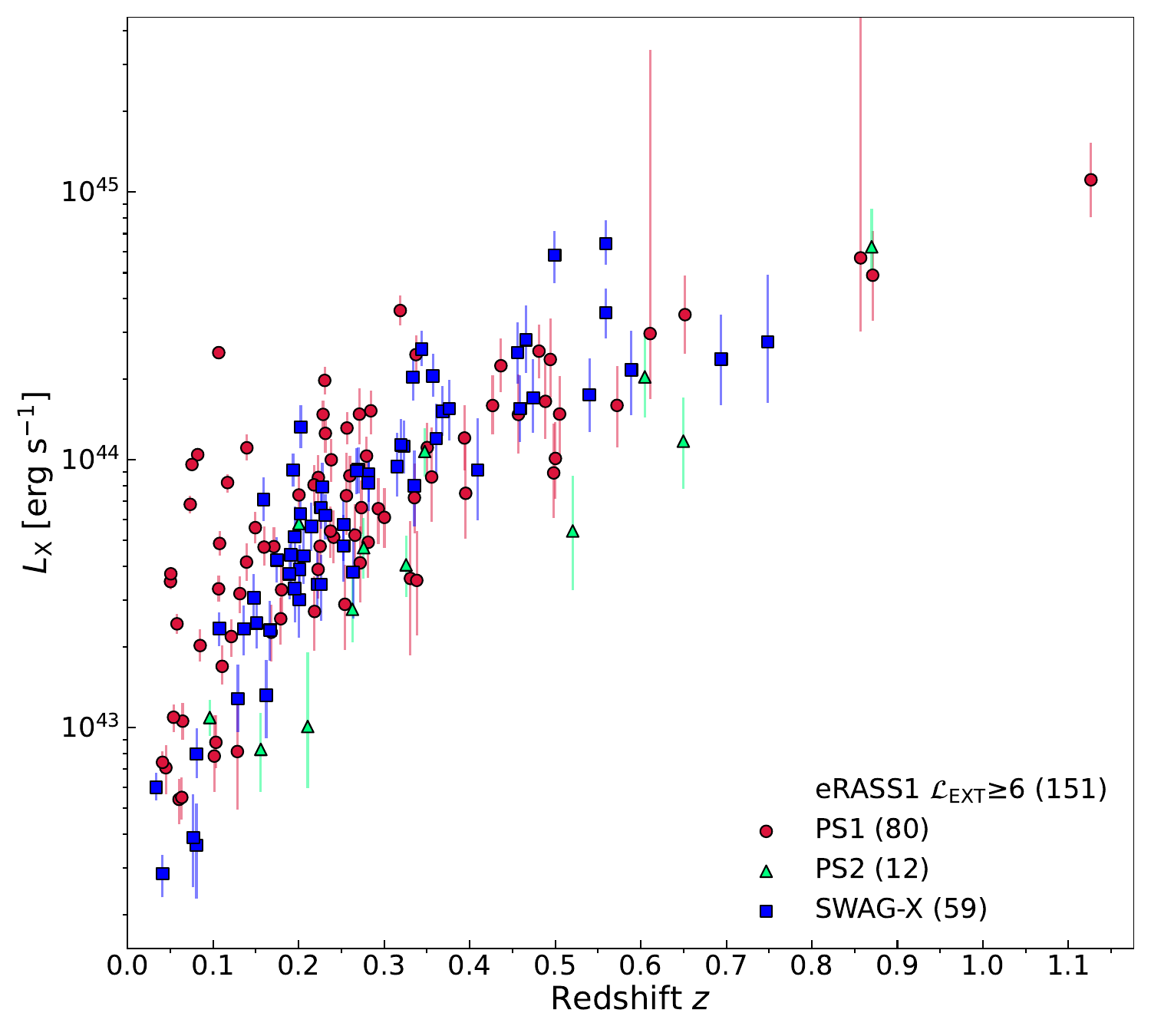}
\includegraphics[width=0.49\textwidth]{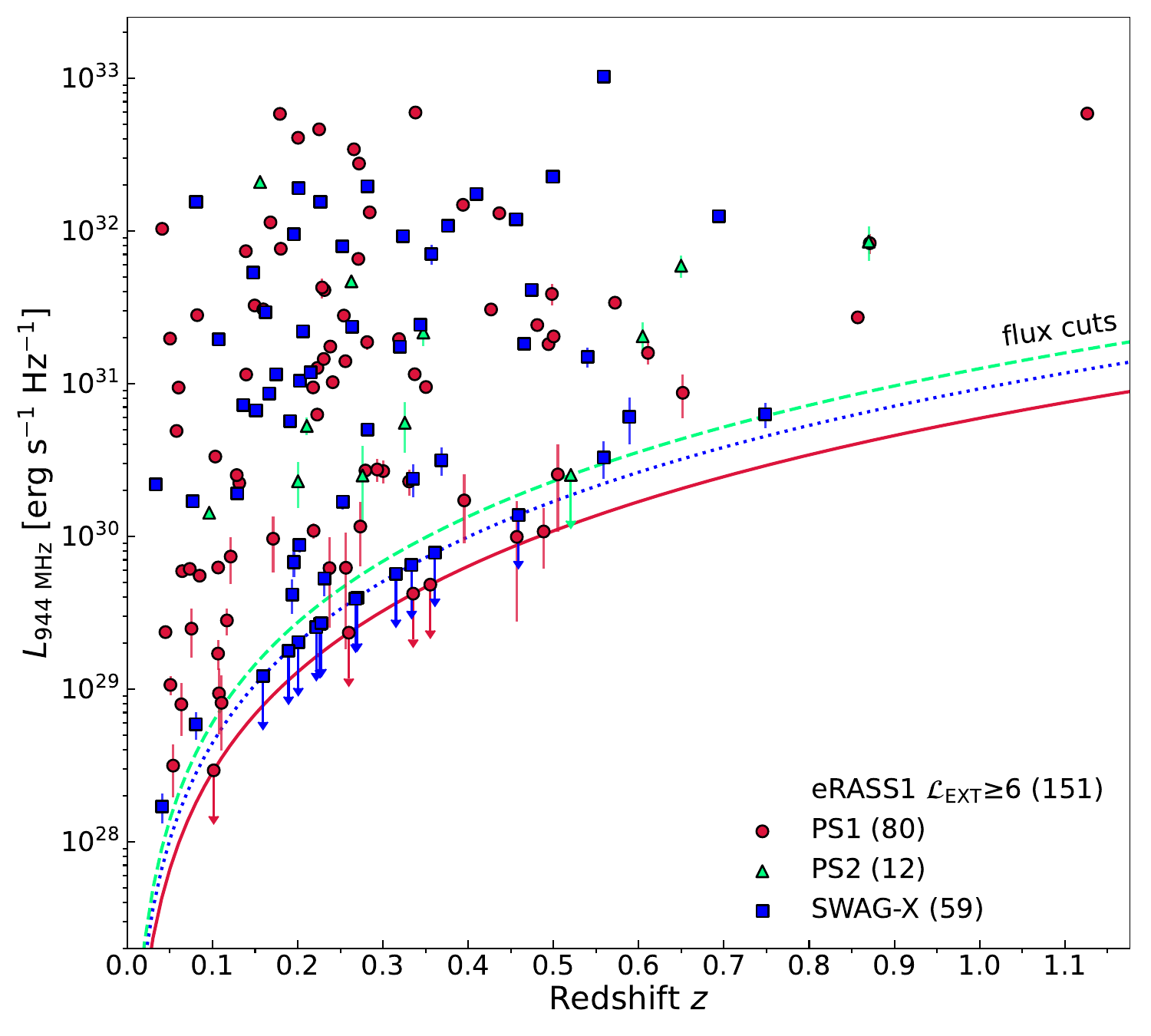}
\caption{X-ray and radio luminosities of the eRASS1/ASKAP cluster sample. In each plot, PS1, PS2, and SWAG-X subsamples are shown by red circles, green triangles, and blue squared, respectively. \textit{Left:} Integrated X-ray luminosity in the $0.2-2.3\,\mathrm{keV}$ band within 300\,kpc as a function of redshift. \textit{Right:} Radio luminosity at 944\,MHz against redshift. The downward arrows represent the upper limits. The solid/dotted/dashed lines in red/blue/green mark the flux limits in the PS1/SWAG-X/PS2 ASKAP fields.}
\label{fig:LX-z-M}
\end{figure*}

\begin{table*}
\centering
\caption{X-ray property ranges of the eRASS1/ASKAP cluster sample compiled from the eRASS1 primary group and cluster catalogue \citep{Bulbul_2024}. The "eRASS1 area" refers to the overlapping area between eRASS1 and the various ASKAP fields. The "median exp." indicates the median eRASS1 exposure in the corresponding fields, and $N$ represents the number of clusters. $L_\mathrm{X}$ is the X-ray luminosity measured in the $0.2-2.3$\,keV band within 300\,kpc from the cluster center.}
\resizebox{0.95\textwidth}{!}
{\begin{tabular}{c | c c c c c c c}
\hline
\hline
Sample & eRASS1 area & median exp. & $N$ & $z$ & $L_\mathrm{X}$ & $M_{500}$ & $R_{500}$\\
 & $[\mathrm{deg}^2]$ & [s] &  &  & $[10^{42}\,\mathrm{erg\,s^{-1}}]$ & $[10^{14}M_\odot]$ & [kpc]\\
\hline
All & 536 & - & 151 & $0.033-1.126$ & $2.72-1321.19$ & $0.39-11.04$ & $ 503-1396$\\
PS1 & 270 & 93 & 80 & $0.041-1.126$ & $5.39-1321.19$ & $0.58-9.10$ & $569-1396$\\
PS2 & 38 & 165 & 12 & $0.096-0.870$ & $6.61-522.55$ & $0.59-5.35$ & $563-1028$\\
SWAG-X & 228 & 79 & 59 & $0.033-0.748$ & $2.72-715.18$ & $0.39-11.04$ & $503-1313$\\
\hline
\hline
\end{tabular}}
\label{tab:xray_prop}
\end{table*}
We compiled the X-ray properties from the eRASS1 primary cluster catalogue, including the X-ray luminosity in the $0.2-2.3\,\mathrm{keV}$ band within 300\,kpc ($L_\mathrm{X}$; \texttt{L300kpc} in the catalogue), the characteristic radius ($R_{500}$; \texttt{R500}), and mass within $R_{500}$ ($M_{500}$; \texttt{M500}). We note that 300\,kpc is on average about $0.34R_{500}$ in our sample. The left panel of Figure~\ref{fig:LX-z-M} illustrates the $L_\mathrm{X}$ distribution as a function of the redshift. The $L_\mathrm{X}$ ranges between $2.72\times10^{42}$ and $1.32\times10^{45}\,\mathrm{erg\,s^{-1}}$. In the plot, clusters from various ASKAP fields are represented with different markers: eRASS1 clusters in PS1 are shown as red circles, clusters in PS2 as green triangles, and clusters in SWAG-X as blue squares. This colour and marker convention will be consistently used throughout the different plots in this work to distinguish between eRASS1 clusters from the various ASKAP fields. Compared to the clusters in the PS1 and SWAG-X fields, those in the PS2 field generally show lower $L_\mathrm{X}$ values and higher $z$. We attribute this trend to the longer median exposure time in the PS2 field, which allows for the detection of less luminous and more distant clusters (see Table~\ref{tab:xray_prop}).
The $M_{500}$ range between $3.9\times10^{13}M_\odot$ and $1.1\times10^{15}M_\odot$, with a median value of $3.03\times10^{14}M_\odot$ (see top panel of Figure~\ref{fig:z-M}). We summarise the X-ray properties of the sample in Table~\ref{tab:xray_prop} and show the distributions of $M_{500}$, $z$, and $R_{500}$ in \ref{App:A}, Figure~\ref{fig:z-M}.
\par

We also obtained the concentration parameter from the eRASS1 cluster morphology catalogue \citep{Sanders_2025}. The concentration parameter is defined as the ratio of surface brightness between the core aperture and a larger ambient aperture \citep{Santos_2008}. This parameter indicates how centrally peaked a cluster is and is, therefore, sensitive to the presence of a cool core. We selected the measurement of the concentration parameter expressed as a fraction of $R_{500}$, that is
\begin{equation}
    c_{R_{500}} = \frac{\mathrm{SB}_{<0.1R_{500}}}{\mathrm{SB}_{<R_{500}}},
\label{eq:cR500}
\end{equation}
where the apertures are centred around the cluster peak centre (\texttt{C\_R500\_P}). We note that in \cite{Sanders_2025}, $c_{R_{500}}$ is listed in logarithmic form, and we converted them to linear form.

\subsubsection{Radio properties}\label{sec:radio_prop}
\defcitealias{Boeckmann_2023}{B23}
First, we cross-matched the clusters in the PS1 field with the radio cluster property catalogue compiled by \citet[][hereafter B23]{Boeckmann_2023}. These authors measured the properties of the central radio source of 75 clusters in the PS1 field, which were identified using the eRASS1 preliminary cluster catalogue. The BCGs associated with each cluster in \citetalias{Boeckmann_2023} were visually identified using WISE and legacy optical data. We cross-matched the positions of our eRASS1 BCGs from the PS1 sub-sample with the radio centres listed in \citetalias{Boeckmann_2023}. The search radius for this match was set to the size of the synthesised radio beam of the survey of $18''$ (see Table~\ref{tab:askap}). In total, we matched 32 clusters. Similarly, for the SWAG-X subsample, we cross-matched the BCG positions with the source list generated by \texttt{PyBDSF} (see Section~\ref{sect:askap}), identifying the corresponding radio sources for 25 BCGs. We note that no cross-matching with any published catalog was performed for the PS2 and SWAG-X fields.
\par
For the non-match clusters, we manually inspected and measured their radio properties following the steps in \citetalias{Boeckmann_2023}. We identified the central radio sources as the emission within a distance $\theta$ from the positions of the BCGs, where $\theta$ corresponds to the major axis of the synthesised radio beam size (see Table~\ref{tab:askap}). The radio flux at a central frequency $\nu$ of a source, $S_\nu$, is measured within an aperture where the emission is above three times the RMS noise ($3\sigma$) of the image. We took the major axis of the $3\sigma$ isophote as the largest angular size (LAS) of the source, which is then converted to the largest linear size (LLS) using the cluster redshift associated with the source and the assumed cosmology. The minimum LAS from which the flux is measured is limited by the beam size, and any central radio sources smaller than this threshold are considered unresolved. The radio luminosity is subsequently calculated from the radio flux using

\begin{equation}
    L_\nu = 4\pi D_\mathrm{L}^2 S_\nu (1+z)^{\alpha-1},
\label{eq:Lradio}
\end{equation}
where $D_\mathrm{L}$ is the luminosity distance at redshift $z$, and $\alpha$ is the spectral index, taken to be 1.0, which is the average spectral index of the cluster centre radio sources in the HIghest X-ray FLUx Galaxy Cluster Sample \citep[\textit{HIFLUGCS},][]{Reiprich_2002} estimated in \cite{Mittal_2009}. Notably, adopting $\alpha=0.8$ or 1.2 does not impact the findings and conclusions presented below. To compare the radio measurements, a conversion of the luminosity into one central frequency is required, for instance, for SWAG-X from 888\,MHz to 944\,MHz. The calculation is done using the equation below
\begin{equation}
    L_{944\,\mathrm{MHz}}= L_{888\,\mathrm{MHz}} \times \left(\frac{944\,\mathrm{MHz}}{888\,\mathrm{MHz}} \right)^{-\alpha},
\label{eq:lr_conv}
\end{equation}
with the same spectral index $\alpha$ as in Equation~\ref{eq:Lradio}. For clusters without radio source counterparts, we categorise them as non-detections and assign an upper limit of $3\sigma$: 0.105\,mJy, 0.222\,mJy, and 0.174\,mJy for the PS1, PS2, and SWAG-X surveys, respectively. In our sample, all sources with radio measurements are resolved, while those with upper limits are deemed unresolved, and thus, they are assigned the beam size as their LAS. In total, we identified 4/1/12 non-detected sources in the PS1/PS2/SWAG-X fields, respectively. This gives us a high detection rate of $89\%$, which can be attributed to the depth of the surveys that results in lower RMS ($\leq74\,\upmu\mathrm{Jy}$). Assuming the detected sources are uniformly distributed within the survey field, the probability of having at least one radio source within a distance $\theta$ from a specific point (the BCG position) can be estimated by
\begin{equation}
    P(\geq1) = 1 - \mathrm{e}^{-\rho\pi\theta^2},
\label{eq:p_random}
\end{equation}
where $\rho$ represents the number density of the detected radio sources and $\theta$ is the beam size. Given that a complete source list for the PS2 survey is not yet available, we estimate the value for the PS1 and SWAG-X surveys, where $\rho = 330\,000/498 \approx 663\,\mathrm{source\,deg^{-2}}$, and $14.2'' \leq \theta \leq 18.0''$, resulting in contamination fraction between 3\% and 5\%. Therefore, out of 134 clusters, we expect four to seven false detections.
\par
The distribution of radio luminosity at 944\,MHz as a function of redshift is illustrated in the right panel of Figure~\ref{fig:LX-z-M}. The flux limits for the PS1, PS2, and SWAG-X surveys are represented by solid, dashed, and dotted lines, respectively. The range of radio luminosities covered spans from $1.70\times10^{28}$ to $1.02\times10^{33}\,\mathrm{erg\,s^{-1}\,Hz^{-1}}$, with a median of $9.40\times10^{30}\,\mathrm{erg\,s^{-1}\,Hz^{-1}}$. The radio properties of the BCGs are listed in Table~\ref{tab:long_tabl} in \ref{App:A}.

\section{Results}\label{sect:results}
\subsection{Largest linear size of the BCGs}\label{sect:lls}
\begin{figure}
\centering
\includegraphics[width=\columnwidth]{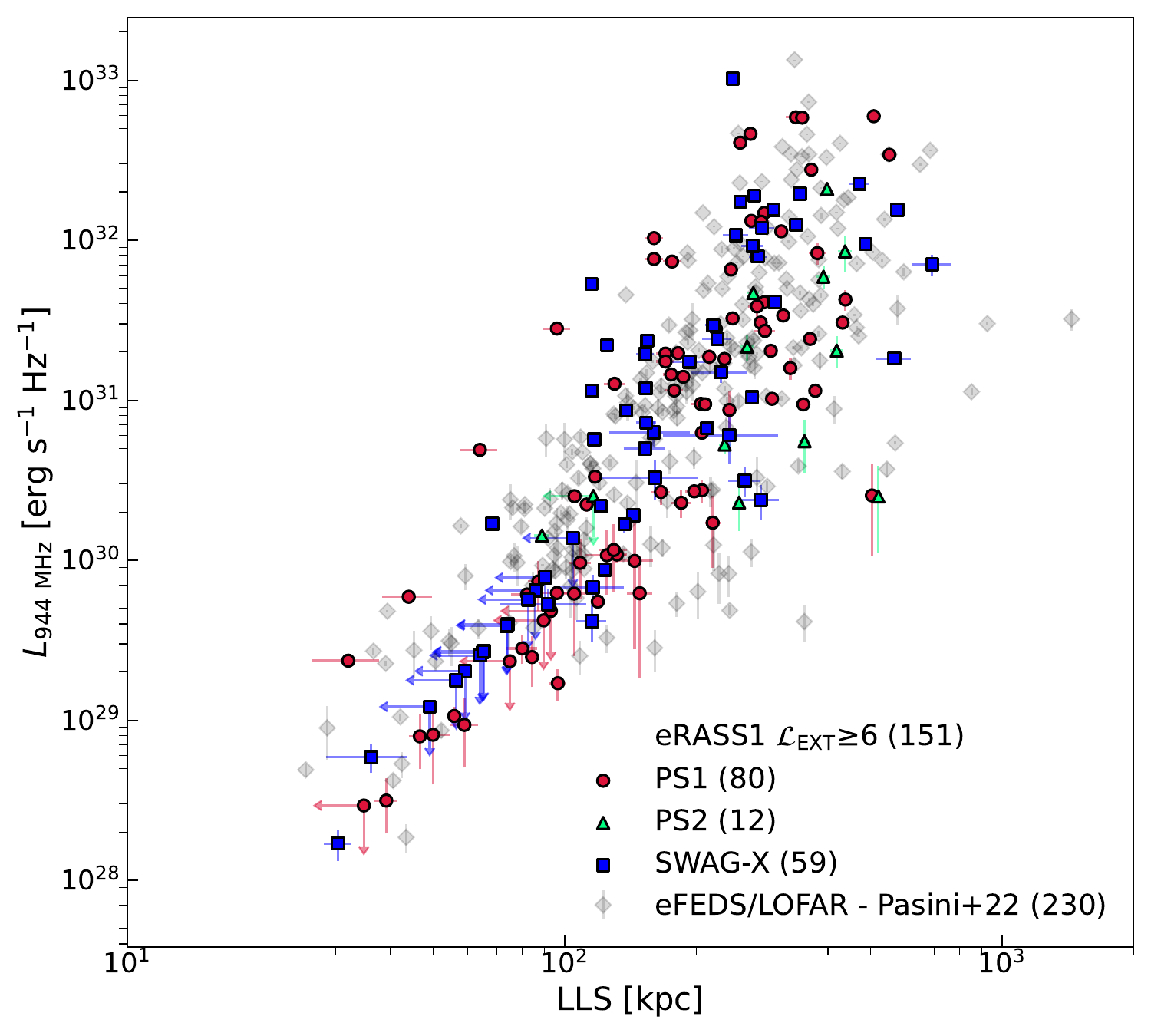}
\caption{Central radio luminosity at 944\,MHz versus largest linear size ($P-D$ diagram). The gray diamonds are the group and cluster central radio sources in the eFEDS field measured by LOFAR at 144\,MHz \citep{Pasini_2022}. The rescaling to 944\,MHz luminosity is done by adopting $\alpha=1.0$.}
\label{fig:LR-z}
\end{figure}

We present the plot of radio luminosity at 944\,MHz versus the LLS, also known as the $P-D$ diagram, in Figure \ref{fig:LR-z}. Similar to the Hertzsprung-Russell diagram for stars, the position of a source in the $P-D$ diagram indicates its initial conditions and evolutionary states \citep{Baldwin_1982, Blundell_1999}.
\par
The LLS of the eRASS1/ASKAP sample ranges from 30 kpc to 692\,kpc, with an average value of 210\,kpc. In Figure~\ref{fig:LR-z} we also included the eFEDS/LOFAR central radio sources \citep[]{Pasini_2022} as grey diamonds. The LOFAR radio luminosity at 144\,MHz, $L_{144\,\mathrm{MHz}}$, was converted to $L_{944\,\mathrm{MHz}}$ using Equation~\ref{eq:lr_conv}. Both samples span the same luminosity range. The LLS values for the eFEDS/LOFAR sample span a similar range, from approximately 10\,kpc to 1500\,kpc, with a mean value of 228\,kpc. We note that the apparent positive correlation is known to be driven by a selection effect against large, low-luminosity sources (i.e., those in the bottom right region of the plot) due to surface brightness limitations \citep{Shabala_2008, Hardcastle_2016, Hardcastle_2019}.

\subsection{BCG offsets}\label{sect:bcgoffset}
\begin{figure*}
\centering
\includegraphics[width=0.49\columnwidth]{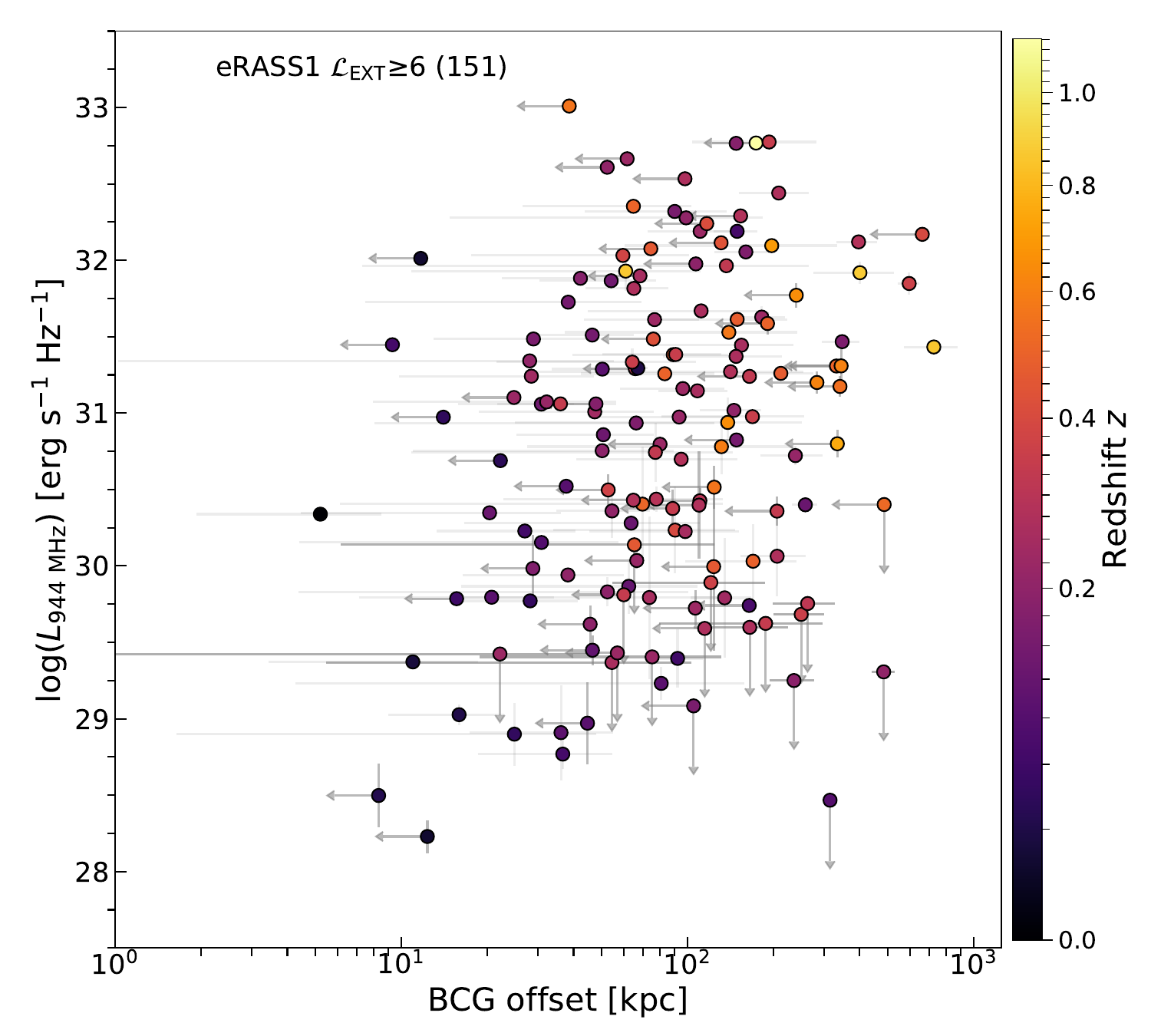}
\includegraphics[width=0.49\columnwidth]{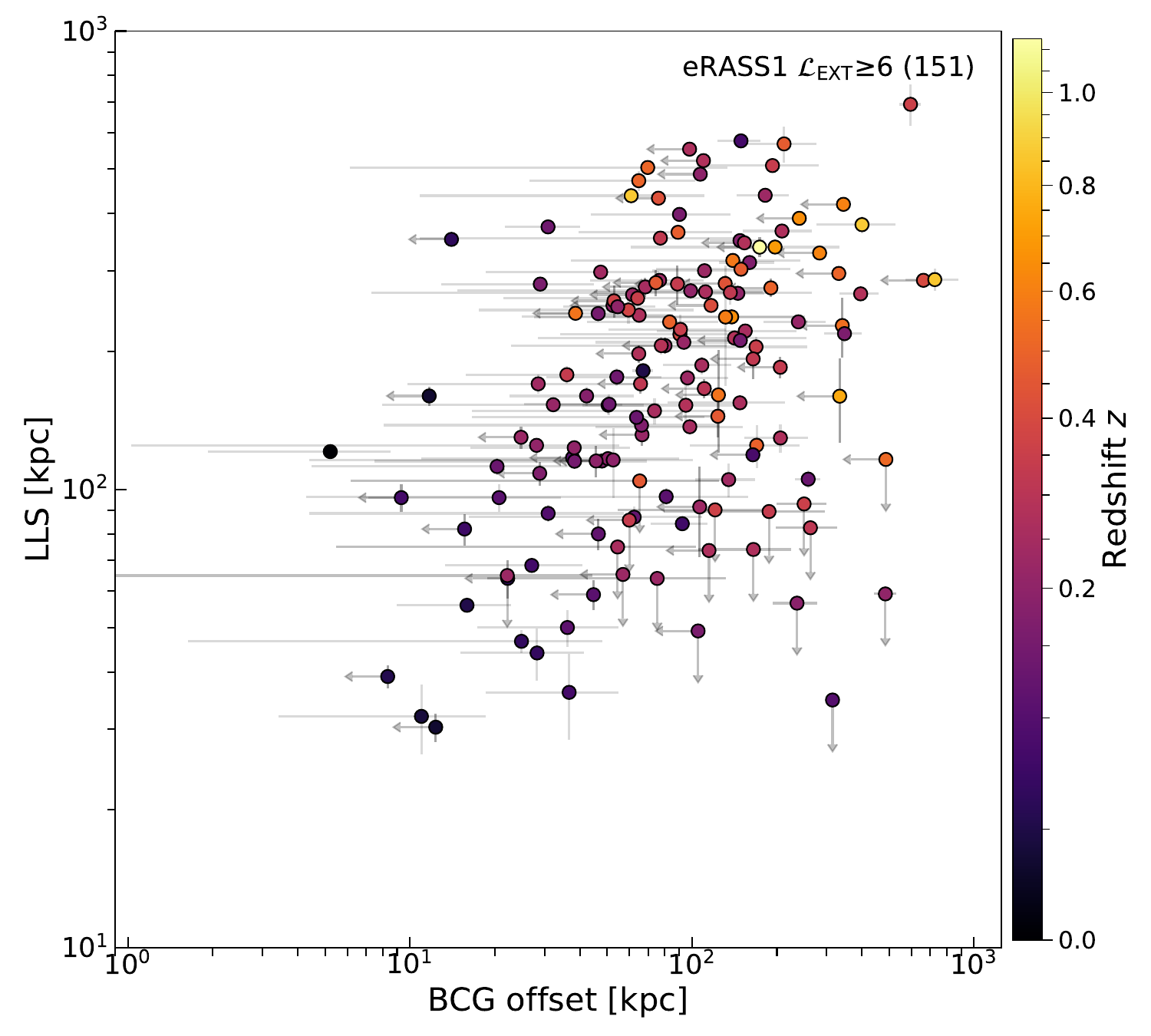}
\caption{Physical separation of the BCGs from the X-ray centres in units of kpc. \textit{Left:} Radio luminosity of the BCGs at 944\,MHz versus BCG offsets. \textit{Right:} Largest linear size of the BCGs versus BCG offsets. In both plots, the data points are colour-coded by redshift $z$, and the arrows indicate upper limits.}
\label{fig:bcg_offset}
\end{figure*}

\begin{figure*}
\centering
\includegraphics[width=0.49\columnwidth]{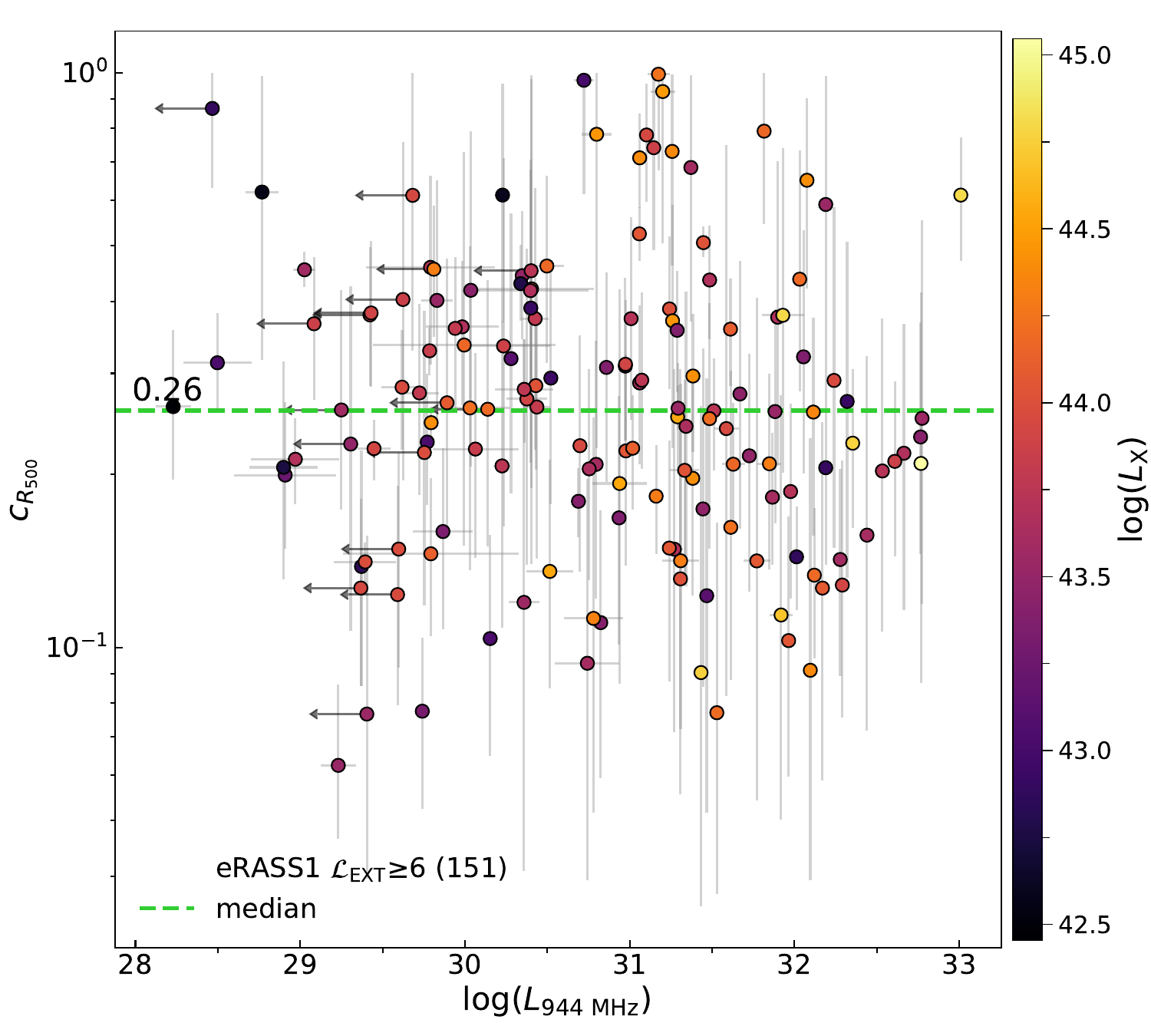}
\includegraphics[width=0.49\columnwidth]{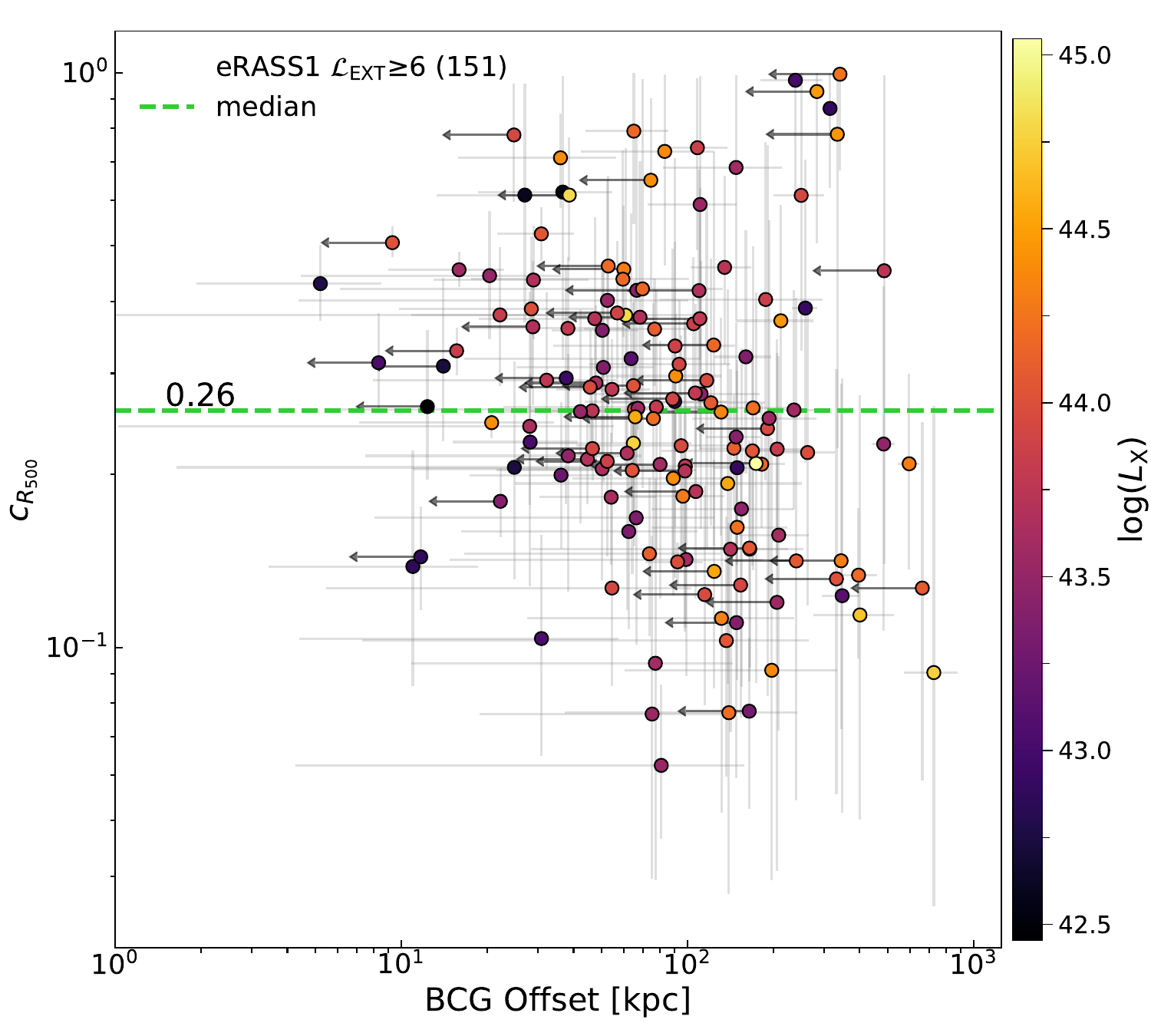}
\caption{Concentration parameter of the eRASS1/ASKAP clusters. \textit{Left:} Concentration parameter ($c_{R_{500}}$) against the 944\,MHz radio luminosity ($L_\mathrm{R}$). \textit{Right:} Concentration parameter as a function of BCG offset. The green dashed horizontal line in each plot indicates the median $c_{R_{500}}$ value of the sample.}
\label{fig:w-c}
\end{figure*}

BCGs typically reside at the bottom of a cluster's gravitational potential well. Therefore, measuring the spatial offset between the BCG and the cluster centre provides an initial indication of the cluster's dynamical state.
\par
We calculated the BCG offsets using the BCG position and the X-ray centre of the eRASS1 primary cluster catalogues (see Section~\ref{sect:sample}). Then, we performed visual inspections on the corrected image of eRASS1 in the $0.2-2.3\,\mathrm{keV}$ band, the DESI Legacy Survey DR10 image and the ASKAP radio image of each cluster to ensure the correct position of the BCG (see two examples in \ref{App:A}, Figures~\ref{fig:example_cl1} and \ref{fig:example_cl2}), as well as the information of the galaxy cluster members provided in \texttt{eRASS Cluster Inspector}.\footnote{\url{https://erass-cluster-inspector.com/}} 
In total, we identified 30 clusters (20\% of the sample) that were assigned incorrect BCGs due to, e.g., undetected BCGs or contamination by bright sources (e.g., stars). For these clusters, we reassigned the BCG to the visually largest and brightest galaxy member located closest to the X-ray peak.
Furthermore, we utilised the X-ray combined positional uncertainties obtained from PSF fitting (\texttt{RADEC\_ERR} in the catalogue) as estimates for the uncertainties of the BCG offset. If the X-ray positional uncertainty exceeds the measured BCG offset, we assigned the offset as an upper limit. In total, there are 54 clusters with BCG offset upper limits. We list the BCG position of our sample in Table~\ref{tab:long_tabl} in \ref{App:A}.
\par
The distribution of the BCG offsets of the eRASS1/ASKAP cluster sample is presented in \ref{App:A}, Figure~\ref{fig:bcg_offset_hist} (black bar). The median value of the BCG offset of the sample is 88.8\,kpc (yellow dashed line) with a dispersion of 121.8\,kpc. There are 59 clusters (39\%) with BCG offsets exceeding the median value and the PSF cut of $15''$. We do not observe any particular trend of the BCG offsets among the different ASKAP fields (coloured bars). While we note that the presented BCG offset distribution includes upper limits, the yielded median value is comparable to the average BCG offset found in a subsample of eFEDS clusters with high counts ($76.3_{-27.1}^{+30.1}\,\mathrm{kpc}$), as well as in TNG300 (57.2\,kpc) and Magneticum Box2/hr (87.1\,kpc) simulations \citep{Seppi_2023}.
\par
Smaller offsets ($\lesssim70\,\mathrm{kpc}$) are commonly observed in cool core clusters, where BCGs often exhibit AGN activity \citep{Edwards_2007, Mittal_2009} and strong radio emissions \citep{Pasini_2019, PasiniA1668_2021}. This aligns with the AGN feedback scenario, in which the accretion of cool gas onto the central supermassive black hole (SMBH) fuels AGN activity. To investigate this, we plot the 944\,MHz radio luminosity of the BCG ($L_\mathrm{R}$) against the BCG offset (left panel of Figure~\ref{fig:bcg_offset}). The data points are colour-coded to the cluster redshift $z$. Downward arrows indicate radio upper limits (non-detected sources), while leftward arrows denote upper limits for the BCG offsets. There is no clear correlation between $L_\mathrm{R}$ and BCG offsets. Meanwhile, the redshift distribution of the data points shows selection effects.
In the right panel of Figure~\ref{fig:bcg_offset}, we display the position of the clusters in the LLS versus BCG offset diagram.
There appears to be a positive trend, which is difficult to quantify due to upper limits in both directions. Nevertheless, the significance of the correlation can be assessed by, for example, the generalised Kendall's $\tau$ test \citep{kendall}, a non-parametric hypothesis test to evaluate the ordinal association. The resulting null-hypothesis probability ($p$-value) of 0.0008 indicates a statistically significant correlation between the two variables, such that BCGs with a larger radio extent are farther away from the X-ray centre.
This is consistent with previous findings, where sources in denser environments (i.e., closer to the cluster centres) are observed to be less extended than those in a diluted environment \citep{Turner_2015, Yates_2021}. However, similar to the $L_\mathrm{R}-\mathrm{BCG~offset}$ plot, the colour-coded redshift distribution of the $\mathrm{LLS}-\mathrm{BCG~offset}$ plot suggests selection effects, i.e., at high redshift, only massive clusters are detected and they typically host more powerful radio sources and have larger BCG offsets.

\subsection{Dynamical states from morphological parameters}\label{sec:dynamic_state}

The morphological parameters derived from X-ray data are a powerful tool for examining the core properties and dynamical states of clusters. In the left panel of Figure~\ref{fig:w-c}, we show the positions of the eRASS1/ASKAP clusters in the $c_{R_{500}}-L_\mathrm{R}$ diagram. The data points are colour-coded by $L_\mathrm{X}$. The median value of the $c_{R_{500}}$ of our sample (0.26) is indicated by the horizontal green dashed lines, and by using this median value as a threshold, 75 clusters are classified as cool cores (CCs) and 76 as non-cool cores (NCCs). As apparent from the plot, we do not observe any obvious trend between $c_{R_{500}}$ and $L_\mathrm{R}$ (see Section~\ref{sect:lx-lr}).
\par
Additionally, we present the concentration parameter $c_{R_{500}}$ as a function of BCG offset in the right panel of Figure~\ref{fig:w-c}, where the data points are colour-coded by $L_{X}$.
We observed a flat negative correlation with a $p$-value from the generalised Kendall's $\tau$ test (see Section~\ref{sect:bcgoffset}) of 0.0055, indicating a statistically significant correlation: clusters with higher $c_{R_{500}}$ values tend to have smaller BCG offsets, while those with lower $c_{R_{500}}$ exhibit larger BCG offsets. This anti-correlation is expected, as a large BCG offset indicates a dynamically disturbed system, which is less likely to host a strongly peaked, cool-core structure \citep[e.g.,][]{Hudson_2010}. Utilizing the median values of $c_{R_{500}}$ and BCG offset as the dynamical state thresholds, we classified 45 clusters ($\sim\!30\%$) as relaxed, 46 ($\sim\!30\%$) as disturbed, and 60 ($\sim\!40\%$) as in intermediate state (clusters with a smaller BCG offset and lower concentration than the medians, as well as clusters with a higher concentration but a larger BCG offset than the medians).
We recall that there are 54 clusters with BCG offset upper limits (see Section~\ref{sect:bcgoffset}) that can significantly affect the results of this $c_{R_{500}}-\mathrm{BCG~offset}$ classification. Additionally, we do not observe any obvious trend between $c_{R_{500}}$, BCG offset, and $L_{X}$. \cite{Lovisari_2017} analysed the X-ray morphological parameters of the Planck Early Sunyaev–Zeldovich (ESZ) clusters observed with \textit{XMM-Newton}. They investigated combinations of eight parameters sensitive to the presence of substructures, which indicate how active a system is, as well as those sensitive to core properties to assess relaxation states. Among them, they found that the centroid shift and concentration parameter are the most effective for identifying relaxed systems. However, due to the low counts, parameters such as centroid shift and power ratios cannot reliably be determined \citep{Sanders_2025}.

\subsection{Radio and X-ray luminosity correlation}\label{sect:lx-lr}
\begin{figure}
\centering
\includegraphics[width=\columnwidth]{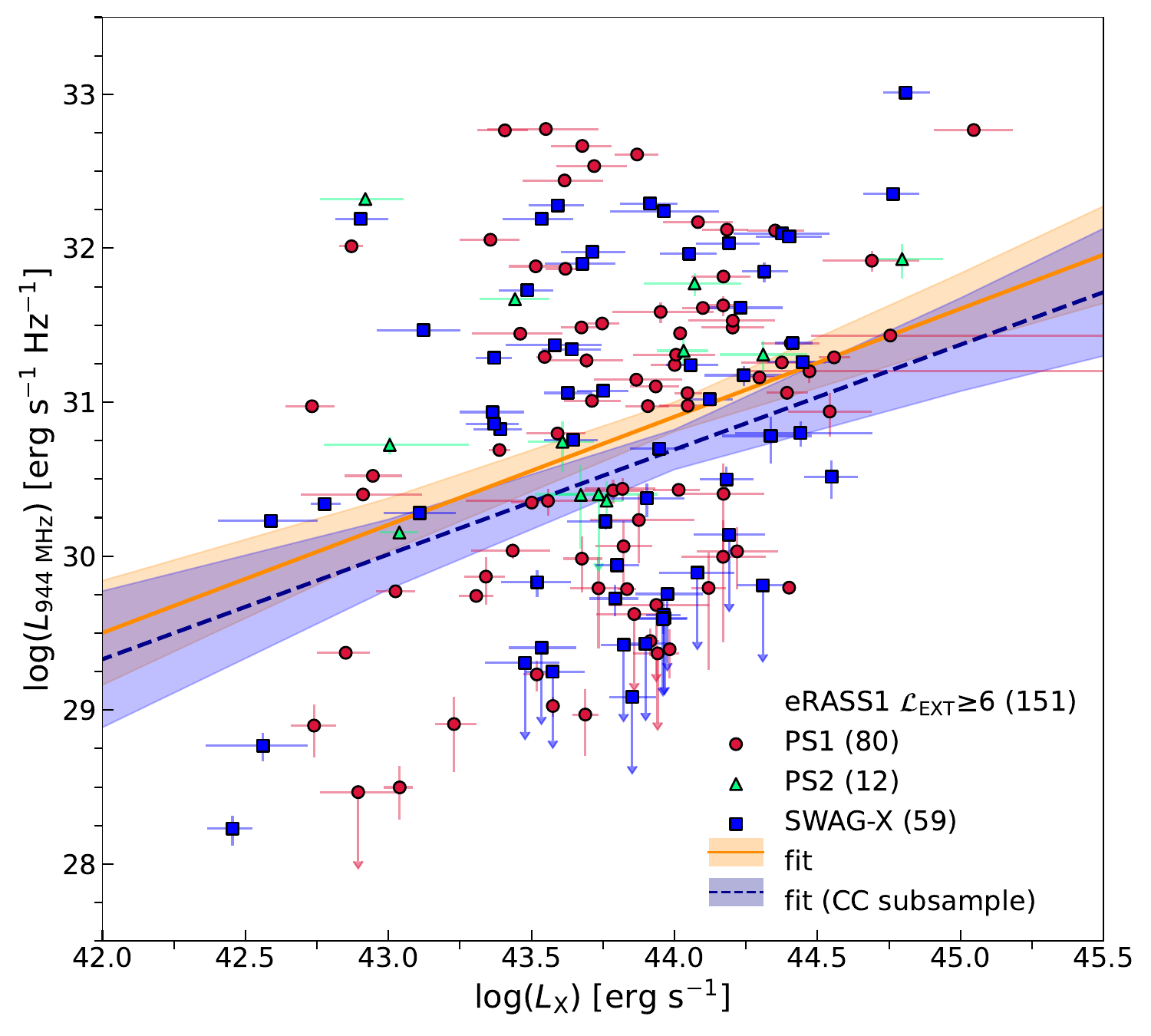}
\caption{Radio luminosity of the BCGs at 944\,MHz against the X-ray luminosity of the host clusters. The dark orange solid line and shaded area are the $\log L_\mathrm{R} - \log L_\mathrm{X}$ relation and its $1\sigma$ confidence band from the entire sample, while the blue dashed line and shaded area are from the CC subsample. The parameters of the correlation are listed in Table~\ref{tab:lr-lx}.}
\label{fig:lr-lx}
\end{figure}

We display the 944\,MHz radio luminosity of the BCGs against the X-ray luminosity of the host clusters in Figure~\ref{fig:lr-lx} (in log-log space). The data points with downward arrows are the upper limits assigned to the 17 non-detected clusters. As also found in \cite{Kolokythas_2018}, these faint radio sources are found only in faint X-ray hosts.
\par
Furthermore, despite the large scatter, there seems to be a trend suggesting that more luminous central radio galaxies are found in clusters with greater X-ray luminosity. To quantify the correlation between these two variables, we performed a linear regression fit using the parametric EM algorithm regression in the Astronomy SURVival analysis package \citep[\texttt{ASURV,}][]{Isobe_1986, asurv}.\footnote{\url{https://github.com/rsnemmen/asurv}} We also employed the Bayesian inference approach from \cite{linmix} using the \texttt{Python} implementation, \texttt{LinMix} package.\footnote{\url{https://linmix.readthedocs.io/en/latest/}}
We find that the results obtained from both methods are consistent with each other within $1\sigma$ of their uncertainties. We adopt the results from \texttt{ASURV} as our default for comparison with previous studies, while employing \texttt{LinMix} as an independent Bayesian cross-check.
The fit is performed in log-log space. For the $\log Y - \log X$, the relation is in the following form
\begin{equation}
    \log \left(\frac{Y}{Y_\mathrm{piv}} \right) = A\cdot\log\left( \frac{X}{X_\mathrm{piv}} \right) + B,
\label{eq:relation}
\end{equation}
where $A$ and $B$ are the slope and the intercept of the relation, respectively, while $Y_\mathrm{piv}$ and $X_\mathrm{piv}$ are the pivot points, which are the median values of the variables.
We show the linear fit and its $68\%$ confidence interval as the orange solid and shaded area, respectively. The best-fit parameters are listed in Table~\ref{tab:lr-lx}.
\par
The $p$-value from the generalised Kendall's $\tau$ test is 0.0018, suggesting that the correlation is statistically significant, that is, it is unlikely that the association is due to random fluctuations. The $\log L_\mathrm{R} - \log L_\mathrm{X}$ slope of the eRASS1/ASKAP sample, $A=0.70\pm0.18$, is in good agreement within the $1\sigma$ uncertainty with other works. For instance, \citetalias{Boeckmann_2023} with the eRASS1/PS1 sample of 75 clusters found a slope of $0.89\pm0.04$, \cite{Pasini_2022} with the eFEDS/LOFAR sample of 542 groups and clusters (including upper limits) found $0.84\pm0.09$, as well as \cite{Pasini_2021} with the VLA-COSMOS sample of 79 groups found $0.94\pm0.43$.
\par
\cite{Mittal_2009} also reported a trend between $L_\mathrm{R}$ and $L_\mathrm{X}$ for the strong cool core (SCC) clusters in the \textit{HIFLUGCS} sample. They found a slope of $1.38\pm0.16$.
From our CC subsample, we determined a slope of $0.68\pm0.23$, which is consistent within $2.5\sigma$ uncertainties with the slope reported by \cite{Mittal_2009}.

\begin{table*}
\centering
\caption{Radio and X-ray luminosity correlations of the eRASS1/ASKAP cluster sample. The relation is formulated in Equation~\ref{eq:relation}.}
\resizebox{0.95\textwidth}{!}
{\begin{tabular}{c | c c c c c c}
\hline
\hline
Correlation$^\star$ & $A$ & $B$ & $Y_\mathrm{piv}$ & $X_\mathrm{piv}$ & $\sigma^\ddagger$ & $p^\dagger$\\
\hline
$\log L_\mathrm{R} - \log L_\mathrm{X}$ (all) & $0.70\pm0.18$ & $-0.19\pm0.09$ & $9.40\times10^{30}\,\mathrm{erg\,s^{-1}\,Hz^{-1}}$ & $6.64\times10^{43}\,\mathrm{erg\,s^{-1}}$ & 1.0956 & 0.0018\\
$\log L_\mathrm{R} - \log L_\mathrm{X}$ (CC) & $0.68\pm0.23$ & $0.13\pm0.12$ & $2.73\times10^{30}\,\mathrm{erg\,s^{-1}\,Hz^{-1}}$ & $6.58\times10^{43}\,\mathrm{erg\,s^{-1}}$ & 1.0322 & 0.0332\\
$\log L_\mathrm{mech,HB+14} - \log L_{\mathrm{X},~r<R_{cool}}$ (CC) & $0.59\pm0.20$ & $0.11\pm0.11$ & $7.49\times10^{42}\,\mathrm{erg\,s^{-1}}$ & $5.59\times10^{43}\,\mathrm{erg\,s^{-1}}$ & 0.8877 & 0.0336 \\[0.5ex]
\hline
\multicolumn{7}{l}{$^\star$Calculated using \texttt{ASURV} package \citep{asurv}, $\ddagger$scatter, $^\dagger$generalised Kendall's $\tau$ null-hypothesis probability.}\\
\hline
\hline
\end{tabular}}
\label{tab:lr-lx}
\end{table*}

\subsection{Mechanical power of the radio jets}\label{sect:lkin}
Radio luminosity measures the synchrotron radiation from the central radio sources, which contributes only a small fraction of the energy outflows produced by the central SMBH. Most of the mechanical (kinetic) AGN jet power is actually stored in the lobes and/or deposited into the ICM during the expansion of the radio sources \citep{Scheuer_1974}. Although it is possible to directly estimate the mechanical jet power, this requires multi-frequency radio data covering the radio-emitting region, which also includes resolving some areas of the radio lobes \citep[e.g.,][]{ODea_2009}. Nonetheless, numerous studies have shown that the mechanical jet power scales well with the radio luminosity \citep[e.g.,][]{Willott_1999, Birzan_2004, Birzan_2008, Daly_2012}.
As expressed in \cite{Heckman_2014}, the formula in terms of 1.4\,GHz radio luminosity, $L_\mathrm{1.4\,GHz}$, is

\begin{equation}
L_\mathrm{mech,HB+14} = 4\times10^{42}\,\mathrm{erg\,s^{-1}}\cdot f_W^{3/2}\cdot\left(\frac{L_\mathrm{1.4\,GHz}}{10^{32}\,\mathrm{erg\,s^{-1}\,Hz^{-1}}}\right)^{0.86},
\label{eq:Pmech}
\end{equation}
where $f_W$ is a factor to represent all the uncertainties due to limited understanding of the physics of radio sources, such as the composition of the radio jet plasma and low energy cutoff of the electron energy distribution. The mechanical AGN  power in radio lobes can also be measured directly through X-ray cavities \citep[e.g.,][]{Boehringer_1993, McNamara_2000}. Although finding the cavities requires deep and spatially resolved observations, those that were observed had been employed to infer the mechanical power of the AGN \citep[e.g.,][]{Birzan_2004} by correlating their power with the radio luminosity. The uncertainty arises from estimating the cavity energy, $E_\mathrm{cav}=f_\mathrm{cav}pV$, where $p$ is the pressure of the surrounding medium and $V$ is the cavity volume. \cite{Cavagnolo_2010} investigated the relationship between the mechanical jet power and radio luminosity using \textit{Chandra} X-ray and Very Large Array radio data, which is given as

\begin{equation}
P_\mathrm{cav} = 7\times10^{43}\,\mathrm{erg\,s^{-1}}\cdot f_\mathrm{cav}\cdot\left(\frac{L_\mathrm{1.4\,GHz}}{10^{32}\,\mathrm{erg\,s^{-1}\,Hz^{-1}}}\right)^{0.68}.
\label{eq:Pcav}
\end{equation}
Typically, $f_\mathrm{cav}=4$ is adopted, such that $4pV$ is the enthalpy of a cavity filled with relativistic plasma. This value is also consistent with a general balance between AGN heating and radiative cooling in massive clusters. For $f_\mathrm{cav}=4$ and $f_W=15$, the normalizations of Equation~\ref{eq:Pmech} and \ref{eq:Pcav} agree with a typical source of $L_\mathrm{1.4\,GHz}\sim10^{25}\,\mathrm{W\,Hz^{-1}}\approx10^{32}\,\mathrm{erg\,s^{-1}\,Hz^{-1}}$ \citep{Heckman_2014}. This factor is also in good agreement with observational findings \citep[e.g.,][]{Merloni_2007}. With this, we used the final form of Equation~\ref{eq:Pmech} to calculate the radio mechanical luminosity, which is

\begin{equation}
    L_\mathrm{mech,HB+14} = 4\times10^{42}\,\mathrm{erg\,s^{-1}}\cdot 15^{3/2}\cdot\left(\frac{L_\mathrm{1.4\,GHz}}{10^{32}\,\mathrm{erg\,s^{-1}\,Hz^{-1}}}\right)^{0.86}.
\label{eq:Lkin}
\end{equation}
We computed the radio luminosities of the sample (rescaled to 1.4\,GHz from their observed central frequency using Equation~\ref{eq:lr_conv}) into this equation. 
\par
Additionally, we also utilised the mechanical luminosity determined by \cite{Shabala_2013} (their Equation~8), which accounts for the size of the source, serving as a proxy for the age of the radio source. The equation is given as

\begin{align}
    L_\mathrm{mech,SG+13} 
    &= 10^{43}\,\mathrm{erg\,s^{-1}} \cdot 1.5_{-0.8}^{+1.8} 
    \left( \frac{L_{151\,\mathrm{MHz}}}{10^{34}\,\mathrm{erg\,s^{-1}\,Hz^{-1}}} \right)^{0.8} \nonumber \\
    &\quad \times (1+z)^{1.0} 
    \left( \frac{\mathrm{LLS}}{\mathrm{kpc}} \right)^{0.58\pm0.17},
\label{eq:Lmech_SG}
\end{align}
where $L_{151\,\mathrm{MHz}}$ is the 151\,MHz radio luminosity, which we obtained by scaling the radio luminosity of our sources by using Equation~\ref{eq:lr_conv}.

\subsubsection{AGN mechanical feedback in the eRASS1/ASKAP CC subsample}\label{sect:lkin-lx_cc}
Without any heating mechanisms, the dense ICM in the cluster cores would cool rapidly, leading to a short central cooling time ($t_\mathrm{cool}$). Observations indicate that heating regulated by AGN is particularly significant in these CC systems. For instance, studies have shown that the fraction of AGN increases with decreasing $t_\mathrm{cool}$ \citep{Mittal_2009}. Enhanced star formation near the cluster center seems to occur only in systems where $t_\mathrm{cool}<1\,\mathrm{Gyr}$ \citep{Rafferty_2008}.
Using the \textit{HIFLUGCS} sample, which consists of 64 clusters, \cite{Main_2017} observed a distinction between the CC and NCC clusters in the AGN mechanical feedback power versus cluster mass diagram (see their Figure~9). They discovered that AGN feedback in CC clusters ($t_\mathrm{cool}<1\,\mathrm{Gyr}$ within an aperture of $0.004R_{500}$ from the cluster centre) is more powerful, and there is a notable correlation with the mass of the cluster. Measurement of a gas property in a small aperture requires deep, high-resolution observations.
Since spatially-resolved spectral analysis is not possible with our sample, we utilised $c_{R_{500}}$, which is also a sensitive identifier of cool cores (see Section~\ref{sec:dynamic_state}). We constructed a CC subsample by using a threshold of $c_{R_{500}}>0.26$. This subsample consists of 75 clusters, including 11 radio upper limits. Although there is no clear distinction between CC and NCC based on their $L_\mathrm{mech}$ and masses, we focus our analysis on the CC subsample where radiative cooling is most significant.
\par
Additionally, we recall that the used $L_\mathrm{X}$ from the eRASS1 primary cluster catalogue was calculated within 300\,kpc, which corresponds to an average of $0.34R_{500}$ in our sample. Previous studies that have utilised high-resolution, deeper X-ray data indicate that estimations should be taken from a much smaller radius, where gas cools to lower temperatures (referred to as the cooling radius, $R_\mathrm{cool}$), and the effect of AGN feedback is more relevant. To have a more meaningful comparison, a comparison with integrated X-ray luminosity at $R_\mathrm{cool}$ ($L_{\mathrm{X},~r<R_\mathrm{cool}}$) should be done. In our attempt, we adopted $R_\mathrm{cool}=0.08R_{500}$, as found in \citet{Hudson_2010} for the CC clusters of the \textit{HIFLUGCS} sample. The scaling factor to convert $L_\mathrm{X}$ to $L_{\mathrm{X},~r<R_\mathrm{cool}}$ ($f_{R_\mathrm{cool}}$) was calculated by taking the ratio of the integrated surface brightness profile value at $R_\mathrm{cool}$ to the value at 300\,kpc. The integrated surface brightness profile was constructed from a single $\beta$-model profile \citep{Cavaliere_1976}, assuming a core radius of $r_\mathrm{c}=0.0035R_{500}$, which was the value found by the analysis of the inner regions of the \textit{HIFLUGCS} CC clusters in \cite{Hudson_2010}. Even when assuming a larger core radius, $r_\mathrm{c}\approx0.05R_{500}$, with the worst case of the slope $\beta=0.5$, the ratio would only decrease to 0.35. Thus, for $0.5\leq\beta\leq1.0$ and $r_\mathrm{c}=0.0035R_{500}$, we obtained $0.7\leq f_{R_\mathrm{cool}}\leq 1.0$, such that $f_{R_\mathrm{cool}}=0.85\pm0.15$ was assumed. By multiplying $L_\mathrm{X}$ by $f_{R_\mathrm{cool}}$, we obtained $L_{\mathrm{X},~r<R_\mathrm{cool}}$. For the CC subsample, the X-ray luminosity ranges are $2.85\times10^{42} \leq L_\mathrm{X}/\mathrm{erg\,s^{-1}} \leq 6.43\times10^{44}$ and $2.42\times10^{42} \leq L_{\mathrm{X},~r<R_\mathrm{cool}}/\,\mathrm{erg\,s^{-1}} \leq 5.46\times10^{44}$, respectively.

\begin{figure}
\centering
\includegraphics[width=\columnwidth]{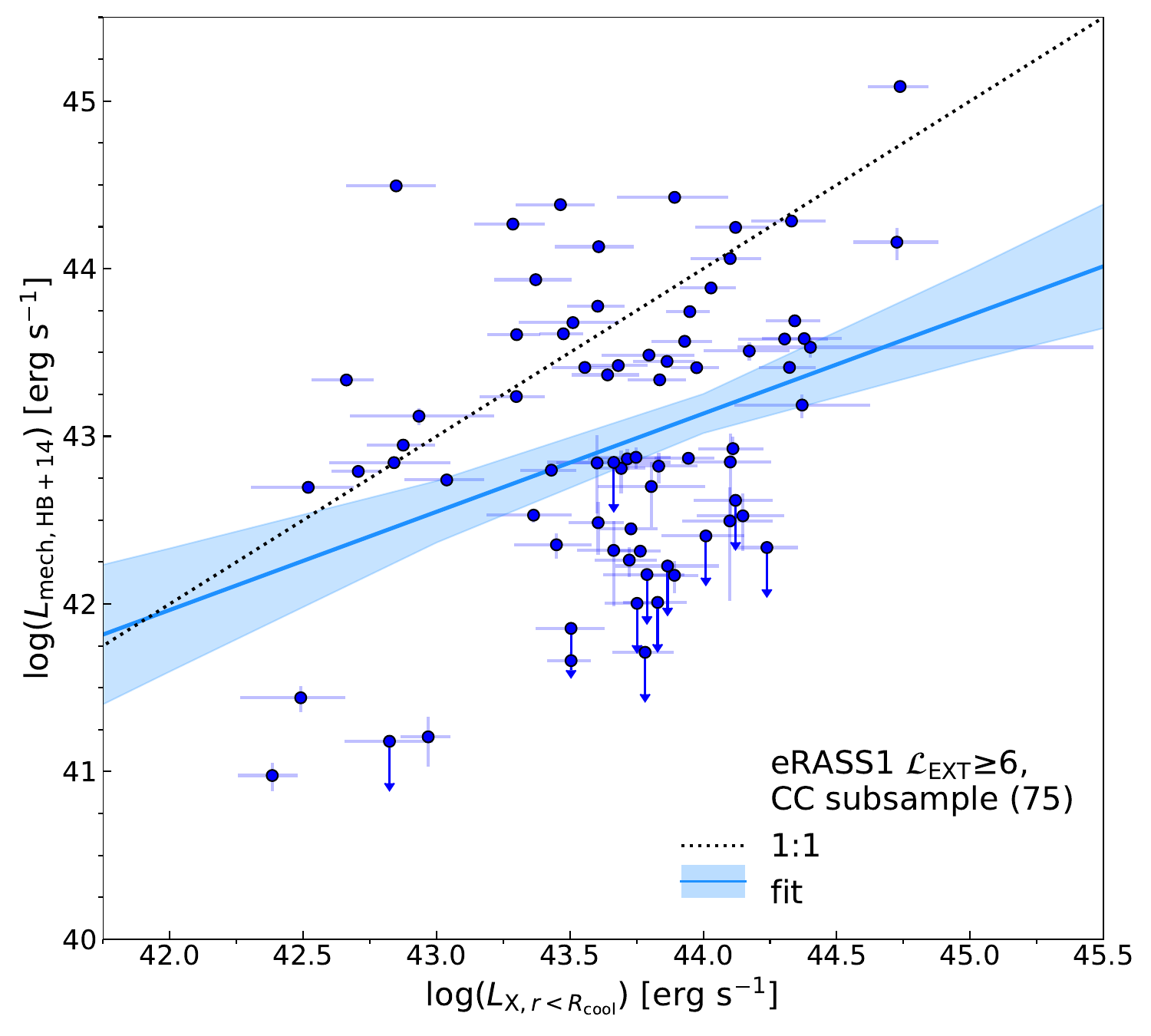}
\caption{Central AGN mechanical luminosity scaled from the monochromatic radio luminosity using Equation~\ref{eq:Lkin} from \cite{Heckman_2014} against X-ray luminosity within the cooling radius for the CC subsample ($c_{R_{500}}>0.26$). The blue solid line and shaded area are the linear fit and the $1\sigma$ band constrained from the CC subsample. The dotted line marks the 1-to-1 line.}
\label{fig:lkin}
\end{figure}

\par
In Figure~\ref{fig:lkin}, the $\log L_\mathrm{mech,HB+14}-\log L_{\mathrm{X},~r<R_\mathrm{cool}}$ plot for the CC subsample is shown, where $L_\mathrm{mech,HB+14}$ was calculated using Equation~\ref{eq:Lkin} from \cite{Heckman_2014}. The $L_\mathrm{mech,HB+14}$ ranges from $9.47\times10^{40}\,\mathrm{erg\,s^{-1}}$ to $1.22\times10^{45}\,\mathrm{erg\,s^{-1}}$, which is consistent with other works \citep[e.g.,][]{Birzan_2004, Merloni_2007, Pasini_2022}.
\par
We performed a linear regression on this subsample using the \texttt{ASURV} package, which yielded a slope of $0.59\pm0.20$ (solid blue line and blue shaded area). The slope is in agreement with the slope estimated from the sample in \cite{Merloni_2007}, $0.81\pm0.11$ and within $2.1\sigma$ from the slope from the eFEDS/LOFAR sample of $1.07\pm0.11$ \citep{Pasini_2022}. The correlation obtained from the CC sub-sample is moderately significant, with a $p$-value of 0.0336 from the generalised Kendall's $\tau$ test and seems to be mainly driven by high redshift clusters. By excluding the non-detected sources, a steeper slope of $A=0.65\pm0.17$ is obtained. For this subsample, we found that the correlation has higher statistical significance with a $p$-value of 0.0045. However, it is important to note that there is significant scatter in the data, with $\sigma=0.89$ for the CC subsample and $\sigma=0.72$ for the CC-detected-only subsample. As shown in Figure~\ref{fig:lkin}, the best-fit line intersects with the 1-to-1 line at $L_{\mathrm{X},~r<R_\mathrm{cool}}\approx8.18\times10^{41}\,\mathrm{erg\,s^{-1}}$, which naively suggests that above these cuts, the central AGN feedback in our sample is ineffective in countering the ICM radiative losses (see Section~\ref{sect:eff}).
\par
Additionally, we present the $\log L_\mathrm{mech,SG+13}-\log L_{\mathrm{X},~r<R_\mathrm{cool}}$ plot, where $L_\mathrm{mech,SG+13}$ was calculated using Equation~\ref{eq:Lmech_SG}, in Figure~\ref{fig:lkin_SG}. The $L_\mathrm{mech,SG+13}$ values obtained from this equation are consistent with the $L_\mathrm{mech,HB+14}$ values within their uncertainties (more discussion in \ref{app:Lmech_SG}).

\subsubsection{Efficiency of AGN feedback}\label{sect:eff}
We used two methods to calculate the efficiency, which is defined as the AGN energy input required to balance the ICM radiative losses. First, we computed the predicted mechanical luminosity (calculated using Equation~\ref{eq:Lkin}) at an integrated X-ray luminosity value, $L_\mathrm{mech,HB+14,\,pred}$, using the linear form of Equation~\ref{eq:relation}. The efficiency is then obtained by taking the ratio between $L_\mathrm{mech,HB+14,\,pred}$ and the given $L_{\mathrm{X},~r<R_\mathrm{cool}}$ ($L_\mathrm{X}$ multiplied by $f_{R_\mathrm{cool}}$, as defined in Section~\ref{sect:lkin-lx_cc}). We evaluated the efficiency at $L_{\mathrm{X},~r<R_\mathrm{cool}}=10^{42}\,\mathrm{erg\,s^{-1}}$, $10^{43.74}\,\mathrm{erg\,s^{-1}}$, and $10^{45}\,\mathrm{erg\,s^{-1}}$. The estimated efficiency at these given values is $92.0\pm77.8\%$, $17.5\pm4.3\%$, and $5.3\pm3.3\%$. Second, we utilised the method described in \cite{Israel_2014}. First, we calculated the logarithmic bias $b=\langle \log L_\mathrm{mech,HB+14} - \log L_{\mathrm{X},~r<R_\mathrm{cool}} \rangle$, which is the average logarithmic difference between $L_\mathrm{mech,HB+14}$ and $L_{\mathrm{X},~r<R_\mathrm{cool}}$. The error of $b$ is taken to be the standard error of $(\log L_\mathrm{mech,HB+14} - \log L_{\mathrm{X},~r<R_\mathrm{cool}})$, and the efficiency is estimated by $10^b\times100\%$. For the CC subsample, a logarithmic bias of $b=-0.66\pm0.09$ corresponds to an efficiency of $21.8_{-4.2}^{+5.2}\%$.
\par
We show the distribution of the efficiency (second method) calculated using $L_\mathrm{mech,HB+14}$ and $L_{\mathrm{X}~r<R_\mathrm{cool}}$ in Figure~\ref{fig:eff}. Furthermore, if the radio upper limits are not considered when estimating the correlation, the efficiency at the above mentioned $L_{\mathrm{X},~r<R_\mathrm{cool}}$ values are $125.0\pm87.3\%$, $30.2\pm6.3\%$, and $10.8\pm5.8\%$, respectively. While using the second method, a bias of $b=-0.49\pm0.09$ was calculated, which corresponds to an efficiency of $32.1_{-6.2}^{+7.7}\%$.

\subsection{Statistical robustness and selection effect}\label{sect:statisticaltests}
To further evaluate the statistical robustness of the observed correlations, we performed a bootstrap resampling analysis. We generated 5000 bootstrap realisations by randomly sampling with replacement from the original dataset. In each realisation, some elements can be repeated while others may be excluded, but the size of each realization remains the same as the original sample size. We then performed linear fitting on each realisation using the \texttt{ASURV} package. Afterward, we compared the distributions of the correlation coefficients from the resampled datasets with the values obtained from the original dataset.
\par
Given that our sample spans a wide redshift range, the apparent correlations may be affected by selection effects. To investigate this, first, we used the partial correlation Kendall's $\tau$ test \citep{Akritas_1996}. The test has been used in other works \citep[e.g.,][]{Ineson_2015, Pasini_2022} to confirm a correlation in the presence of redshift dependence and upper limits. A lower null-hypothesis probability value indicates a high likelihood that the observed correlation is real and not driven by selection effects. Secondly, we examined the corresponding flux–flux correlations to further assess the potential impact of observational biases, including selection effects and detection thresholds. The results of bootstrap, partial correlation Kendall's $\tau$ test, and flux-flux analyses are presented in the following subsections.

\subsubsection{$\log L_\mathrm{R}$–$\log L_{\mathrm{X}}$}\label{sect:test_lxlr}
We performed the bootstrap resampling on the full sample. The mean and standard deviation derived from the slope distribution of the resampled datasets is $A_\mathrm{boot}=0.70\pm0.18$, which is consistent with the slope from the original dataset, $A=0.70\pm0.18$ (see Table~\ref{tab:lr-lx} for the correlation coefficients of the original dataset). The mean normalization obtained from the resampled datasets is $B_\mathrm{boot}=-0.19\pm0.08$, also in good agreement with $B=-0.19\pm0.09$. The median of the scatters and Kendall's $\tau$ $p$-values across the resamplings are $\sigma_\mathrm{boot}=1.0865$ and $p_\mathrm{boot}=0.0019$, respectively.
These results support the statistical significance of the observed $\log L_\mathrm{R}$–$\log L_{\mathrm{X}}$ correlation and suggest that it is not strongly dependent on any particular data point, i.e., it is unlikely to have occurred by chance.
\par
However, the partial correlation Kendall's $\tau$ test of the $\log L_\mathrm{R}$–$\log L_{\mathrm{X}}$ correlation returns a high null-hypothesis probability value, $p>0.2$ ($\tau=0.0459$ and $\sigma=0.0433$), suggesting that the observed correlation is influenced by selection effects.
From the flux-flux analysis ($\log f_\mathrm{R}$–$\log f_{\mathrm{X}}$), we determined a slope of $A_\mathrm{f-f}=0.51\pm0.22$ and a normalization of $B_\mathrm{f-f}=-0.12\pm0.09$, both of which are consistent with the values from the original dataset within $1\sigma$ (Table~\ref{tab:lr-lx}). The scatter, $\sigma_\mathrm{f-f}=1.0872$, is also at a similar order of magnitude. The generalised Kendall's $\tau$ statistic test gives a null-hypothesis probability value of $p_\mathrm{f-f}=0.0328$. Nevertheless, when the highest flux clusters ($f_\mathrm{X}>10^{-11.5}\,\mathrm{erg\,s^{-1}\,cm^{-2}}$) are excluded, the $p$-value increases to $\gtrsim 5\%$, indicating that the correlation becomes statistically insignificant. This implies that the observed $\log f_\mathrm{R}-\log f_{\mathrm{X}}$ correlation is primarily driven by a small number of high flux clusters.

\subsubsection{$\log L_\mathrm{mech,HB+14}$–$\log L_{\mathrm{X},~r<R_\mathrm{cool}}$}\label{sect:test_lkin}
The bootstrap resampling analysis of the CC subsample for the $\log L_\mathrm{mech,HB+14}$–$\log L_{\mathrm{X},~r<R_\mathrm{cool}}$ correlation results in a slope of $A_\mathrm{boot}=0.58\pm0.22$ and a normalization of $B_\mathrm{boot}=0.11\pm0.11$, which are consistent with the values from the original dataset, $A=0.59\pm0.20$ and $B=0.11\pm0.11$. The median of the scatters and Kendall's $\tau$ $p$-values are $\sigma_\mathrm{boot}=0.0.8711$ and $p_\mathrm{boot}=0.0323$, respectively, both in good agreement with the values determined from the original dataset. These results confirm that the correlation found in the CC subsample is robust against outliers.
\par
We further validated the correlation using the partial correlation Kendall's $\tau$ test and flux-flux analysis. From the former, we obtained $\tau=0.0839$ and $\sigma=0.0698$, subsequently, $p=0.2292$. This implies that the $\log L_\mathrm{mech,HB+14}$–$\log L_{\mathrm{X},~r<R_\mathrm{cool}}$ correlation is dependent on the redshift. In contrast, the flux-flux analysis yielded $A_\mathrm{f-f}=0.86\pm0.26$, $B_\mathrm{f-f}=-0.03\pm0.11$, and $\sigma_\mathrm{f-f}=0.9129$. The generalised Kendall's $\tau$ test showed a smaller null-hypothesis probability of $p_\mathrm{f-f}=0.0076$ compared to the luminosity-luminosity analysis. This value increases to 0.0391 when excluding the highest flux clusters ($f_{\mathrm{X},~r<R_\mathrm{cool}}>10^{-11.5} \,\mathrm{erg\,s^{-1}\,cm^{-2}}$), but it still indicates a statistically significant correlation. The findings from the bootstrapping and flux-flux analyses may suggest that the correlation is intrinsic to the $L_\mathrm{mech,HB+14}$ and $L_{\mathrm{X},~r<R_\mathrm{cool}}$.

\section{Discussions}\label{sec:discussion}
\subsection{Radio and X-ray luminosity correlation}
Although the apparent $\log L_\mathrm{R}$–$\log L_{\mathrm{X}}$ correlation (Section~\ref{sect:lx-lr}) appears to be driven by selection effects (Section~\ref{sect:test_lxlr}), we note that our sample is based on eRASS1-only data, which is not complete \citep[see Figure~8 of][]{Bulbul_2024}. Hence, some clusters, especially the fainter ones, are missing from our analysis. With deeper data (e.g., eRASS:5) combined with radio data with higher spatial resolution, we expect more sources to populate the lower left area of the $\log L_\mathrm{R}$–$\log L_{\mathrm{X}}$ space (Figure~\ref{fig:lr-lx}). This could potentially strengthen the correlation and result in a steeper slope. Nonetheless, we analysed the impact of the selection threshold, i.e., the flux cut, on both the significance of the correlation and the best-fit parameters. We created subsamples by applying flux cuts that range from $1\times10^{-13}\,\mathrm{erg\,s^{-1}\,cm^{-2}}$ to $3\times10^{-13}\,\mathrm{erg\,s^{-1}\,cm^{-2}}$, resulting in a number of clusters between 84 and 145. We recall that the flux cut from the eRASS1 primary cluster catalog is $4\times10^{-14}\,\mathrm{erg\,s^{-1}\,cm^{-2}}$ \citep{Bulbul_2024}. Our findings indicate that the $\log L_\mathrm{R}$–$\log L_{\mathrm{X}}$ correlations from these subsamples are consistent with those obtained from the full sample, and they are statistically significant, with generalised Kendall’s $\tau$ $p$-values always staying below 0.05.
\par
Nevertheless, the indication of a weak positive correlation between the X-ray luminosity of host clusters and the radio power of BCGs may come from the fact that the X-ray luminosity correlates directly with the mass of the cluster \citep[e.g.,][]{Lovisari_2020}. Therefore, more massive and luminous clusters contain a larger gas reservoir to fuel the central SMBHs \citep{Stott_2012, Kolokythas_2018, Gaspari_2019}, triggering AGN activity and thus, increasing the radio power of the BCGs. Various studies have also investigated the correlation between the mass of SMBHs and the radio luminosity of the host BCGs. For example, \cite{Mittal_2009} reported a correlation between ($M_\mathrm{BH}$) and $L_\mathrm{R}$ for their SCC subsample (see their Figure~10). Additionally, the authors found that the mass of the BCG increases with the size of the cluster, indicating that $M_\mathrm{BH}$ scales with $M_{500}$ \citep[see Figure~11 of][]{Mittal_2009}. This suggests that more massive clusters may host more powerful radio sources, even when assuming a constant SMBH energy conversion rate \citep[e.g.,][]{Fujita_2004}.
\par
As our sample did not reveal a positive correlation between the concentration parameter $c_{R_{500}}$ and $L_\mathrm{R}$ (left panel of Figure~\ref{fig:w-c}), this may suggest that clusters with higher central gas density do not necessarily harbour more powerful radio sources. Therefore, it might be that gas supply is not the primary mechanism driving the $L_\mathrm{R}-L_\mathrm{X}$ correlation. However, to explore the other scenario, an investigation on the optical properties of the host galaxies of the SMBHs is required, which is beyond the scope of this study.

\subsection{Efficiency of AGN feedback in the eRASS1/ASKAP CC subsample}
The linear fit from our measurements suggests that while AGN feedback might be able to counterbalance ICM cooling in low-luminosity systems (i.e., $L_{\mathrm{X},~r<R_\mathrm{cool}} \lesssim 8.18 \times 10^{41}\,\mathrm{erg\,s^{-1}}$, see Section~\ref{sect:lkin-lx_cc} and Figure~\ref{fig:lkin}), its efficiency appears to decrease as the X-ray cooling luminosity increases. For instance, at $L_{\mathrm{X},~r<R_\mathrm{cool}}\approx5.50\times10^{43}\,\mathrm{erg\,s^{-1}}$, the AGN feedback seems to only supply, on average, around $13.3-21.8\%$ of the energy required to counterbalance the radiative losses of the ICM. This may imply that in more luminous (high mass) systems, AGN feedback alone may not be sufficient to completely compensate for radiative losses (see Section~\ref{sect:intro}). One notable case is the cooling flow cluster, Abell 1068 \citep{Wise_2004, McNamara_2004}, where a starburst is observed in its cD galaxy, along with an increase in gas metallicity toward the cluster core. The authors claimed that the heating from the central radio source and thermal conduction in this cluster is minor compared to the contributions from supernovae. Recent results from Resolve, the high-spectral-resolution X-ray microcalorimeter on board the XRISM satellite, also indicate that gas turbulence and bulk motions induced by sloshing play a more significant role in thermal heating than cavities formed by AGN \citep{Xrism_2025}.
\par
The insufficiency of AGN feedback in offsetting cooling losses has also been noted in other statistical studies. For example, \cite{Igo_2025} found that in large groups and clusters ($M_\mathrm{halo}>3\times10^{13}M_\odot$) from their LOFAR-eFEDS sample, the ratio of AGN jet kinetic energy to gas cooling energy drops to approximately $3-60\%$ (see their Figure~7). Our finding is also consistent with that of \cite{Calzadilla_2024}, where, in their SZ-selected sample, the trend of low core entropy in the radio-detected objects is absent. \cite{Gupta_2020} reported that for their high mass clusters $(M_{500}\gtrsim5\times10^{14}M_\odot)$ the average radio feedback is smaller than the average core radiative losses within $0.1R_{500}$ (see their right plot of Figure~11). Their ratio between the AGN mechanical feedback energy and the central X-ray luminosities goes as low as 0.15. These authors included the entire radio AGN population and not exclusively those hosted by the BCGs. Furthermore, their sample consisted of only detected radio AGNs from the Sydney University Molonglo Sky Survey \citep[SUMSS;][]{Mauch_2003} catalogue, which has the minimum flux density of 6\,mJy. This indicates that their sample consisted of brighter radio sources compared to the ones in this study, and their ratio estimates could be lower if the non-detection sources were considered.
\par
While the age of the radio sources was considered using Equation~\ref{eq:Lmech_SG}, other assumptions came into the conversion of the 944\,MHz radio emission to the mechanical luminosity (e.g., environment). Follow-up studies aimed at better defining the relationship between $L_{\mathrm{mech}}$ and $L_{\mathrm{R}}$ will help to reduce these uncertainties. There are also assumptions when estimating $f_{R_\mathrm{cool}}$ that might contribute further to the distribution of the data points. Adding to the scatter in the y-axis direction is the fact that AGN activity is variable, and therefore, the AGN jet power is also variable \citep{McNamara_2012, Hardcastle_2019}.

\section{Conclusions}\label{sect:conclusions}
We conducted a statistical study of eRASS1 galaxy clusters in the ASKAP fields (PS1, PS2, and SWAG-X). We identified the radio sources associated with the BCGs of the clusters and analysed the correlation between their radio properties and the X-ray properties of the host clusters. Below, we summarise our findings:
   \begin{itemize}
       \item Our sample consists of 151 clusters. The sample spans a redshift range of $0.033\leq z\leq1.126$ and X-ray luminosity range of $2.72\times10^{42}\,\mathrm{erg\,s^{-1}}\leq L_\mathrm{X}\leq1.32\times10^{45}\,\mathrm{erg\,s^{-1}}$. We identified radio sources associated with 134 BCGs, resulting in a detection rate of 89\%. The 944\,MHz radio luminosities of the corresponding central sources lie between $\sim\!10^{28}$ and $10^{33}\,\mathrm{erg\,s^{-1}\,Hz^{-1}}.$
       \item We identified a potential positive trend between LLS and BCG offset, which may hint at an environmental influence on the morphology of central radio sources. However, we notice that selection effects are present due to the fact that at high redshifts, only massive clusters with more powerful radio sources are detected.
       \item We observed a weak correlation between the 944\,MHz radio luminosity of the central sources and the X-ray luminosity of the host clusters, suggesting that more luminous central radio galaxies are found in more luminous clusters.
       \item We estimated the mechanical power of the central radio sources from their monochromatic radio luminosities by using well-established scaling relations. We also scaled the integrated X-ray luminosity from within 300\,kpc into a cooling radius of 8\%$R_{500}$. The determined $\log L_\mathrm{mech,HB+14}$–$\log L_{\mathrm{X},~r<R_\mathrm{cool}}$ correlation from our CC subsample shows an indication that the AGN feedback is not efficient for high luminosity (high mass) clusters. At $L_{\mathrm{X},~r<R_\mathrm{cool}}\approx5.50\times10^{43}\,\mathrm{erg\,s^{-1}}$, the AGN feedback seems to only supply, on average, around $13-22\%$ of the energy required to counterbalance the radiative losses of the ICM. Uncertainties in the scaling of $L_\mathrm{mech}$ and $L_{\mathrm{X},~r<R_\mathrm{cool}}$ may contribute to this observed inefficiency.
   \end{itemize}
Future studies utilising deeper X-ray data (e.g., eRASS:5) and broader sky coverage \citep[e.g., the upcoming ASKAP/EMU survey,][]{Hopkins_2025} will include a larger sample of galaxy groups and clusters, enabling a more comprehensive investigation of the interplay between central AGNs and ICM.

\begin{acknowledgement}
We thank the anonymous referee for their valuable feedback that helped improve the manuscript.
We thank Christos Karoumpis for his valuable guidance on radio flux measurements and Heinz Andernach for his help in the early phase of this work, as well as for alerting us about issues in Table 4 which were corrected before publication.
      Funded by the Deutsche Forschungsgemeinschaft (DFG, German Research Foundation) – 450861021.
      A.V. and T.R. acknowledge support from the German Federal Ministry of Economics and Technology (BMWi) provided through the German Space Agency (DLR) under project 50 OR 2112.
      MB acknowledges funding by the Deutsche Forschungsgemeinschaft (DFG, German Research Foundation) under Germany's Excellence Strategy -- EXC 2121 ``Quantum Universe'' -- 390833306 and the DFG Research Group "Relativistic Jets".
      JEMC acknowledges funding from an STFC studentship.
      SD acknowledges support from the Collaborative Research Center 1601 (SFB 1601 sub-project B2) funded by the Deutsche Forschungsgemeinschaft (DFG, German Research Foundation) – 500700252.
      This work is based on data from eROSITA, the soft X-ray instrument aboard SRG, a joint Russian-German science mission supported by the Russian Space Agency (Roskosmos), in the interests of the Russian Academy of Sciences represented by its Space Research Institute (IKI), and the Deutsches Zentrum für Luft- und Raumfahrt (DLR). The SRG spacecraft was built by Lavochkin Association (NPOL) and its subcontractors, and is operated by NPOL with support from the Max Planck Institute for Extraterrestrial Physics (MPE). The development and construction of the eROSITA X-ray instrument was led by MPE, with contributions from the Dr. Karl Remeis Observatory Bamberg and ECAP (FAU Erlangen-Nuernberg), the University of Hamburg Observatory, the Leibniz Institute for Astrophysics Potsdam (AIP), and the Institute for Astronomy and Astrophysics of the University of Tübingen, with the support of DLR and the Max Planck Society. The Argelander Institute for Astronomy of the University of Bonn and the Ludwig Maximilians Universität Munich also participated in the science preparation for eROSITA. The eROSITA data shown here were processed using the eSASS software system developed by the German eROSITA consortium.
      This scientific work uses data obtained from Inyarrimanha Ilgari Bundara / the Murchison Radio-astronomy Observatory. We acknowledge the Wajarri Yamaji People as the Traditional Owners and native title holders of the Observatory site. CSIRO’s ASKAP radio telescope is part of the Australia Telescope National Facility (\url{https://ror.org/05qajvd42}). Operation of ASKAP is funded by the Australian Government with support from the National Collaborative Research Infrastructure Strategy. ASKAP uses the resources of the Pawsey Supercomputing Research Centre. Establishment of ASKAP, Inyarrimanha Ilgari Bundara, the CSIRO Murchison Radio-astronomy Observatory and the Pawsey Supercomputing Research Centre are initiatives of the Australian Government, with support from the Government of Western Australia and the Science and Industry Endowment Fund.
      The Legacy Surveys consist of three individual and complementary projects: the Dark Energy Camera Legacy Survey (DECaLS; Proposal ID \#2014B-0404; PIs: David Schlegel and Arjun Dey), the Beijing-Arizona Sky Survey (BASS; NOAO Prop. ID \#2015A-0801; PIs: Zhou Xu and Xiaohui Fan), and the Mayall z-band Legacy Survey (MzLS; Prop. ID \#2016A-0453; PI: Arjun Dey). DECaLS, BASS and MzLS together include data obtained, respectively, at the Blanco telescope, Cerro Tololo Inter-American Observatory, NSF’s NOIRLab; the Bok telescope, Steward Observatory, University of Arizona; and the Mayall telescope, Kitt Peak National Observatory, NOIRLab. Pipeline processing and analyses of the data were supported by NOIRLab and the Lawrence Berkeley National Laboratory (LBNL). The Legacy Surveys project is honored to be permitted to conduct astronomical research on Iolkam Du’ag (Kitt Peak), a mountain with particular significance to the Tohono O’odham Nation.
      NOIRLab is operated by the Association of Universities for Research in Astronomy (AURA) under a cooperative agreement with the National Science Foundation. LBNL is managed by the Regents of the University of California under contract to the U.S. Department of Energy.
      This project used data obtained with the Dark Energy Camera (DECam), which was constructed by the Dark Energy Survey (DES) collaboration. Funding for the DES Projects has been provided by the U.S. Department of Energy, the U.S. National Science Foundation, the Ministry of Science and Education of Spain, the Science and Technology Facilities Council of the United Kingdom, the Higher Education Funding Council for England, the National Center for Supercomputing Applications at the University of Illinois at Urbana-Champaign, the Kavli Institute of Cosmological Physics at the University of Chicago, Center for Cosmology and Astro-Particle Physics at the Ohio State University, the Mitchell Institute for Fundamental Physics and Astronomy at Texas A\&M University, Financiadora de Estudos e Projetos, Fundacao Carlos Chagas Filho de Amparo, Financiadora de Estudos e Projetos, Fundacao Carlos Chagas Filho de Amparo a Pesquisa do Estado do Rio de Janeiro, Conselho Nacional de Desenvolvimento Cientifico e Tecnologico and the Ministerio da Ciencia, Tecnologia e Inovacao, the Deutsche Forschungsgemeinschaft and the Collaborating Institutions in the Dark Energy Survey. The Collaborating Institutions are Argonne National Laboratory, the University of California at Santa Cruz, the University of Cambridge, Centro de Investigaciones Energeticas, Medioambientales y Tecnologicas-Madrid, the University of Chicago, University College London, the DES-Brazil Consortium, the University of Edinburgh, the Eidgenossische Technische Hochschule (ETH) Zurich, Fermi National Accelerator Laboratory, the University of Illinois at Urbana-Champaign, the Institut de Ciencies de l’Espai (IEEC/CSIC), the Institut de Fisica d’Altes Energies, Lawrence Berkeley National Laboratory, the Ludwig Maximilians Universitat Munchen and the associated Excellence Cluster Universe, the University of Michigan, NSF’s NOIRLab, the University of Nottingham, the Ohio State University, the University of Pennsylvania, the University of Portsmouth, SLAC National Accelerator Laboratory, Stanford University, the University of Sussex, and Texas A\&M University.
      BASS is a key project of the Telescope Access Program (TAP), which has been funded by the National Astronomical Observatories of China, the Chinese Academy of Sciences (the Strategic Priority Research Program ”The Emergence of Cosmological Structures” Grant \# XDB09000000), and the Special Fund for Astronomy from the Ministry of Finance. The BASS is also supported by the External Cooperation Program of Chinese Academy of Sciences (Grant \# 114A11KYSB20160057), and Chinese National Natural Science Foundation (Grant \# 12120101003, \# 11433005).
      The Legacy Survey team makes use of data products from the Near-Earth Object Wide-field Infrared Survey Explorer (NEOWISE), which is a project of the Jet Propulsion Laboratory/California Institute of Technology. NEOWISE is funded by the National Aeronautics and Space Administration.
      The Legacy Surveys imaging of the DESI footprint is supported by the Director, Office of Science, Office of High Energy Physics of the U.S. Department of Energy under Contract No. DE-AC02-05CH1123, by the National Energy Research Scientific Computing Center, a DOE Office of Science User Facility under the same contract; and by the U.S. National Science Foundation, Division of Astronomical Sciences under Contract No. AST-0950945 to NOAO.

\end{acknowledgement}


\bibliography{bibtemplate}

\newcommand{\noop}[1]{}
\begin{thebibliography}{}
\expandafter\ifx\csname natexlab\endcsname\relax\def\natexlab#1{#1}\fi

\bibitem[{{Akritas} \& {Siebert}(1996)}]{Akritas_1996}
{Akritas}, M.~G., \& {Siebert}, J. 1996, \mnras, 278, 919

\bibitem[{{Baldwin}(1982)}]{Baldwin_1982}
{Baldwin}, J.~E. 1982, in IAU Symposium, Vol.~97, Extragalactic Radio Sources, ed. D.~S. {Heeschen} \& C.~M. {Wade}, 21--24

\bibitem[{{Bertin} {et~al.}(2002){Bertin}, {Mellier}, {Radovich}, {Missonnier}, {Didelon}, \& {Morin}}]{swarp}
{Bertin}, E., {Mellier}, Y., {Radovich}, M., {et~al.} 2002, in Astronomical Society of the Pacific Conference Series, Vol. 281, Astronomical Data Analysis Software and Systems XI, ed. D.~A. {Bohlender}, D.~{Durand}, \& T.~H. {Handley}, 228

\bibitem[{{B{\^\i}rzan} {et~al.}(2008){B{\^\i}rzan}, {McNamara}, {Nulsen}, {Carilli}, \& {Wise}}]{Birzan_2008}
{B{\^\i}rzan}, L., {McNamara}, B.~R., {Nulsen}, P.~E.~J., {Carilli}, C.~L., \& {Wise}, M.~W. 2008, \apj, 686, 859

\bibitem[{{B{\^\i}rzan} {et~al.}(2004){B{\^\i}rzan}, {Rafferty}, {McNamara}, {Wise}, \& {Nulsen}}]{Birzan_2004}
{B{\^\i}rzan}, L., {Rafferty}, D.~A., {McNamara}, B.~R., {Wise}, M.~W., \& {Nulsen}, P.~E.~J. 2004, \apj, 607, 800

\bibitem[{{Blundell} {et~al.}(1999){Blundell}, {Rawlings}, \& {Willott}}]{Blundell_1999}
{Blundell}, K.~M., {Rawlings}, S., \& {Willott}, C.~J. 1999, \aj, 117, 677

\bibitem[{{B{\"o}ckmann} {et~al.}(2023){B{\"o}ckmann}, {Br{\"u}ggen}, {Koribalski}, {Veronica}, {Reiprich}, {Bulbul}, {Bahar}, {Balzer}, {Comparat}, {Garrel}, {Ghirardini}, {G{\"u}rkan}, {Kluge}, {Leahy}, {Merloni}, {Liu}, {Ramos-Ceja}, {Salvato}, {Sanders}, {Shabala}, \& {Zhang}}]{Boeckmann_2023}
{B{\"o}ckmann}, K., {Br{\"u}ggen}, M., {Koribalski}, B., {et~al.} 2023, \aap, 677, A188

\bibitem[{{Boehringer} {et~al.}(1993){Boehringer}, {Voges}, {Fabian}, {Edge}, \& {Neumann}}]{Boehringer_1993}
{Boehringer}, H., {Voges}, W., {Fabian}, A.~C., {Edge}, A.~C., \& {Neumann}, D.~M. 1993, \mnras, 264, L25

\bibitem[{{B{\"o}hringer} {et~al.}(2002){B{\"o}hringer}, {Matsushita}, {Churazov}, {Ikebe}, \& {Chen}}]{Boehringer_2002}
{B{\"o}hringer}, H., {Matsushita}, K., {Churazov}, E., {Ikebe}, Y., \& {Chen}, Y. 2002, \aap, 382, 804

\bibitem[{{Bower} {et~al.}(2006){Bower}, {Benson}, {Malbon}, {Helly}, {Frenk}, {Baugh}, {Cole}, \& {Lacey}}]{Bower_2006}
{Bower}, R.~G., {Benson}, A.~J., {Malbon}, R., {et~al.} 2006, \mnras, 370, 645

\bibitem[{{Brunner} {et~al.}(2022){Brunner}, {Liu}, {Lamer}, {Georgakakis}, {Merloni}, {Brusa}, {Bulbul}, {Dennerl}, {Friedrich}, {Liu}, {Maitra}, {Nandra}, {Ramos-Ceja}, {Sanders}, {Stewart}, {Boller}, {Buchner}, {Clerc}, {Comparat}, {Dwelly}, {Eckert}, {Finoguenov}, {Freyberg}, {Ghirardini}, {Gueguen}, {Haberl}, {Kreykenbohm}, {Krumpe}, {Osterhage}, {Pacaud}, {Predehl}, {Reiprich}, {Robrade}, {Salvato}, {Santangelo}, {Schrabback}, {Schwope}, \& {Wilms}}]{Brunner_2022}
{Brunner}, H., {Liu}, T., {Lamer}, G., {et~al.} 2022, \aap, 661, A1

\bibitem[{{Bulbul} {et~al.}(2024){Bulbul}, {Liu}, {Kluge}, {Zhang}, {Sanders}, {Bahar}, {Ghirardini}, {Artis}, {Seppi}, {Garrel}, {Ramos-Ceja}, {Comparat}, {Balzer}, {B{\"o}ckmann}, {Br{\"u}ggen}, {Clerc}, {Dennerl}, {Dolag}, {Freyberg}, {Grandis}, {Gruen}, {Kleinebreil}, {Krippendorf}, {Lamer}, {Merloni}, {Migkas}, {Nandra}, {Pacaud}, {Predehl}, {Reiprich}, {Schrabback}, {Veronica}, {Weller}, \& {Zelmer}}]{Bulbul_2024}
{Bulbul}, E., {Liu}, A., {Kluge}, M., {et~al.} 2024, \aap, 685, A106

\bibitem[{{Calzadilla} {et~al.}(2024){Calzadilla}, {McDonald}, {Benson}, {Bleem}, {Croston}, {Donahue}, {Edge}, {Floyd}, {Garmire}, {Hlavacek-Larrondo}, {Huynh}, {Khullar}, {Kraft}, {McNamara}, {Noble}, {Romero}, {Ruppin}, {Somboonpanyakul}, \& {Voit}}]{Calzadilla_2024}
{Calzadilla}, M.~S., {McDonald}, M., {Benson}, B.~A., {et~al.} 2024, \apj, 976, 169

\bibitem[{{Cavagnolo} {et~al.}(2010){Cavagnolo}, {McNamara}, {Nulsen}, {Carilli}, {Jones}, \& {B{\^\i}rzan}}]{Cavagnolo_2010}
{Cavagnolo}, K.~W., {McNamara}, B.~R., {Nulsen}, P.~E.~J., {et~al.} 2010, \apj, 720, 1066

\bibitem[{{Cavaliere} \& {Fusco-Femiano}(1976)}]{Cavaliere_1976}
{Cavaliere}, A., \& {Fusco-Femiano}, R. 1976, \aap, 49, 137

\bibitem[{{Croton} {et~al.}(2006){Croton}, {Springel}, {White}, {De Lucia}, {Frenk}, {Gao}, {Jenkins}, {Kauffmann}, {Navarro}, \& {Yoshida}}]{Croton_2006}
{Croton}, D.~J., {Springel}, V., {White}, S. D.~M., {et~al.} 2006, \mnras, 365, 11

\bibitem[{{Daly} {et~al.}(2012){Daly}, {Sprinkle}, {O'Dea}, {Kharb}, \& {Baum}}]{Daly_2012}
{Daly}, R.~A., {Sprinkle}, T.~B., {O'Dea}, C.~P., {Kharb}, P., \& {Baum}, S.~A. 2012, \mnras, 423, 2498

\bibitem[{{Dark Energy Survey Collaboration} {et~al.}(2016){Dark Energy Survey Collaboration}, {Abbott}, {Abdalla}, {Aleksi{\'c}}, {Allam}, {Amara}, {Bacon}, {Balbinot}, {Banerji}, {Bechtol}, {Benoit-L{\'e}vy}, {Bernstein}, {Bertin}, {Blazek}, {Bonnett}, {Bridle}, {Brooks}, {Brunner}, {Buckley-Geer}, {Burke}, {Caminha}, {Capozzi}, {Carlsen}, {Carnero-Rosell}, {Carollo}, {Carrasco-Kind}, {Carretero}, {Castander}, {Clerkin}, {Collett}, {Conselice}, {Crocce}, {Cunha}, {D'Andrea}, {da Costa}, {Davis}, {Desai}, {Diehl}, {Dietrich}, {Dodelson}, {Doel}, {Drlica-Wagner}, {Estrada}, {Etherington}, {Evrard}, {Fabbri}, {Finley}, {Flaugher}, {Foley}, {Fosalba}, {Frieman}, {Garc{\'\i}a-Bellido}, {Gaztanaga}, {Gerdes}, {Giannantonio}, {Goldstein}, {Gruen}, {Gruendl}, {Guarnieri}, {Gutierrez}, {Hartley}, {Honscheid}, {Jain}, {James}, {Jeltema}, {Jouvel}, {Kessler}, {King}, {Kirk}, {Kron}, {Kuehn}, {Kuropatkin}, {Lahav}, {Li}, {Lima}, {Lin}, {Maia}, {Makler}, {Manera}, {Maraston}, {Marshall}, {Martini}, {McMahon},
  {Melchior}, {Merson}, {Miller}, {Miquel}, {Mohr}, {Morice-Atkinson}, {Naidoo}, {Neilsen}, {Nichol}, {Nord}, {Ogando}, {Ostrovski}, {Palmese}, {Papadopoulos}, {Peiris}, {Peoples}, {Percival}, {Plazas}, {Reed}, {Refregier}, {Romer}, {Roodman}, {Ross}, {Rozo}, {Rykoff}, {Sadeh}, {Sako}, {S{\'a}nchez}, {Sanchez}, {Santiago}, {Scarpine}, {Schubnell}, {Sevilla-Noarbe}, {Sheldon}, {Smith}, {Smith}, {Soares-Santos}, {Sobreira}, {Soumagnac}, {Suchyta}, {Sullivan}, {Swanson}, {Tarle}, {Thaler}, {Thomas}, {Thomas}, {Tucker}, {Vieira}, {Vikram}, {Walker}, {Wechsler}, {Weller}, {Wester}, {Whiteway}, {Wilcox}, {Yanny}, {Zhang}, \& {Zuntz}}]{DES}
{Dark Energy Survey Collaboration}, {Abbott}, T., {Abdalla}, F.~B., {et~al.} 2016, \mnras, 460, 1270

\bibitem[{{David} {et~al.}(2001){David}, {Nulsen}, {McNamara}, {Forman}, {Jones}, {Ponman}, {Robertson}, \& {Wise}}]{David_2001}
{David}, L.~P., {Nulsen}, P.~E.~J., {McNamara}, B.~R., {et~al.} 2001, \apj, 557, 546

\bibitem[{{Dennis} \& {Chandran}(2005)}]{Dennis_2005}
{Dennis}, T.~J., \& {Chandran}, B. D.~G. 2005, \apj, 622, 205

\bibitem[{{Dey} {et~al.}(2019){Dey}, {Schlegel}, {Lang}, {Blum}, {Burleigh}, {Fan}, {Findlay}, {Finkbeiner}, {Herrera}, {Juneau}, {Landriau}, {Levi}, {McGreer}, {Meisner}, {Myers}, {Moustakas}, {Nugent}, {Patej}, {Schlafly}, {Walker}, {Valdes}, {Weaver}, {Y{\`e}che}, {Zou}, {Zhou}, {Abareshi}, {Abbott}, {Abolfathi}, {Aguilera}, {Alam}, {Allen}, {Alvarez}, {Annis}, {Ansarinejad}, {Aubert}, {Beechert}, {Bell}, {BenZvi}, {Beutler}, {Bielby}, {Bolton}, {Brice{\~n}o}, {Buckley-Geer}, {Butler}, {Calamida}, {Carlberg}, {Carter}, {Casas}, {Castander}, {Choi}, {Comparat}, {Cukanovaite}, {Delubac}, {DeVries}, {Dey}, {Dhungana}, {Dickinson}, {Ding}, {Donaldson}, {Duan}, {Duckworth}, {Eftekharzadeh}, {Eisenstein}, {Etourneau}, {Fagrelius}, {Farihi}, {Fitzpatrick}, {Font-Ribera}, {Fulmer}, {G{\"a}nsicke}, {Gaztanaga}, {George}, {Gerdes}, {Gontcho}, {Gorgoni}, {Green}, {Guy}, {Harmer}, {Hernandez}, {Honscheid}, {Huang}, {James}, {Jannuzi}, {Jiang}, {Joyce}, {Karcher}, {Karkar}, {Kehoe}, {Kneib}, {Kueter-Young}, {Lan},
  {Lauer}, {Le Guillou}, {Le Van Suu}, {Lee}, {Lesser}, {Perreault Levasseur}, {Li}, {Mann}, {Marshall}, {Mart{\'\i}nez-V{\'a}zquez}, {Martini}, {du Mas des Bourboux}, {McManus}, {Meier}, {M{\'e}nard}, {Metcalfe}, {Mu{\~n}oz-Guti{\'e}rrez}, {Najita}, {Napier}, {Narayan}, {Newman}, {Nie}, {Nord}, {Norman}, {Olsen}, {Paat}, {Palanque-Delabrouille}, {Peng}, {Poppett}, {Poremba}, {Prakash}, {Rabinowitz}, {Raichoor}, {Rezaie}, {Robertson}, {Roe}, {Ross}, {Ross}, {Rudnick}, {Safonova}, {Saha}, {S{\'a}nchez}, {Savary}, {Schweiker}, {Scott}, {Seo}, {Shan}, {Silva}, {Slepian}, {Soto}, {Sprayberry}, {Staten}, {Stillman}, {Stupak}, {Summers}, {Sien Tie}, {Tirado}, {Vargas-Maga{\~n}a}, {Vivas}, {Wechsler}, {Williams}, {Yang}, {Yang}, {Yapici}, {Zaritsky}, {Zenteno}, {Zhang}, {Zhang}, {Zhou}, \& {Zhou}}]{Dey_2019}
{Dey}, A., {Schlegel}, D.~J., {Lang}, D., {et~al.} 2019, \aj, 157, 168

\bibitem[{{Driver} {et~al.}(2022){Driver}, {Bellstedt}, {Robotham}, {Baldry}, {Davies}, {Liske}, {Obreschkow}, {Taylor}, {Wright}, {Alpaslan}, {Bamford}, {Bauer}, {Bland-Hawthorn}, {Bilicki}, {Bravo}, {Brough}, {Casura}, {Cluver}, {Colless}, {Conselice}, {Croom}, {de Jong}, {D'Eugenio}, {De Propris}, {Dogruel}, {Drinkwater}, {Dvornik}, {Farrow}, {Frenk}, {Giblin}, {Graham}, {Grootes}, {Gunawardhana}, {Hashemizadeh}, {H{\"a}u{\ss}ler}, {Heymans}, {Hildebrandt}, {Holwerda}, {Hopkins}, {Jarrett}, {Heath Jones}, {Kelvin}, {Koushan}, {Kuijken}, {Lara-L{\'o}pez}, {Lange}, {L{\'o}pez-S{\'a}nchez}, {Loveday}, {Mahajan}, {Meyer}, {Moffett}, {Napolitano}, {Norberg}, {Owers}, {Radovich}, {Raouf}, {Peacock}, {Phillipps}, {Pimbblet}, {Popescu}, {Said}, {Sansom}, {Seibert}, {Sutherland}, {Thorne}, {Tuffs}, {Turner}, {van der Wel}, {van Kampen}, \& {Wilkins}}]{Gama09}
{Driver}, S.~P., {Bellstedt}, S., {Robotham}, A. S.~G., {et~al.} 2022, \mnras, 513, 439

\bibitem[{{Duchesne} {et~al.}(2024{\natexlab{a}}){Duchesne}, {Botteon}, {Koribalski}, {Loi}, {Rajpurohit}, {Riseley}, {Rudnick}, {Vernstrom}, {Andernach}, {Hopkins}, {Kapinska}, {Norris}, \& {Zafar}}]{Duchesne_2024}
{Duchesne}, S.~W., {Botteon}, A., {Koribalski}, B.~S., {et~al.} 2024{\natexlab{a}}, \pasa, 41, e026

\bibitem[{{Duchesne} {et~al.}(2024{\natexlab{b}}){Duchesne}, {Grundy}, {Heald}, {Lenc}, {Leung}, {McConnell}, {Murphy}, {Pritchard}, {Rose}, {Thomson}, {Wang}, {Wang}, \& {Whiting}}]{Duchesne_2024RACS}
{Duchesne}, S.~W., {Grundy}, J.~A., {Heald}, G.~H., {et~al.} 2024{\natexlab{b}}, \pasa, 41, e003

\bibitem[{{Dunn} {et~al.}(2005){Dunn}, {Fabian}, \& {Taylor}}]{Dunn_2005}
{Dunn}, R.~J.~H., {Fabian}, A.~C., \& {Taylor}, G.~B. 2005, \mnras, 364, 1343

\bibitem[{{Edge}(2001)}]{Edge_2001}
{Edge}, A.~C. 2001, \mnras, 328, 762

\bibitem[{{Edwards} {et~al.}(2007){Edwards}, {Hudson}, {Balogh}, \& {Smith}}]{Edwards_2007}
{Edwards}, L. O.~V., {Hudson}, M.~J., {Balogh}, M.~L., \& {Smith}, R.~J. 2007, \mnras, 379, 100

\bibitem[{Fabian(1994)}]{Fabian_1994}
Fabian, A.~C. 1994, Annual Review of Astronomy and Astrophysics, 32, 277

\bibitem[{{Fabian}(2012)}]{Fabian_2012}
{Fabian}, A.~C. 2012, \araa, 50, 455

\bibitem[{{Fabian} {et~al.}(2005){Fabian}, {Reynolds}, {Taylor}, \& {Dunn}}]{Fabian_2005}
{Fabian}, A.~C., {Reynolds}, C.~S., {Taylor}, G.~B., \& {Dunn}, R.~J.~H. 2005, \mnras, 363, 891

\bibitem[{{Fabian} {et~al.}(2000){Fabian}, {Sanders}, {Ettori}, {Taylor}, {Allen}, {Crawford}, {Iwasawa}, {Johnstone}, \& {Ogle}}]{Fabian_2000}
{Fabian}, A.~C., {Sanders}, J.~S., {Ettori}, S., {et~al.} 2000, \mnras, 318, L65

\bibitem[{{Feigelson} {et~al.}(2014){Feigelson}, {Nelson}, {Isobe}, \& {LaValley}}]{asurv}
{Feigelson}, E.~D., {Nelson}, P.~I., {Isobe}, T., \& {LaValley}, M. 2014, {ASURV: Astronomical SURVival Statistics}, Astrophysics Source Code Library, record ascl:1406.001

\bibitem[{{Fujita} \& {Reiprich}(2004)}]{Fujita_2004}
{Fujita}, Y., \& {Reiprich}, T.~H. 2004, \apj, 612, 797

\bibitem[{Gaspari {et~al.}(2019)Gaspari, Eckert, Ettori, Tozzi, Bassini, Rasia, Brighenti, Sun, Borgani, Johnson, Tremblay, Stone, Temi, Yang, Tombesi, \& Cappi}]{Gaspari_2019}
Gaspari, M., Eckert, D., Ettori, S., {et~al.} 2019, The Astrophysical Journal, 884, 169

\bibitem[{{Ghirardini} {et~al.}(2024){Ghirardini}, {Bulbul}, {Artis}, {Clerc}, {Garrel}, {Grandis}, {Kluge}, {Liu}, {Bahar}, {Balzer}, {Chiu}, {Comparat}, {Gruen}, {Kleinebreil}, {Krippendorf}, {Merloni}, {Nandra}, {Okabe}, {Pacaud}, {Predehl}, {Ramos-Ceja}, {Reiprich}, {Sanders}, {Schrabback}, {Seppi}, {Zelmer}, {Zhang}, {Bornemann}, {Brunner}, {Burwitz}, {Coutinho}, {Dennerl}, {Freyberg}, {Friedrich}, {Gaida}, {Gueguen}, {Haberl}, {Kink}, {Lamer}, {Li}, {Liu}, {Maitra}, {Meidinger}, {Mueller}, {Miyatake}, {Miyazaki}, {Robrade}, {Schwope}, \& {Stewart}}]{erass1cosmo}
{Ghirardini}, V., {Bulbul}, E., {Artis}, E., {et~al.} 2024, \aap, 689, A298

\bibitem[{{Guo} \& {Oh}(2008)}]{Guo_2008}
{Guo}, F., \& {Oh}, S.~P. 2008, \mnras, 384, 251

\bibitem[{{Gupta} {et~al.}(2020){Gupta}, {Pannella}, {Mohr}, {Klein}, {Rykoff}, {Annis}, {Avila}, {Bianchini}, {Brooks}, {Buckley-Geer}, {Bulbul}, {Carnero Rosell}, {Carrasco Kind}, {Carretero}, {Chiu}, {Costanzi}, {da Costa}, {De Vicente}, {Desai}, {Dietrich}, {Doel}, {Everett}, {Evrard}, {Garc{\'\i}a-Bellido}, {Gaztanaga}, {Gruen}, {Gruendl}, {Gschwend}, {Gutierrez}, {Hollowood}, {Honscheid}, {James}, {Jeltema}, {Kuehn}, {Lidman}, {Lima}, {Maia}, {Marshall}, {McDonald}, {Menanteau}, {Miquel}, {Ogando}, {Palmese}, {Paz-Chinch{\'o}n}, {Plazas}, {Reichardt}, {Sanchez}, {Santiago}, {Saro}, {Scarpine}, {Schindler}, {Schubnell}, {Serrano}, {Sevilla-Noarbe}, {Shao}, {Smith}, {Stott}, {Strazzullo}, {Suchyta}, {Swanson}, {Vikram}, \& {Zenteno}}]{Gupta_2020}
{Gupta}, N., {Pannella}, M., {Mohr}, J.~J., {et~al.} 2020, \mnras, 494, 1705

\bibitem[{{Guzman} {et~al.}(2019){Guzman}, {Whiting}, {Voronkov}, {Mitchell}, {Ord}, {Collins}, {Marquarding}, {Lahur}, {Maher}, {Van Diepen}, {Bannister}, {Wu}, {Lenc}, {Khoo}, \& {Bastholm}}]{Guzman_2019}
{Guzman}, J., {Whiting}, M., {Voronkov}, M., {et~al.} 2019, {ASKAPsoft: ASKAP science data processor software}, Astrophysics Source Code Library, record ascl:1912.003

\bibitem[{{Hardcastle} {et~al.}(2016){Hardcastle}, {G{\"u}rkan}, {van Weeren}, {Williams}, {Best}, {de Gasperin}, {Rafferty}, {Read}, {Sabater}, {Shimwell}, {Smith}, {Tasse}, {Bourne}, {Brienza}, {Br{\"u}ggen}, {Brunetti}, {Chy{\.z}y}, {Conway}, {Dunne}, {Eales}, {Maddox}, {Jarvis}, {Mahony}, {Morganti}, {Prandoni}, {R{\"o}ttgering}, {Valiante}, \& {White}}]{Hardcastle_2016}
{Hardcastle}, M.~J., {G{\"u}rkan}, G., {van Weeren}, R.~J., {et~al.} 2016, \mnras, 462, 1910

\bibitem[{{Hardcastle} {et~al.}(2019){Hardcastle}, {Williams}, {Best}, {Croston}, {Duncan}, {R{\"o}ttgering}, {Sabater}, {Shimwell}, {Tasse}, {Callingham}, {Cochrane}, {de Gasperin}, {G{\"u}rkan}, {Jarvis}, {Mahatma}, {Miley}, {Mingo}, {Mooney}, {Morabito}, {O'Sullivan}, {Prandoni}, {Shulevski}, \& {Smith}}]{Hardcastle_2019}
{Hardcastle}, M.~J., {Williams}, W.~L., {Best}, P.~N., {et~al.} 2019, \aap, 622, A12

\bibitem[{{Heckman} \& {Best}(2014)}]{Heckman_2014}
{Heckman}, T.~M., \& {Best}, P.~N. 2014, \araa, 52, 589

\bibitem[{{Hlavacek-Larrondo} {et~al.}(2022){Hlavacek-Larrondo}, {Li}, \& {Churazov}}]{Hlavacek_2022}
{Hlavacek-Larrondo}, J., {Li}, Y., \& {Churazov}, E. 2022, in Handbook of X-ray and Gamma-ray Astrophysics, ed. C.~{Bambi} \& A.~{Sangangelo}, 5

\bibitem[{{Hopkins} {et~al.}(2025){Hopkins}, {Kapinska}, {Marvil}, {Vernstrom}, {Collier}, {Norris}, {Gordon}, {Duchesne}, {Rudnick}, {Gupta}, {Carretti}, {Anderson}, {Dai}, {G{\"u}rkan}, {Parkinson}, {Prandoni}, {Riggi}, {Shekhar Saraf}, {Ma}, {Filipovi{\'c}}, {Umana}, {Bahr-Kalus}, {Koribalski}, {Lenc}, {Ingallinera}, {Afonso}, {Ahmad}, {Ahmed}, {Alexander}, {Andernach}, {Asorey}, {Battisti}, {Bilicki}, {Botteon}, {Brown}, {Br{\"u}ggen}, {Cowley}, {Dage}, {Hale}, {Hardcastle}, {Kothes}, {Lazarevi{\'c}}, {Lin}, {Luken}, {Moss}, {Prathap}, {ur Rahman}, {Reiprich}, {Riseley}, {Salvato}, {Seymour}, {Shabala}, {Smith}, {Vaccari}, {van Loon}, {Wong}, {Zainal Alsaberi}, {Asher}, {Ball}, {Barbosa}, {Biava}, {Bradley}, {Carvajal}, {Crawford}, {Galvin}, {Huynh}, {Leahy}, {Matute}, {Moss}, {Pappalardo}, {Smeaton}, {Velovi{\'c}}, \& {Zafar}}]{Hopkins_2025}
{Hopkins}, A., {Kapinska}, A., {Marvil}, J., {et~al.} 2025, \pasa, 42, e071

\bibitem[{{Hotan} {et~al.}(2021){Hotan}, {Bunton}, {Chippendale}, {Whiting}, {Tuthill}, {Moss}, {McConnell}, {Amy}, {Huynh}, {Allison}, {Anderson}, {Bannister}, {Bastholm}, {Beresford}, {Bock}, {Bolton}, {Chapman}, {Chow}, {Collier}, {Cooray}, {Cornwell}, {Diamond}, {Edwards}, {Feain}, {Franzen}, {George}, {Gupta}, {Hampson}, {Harvey-Smith}, {Hayman}, {Heywood}, {Jacka}, {Jackson}, {Jackson}, {Jeganathan}, {Johnston}, {Kesteven}, {Kleiner}, {Koribalski}, {Lee-Waddell}, {Lenc}, {Lensson}, {Mackay}, {Mahony}, {McClure-Griffiths}, {McConigley}, {Mirtschin}, {Ng}, {Norris}, {Pearce}, {Phillips}, {Pilawa}, {Raja}, {Reynolds}, {Roberts}, {Roxby}, {Sadler}, {Shields}, {Schinckel}, {Serra}, {Shaw}, {Sweetnam}, {Troup}, {Tzioumis}, {Voronkov}, \& {Westmeier}}]{Hotan_2021}
{Hotan}, A.~W., {Bunton}, J.~D., {Chippendale}, A.~P., {et~al.} 2021, \pasa, 38, e009

\bibitem[{{Hudson} {et~al.}(2010){Hudson}, {Mittal}, {Reiprich}, {Nulsen}, {Andernach}, \& {Sarazin}}]{Hudson_2010}
{Hudson}, D.~S., {Mittal}, R., {Reiprich}, T.~H., {et~al.} 2010, \aap, 513, A37

\bibitem[{{Ider Chitham} {et~al.}(2020){Ider Chitham}, {Comparat}, {Finoguenov}, {Clerc}, {Kirkpatrick}, {Damsted}, {Kukkola}, {Capasso}, {Nandra}, {Merloni}, {Bulbul}, {Rykoff}, {Schneider}, \& {Brownstein}}]{Chitham_2020}
{Ider Chitham}, J., {Comparat}, J., {Finoguenov}, A., {et~al.} 2020, \mnras, 499, 4768

\bibitem[{{Igo} \& {Merloni}(2025)}]{Igo_2025}
{Igo}, Z., \& {Merloni}, A. 2025, \aap, 697, A196

\bibitem[{{Ineson} {et~al.}(2015){Ineson}, {Croston}, {Hardcastle}, {Kraft}, {Evans}, \& {Jarvis}}]{Ineson_2015}
{Ineson}, J., {Croston}, J.~H., {Hardcastle}, M.~J., {et~al.} 2015, \mnras, 453, 2682

\bibitem[{{Isobe} {et~al.}(1986){Isobe}, {Feigelson}, \& {Nelson}}]{Isobe_1986}
{Isobe}, T., {Feigelson}, E.~D., \& {Nelson}, P.~I. 1986, \apj, 306, 490

\bibitem[{{Israel} {et~al.}(2014){Israel}, {Reiprich}, {Erben}, {Massey}, {Sarazin}, {Schneider}, \& {Vikhlinin}}]{Israel_2014}
{Israel}, H., {Reiprich}, T.~H., {Erben}, T., {et~al.} 2014, \aap, 564, A129

\bibitem[{{Johnston} {et~al.}(2008){Johnston}, {Taylor}, {Bailes}, {Bartel}, {Baugh}, {Bietenholz}, {Blake}, {Braun}, {Brown}, {Chatterjee}, {Darling}, {Deller}, {Dodson}, {Edwards}, {Ekers}, {Ellingsen}, {Feain}, {Gaensler}, {Haverkorn}, {Hobbs}, {Hopkins}, {Jackson}, {James}, {Joncas}, {Kaspi}, {Kilborn}, {Koribalski}, {Kothes}, {Landecker}, {Lenc}, {Lovell}, {Macquart}, {Manchester}, {Matthews}, {McClure-Griffiths}, {Norris}, {Pen}, {Phillips}, {Power}, {Protheroe}, {Sadler}, {Schmidt}, {Stairs}, {Staveley-Smith}, {Stil}, {Tingay}, {Tzioumis}, {Walker}, {Wall}, \& {Wolleben}}]{Johnston_2008}
{Johnston}, S., {Taylor}, R., {Bailes}, M., {et~al.} 2008, Experimental Astronomy, 22, 151

\bibitem[{{Kelly}(2007)}]{linmix}
{Kelly}, B.~C. 2007, \apj, 665, 1489

\bibitem[{{Kendall}(1938)}]{kendall}
{Kendall}, M.~G. 1938, Biometrika, 30, 81

\bibitem[{{Kirkpatrick} \& {McNamara}(2015)}]{Kirkpatrick_2015}
{Kirkpatrick}, C.~C., \& {McNamara}, B.~R. 2015, \mnras, 452, 4361

\bibitem[{{Kluge} {et~al.}(2024){Kluge}, {Comparat}, {Liu}, {Balzer}, {Bulbul}, {Ider Chitham}, {Ghirardini}, {Garrel}, {Bahar}, {Artis}, {Bender}, {Clerc}, {Dwelly}, {Fabricius}, {Grandis}, {Hern{\'a}ndez-Lang}, {Hill}, {Joshi}, {Lamer}, {Merloni}, {Nandra}, {Pacaud}, {Predehl}, {Ramos-Ceja}, {Reiprich}, {Salvato}, {Sanders}, {Schrabback}, {Seppi}, {Zelmer}, {Zenteno}, \& {Zhang}}]{Kluge_2024}
{Kluge}, M., {Comparat}, J., {Liu}, A., {et~al.} 2024, \aap, 688, A210

\bibitem[{{Kolokythas} {et~al.}(2018){Kolokythas}, {O'Sullivan}, {Raychaudhury}, {Giacintucci}, {Gitti}, \& {Babul}}]{Kolokythas_2018}
{Kolokythas}, K., {O'Sullivan}, E., {Raychaudhury}, S., {et~al.} 2018, \mnras, 481, 1550

\bibitem[{{Liu} {et~al.}(2022){Liu}, {Bulbul}, {Ghirardini}, {Liu}, {Klein}, {Clerc}, {{\"O}zsoy}, {Ramos-Ceja}, {Pacaud}, {Comparat}, {Okabe}, {Bahar}, {Biffi}, {Brunner}, {Br{\"u}ggen}, {Buchner}, {Ider Chitham}, {Chiu}, {Dolag}, {Gatuzz}, {Gonzalez}, {Hoang}, {Lamer}, {Merloni}, {Nandra}, {Oguri}, {Ota}, {Predehl}, {Reiprich}, {Salvato}, {Schrabback}, {Sanders}, {Seppi}, \& {Thibaud}}]{Liu_2022}
{Liu}, A., {Bulbul}, E., {Ghirardini}, V., {et~al.} 2022, \aap, 661, A2

\bibitem[{{Lovisari} {et~al.}(2017){Lovisari}, {Forman}, {Jones}, {Ettori}, {Andrade-Santos}, {Arnaud}, {D{\'e}mocl{\`e}s}, {Pratt}, {Randall}, \& {Kraft}}]{Lovisari_2017}
{Lovisari}, L., {Forman}, W.~R., {Jones}, C., {et~al.} 2017, \apj, 846, 51

\bibitem[{Lovisari {et~al.}(2020)Lovisari, Schellenberger, Sereno, Ettori, Pratt, Forman, Jones, Andrade-Santos, Randall, \& Kraft}]{Lovisari_2020}
Lovisari, L., Schellenberger, G., Sereno, M., {et~al.} 2020, The Astrophysical Journal, 892, 102

\bibitem[{{Main} {et~al.}(2017){Main}, {McNamara}, {Nulsen}, {Russell}, \& {Vantyghem}}]{Main_2017}
{Main}, R.~A., {McNamara}, B.~R., {Nulsen}, P.~E.~J., {Russell}, H.~R., \& {Vantyghem}, A.~N. 2017, \mnras, 464, 4360

\bibitem[{{Mauch} {et~al.}(2003){Mauch}, {Murphy}, {Buttery}, {Curran}, {Hunstead}, {Piestrzynski}, {Robertson}, \& {Sadler}}]{Mauch_2003}
{Mauch}, T., {Murphy}, T., {Buttery}, H.~J., {et~al.} 2003, \mnras, 342, 1117

\bibitem[{{McConnell} {et~al.}(2016){McConnell}, {Allison}, {Bannister}, {Bell}, {Bignall}, {Chippendale}, {Edwards}, {Harvey-Smith}, {Hegarty}, {Heywood}, {Hotan}, {Indermuehle}, {Lenc}, {Marvil}, {Popping}, {Raja}, {Reynolds}, {Sault}, {Serra}, {Voronkov}, {Whiting}, {Amy}, {Axtens}, {Ball}, {Bateman}, {Bock}, {Bolton}, {Brodrick}, {Brothers}, {Brown}, {Bunton}, {Cheng}, {Cornwell}, {DeBoer}, {Feain}, {Gough}, {Gupta}, {Guzman}, {Hampson}, {Hay}, {Hayman}, {Hoyle}, {Humphreys}, {Jacka}, {Jackson}, {Jackson}, {Jeganathan}, {Joseph}, {Koribalski}, {Leach}, {Lensson}, {MacLeod}, {Mackay}, {Marquarding}, {McClure-Griffiths}, {Mirtschin}, {Mitchell}, {Neuhold}, {Ng}, {Norris}, {Pearce}, {Qiao}, {Schinckel}, {Shields}, {Shimwell}, {Storey}, {Troup}, {Turner}, {Tuthill}, {Tzioumis}, {Wark}, {Westmeier}, {Wilson}, \& {Wilson}}]{McConnell_2016}
{McConnell}, D., {Allison}, J.~R., {Bannister}, K., {et~al.} 2016, \pasa, 33, e042

\bibitem[{{McNamara} \& {Nulsen}(2007)}]{McNamara_2007}
{McNamara}, B.~R., \& {Nulsen}, P.~E.~J. 2007, \araa, 45, 117

\bibitem[{{McNamara} \& {Nulsen}(2012)}]{McNamara_2012}
---. 2012, New Journal of Physics, 14, 055023

\bibitem[{{McNamara} \& {O'Connell}(1989)}]{McNamara_1989}
{McNamara}, B.~R., \& {O'Connell}, R.~W. 1989, \aj, 98, 2018

\bibitem[{{McNamara} {et~al.}(2004){McNamara}, {Wise}, \& {Murray}}]{McNamara_2004}
{McNamara}, B.~R., {Wise}, M.~W., \& {Murray}, S.~S. 2004, \apj, 601, 173

\bibitem[{{McNamara} {et~al.}(2000){McNamara}, {Wise}, {Nulsen}, {David}, {Sarazin}, {Bautz}, {Markevitch}, {Vikhlinin}, {Forman}, {Jones}, \& {Harris}}]{McNamara_2000}
{McNamara}, B.~R., {Wise}, M., {Nulsen}, P.~E.~J., {et~al.} 2000, \apjl, 534, L135

\bibitem[{{Merloni} \& {Heinz}(2007)}]{Merloni_2007}
{Merloni}, A., \& {Heinz}, S. 2007, \mnras, 381, 589

\bibitem[{{Merloni} {et~al.}(2024){Merloni}, {Lamer}, {Liu}, {Ramos-Ceja}, {Brunner}, {Bulbul}, {Dennerl}, {Doroshenko}, {Freyberg}, {Friedrich}, {Gatuzz}, {Georgakakis}, {Haberl}, {Igo}, {Kreykenbohm}, {Liu}, {Maitra}, {Malyali}, {Mayer}, {Nandra}, {Predehl}, {Robrade}, {Salvato}, {Sanders}, {Stewart}, {Tub{\'\i}n-Arenas}, {Weber}, {Wilms}, {Arcodia}, {Artis}, {Aschersleben}, {Avakyan}, {Aydar}, {Bahar}, {Balzer}, {Becker}, {Berger}, {Boller}, {Bornemann}, {Br{\"u}ggen}, {Brusa}, {Buchner}, {Burwitz}, {Camilloni}, {Clerc}, {Comparat}, {Coutinho}, {Czesla}, {Dannhauer}, {Dauner}, {Dauser}, {Dietl}, {Dolag}, {Dwelly}, {Egg}, {Ehl}, {Freund}, {Friedrich}, {Gaida}, {Garrel}, {Ghirardini}, {Gokus}, {Gr{\"u}nwald}, {Grandis}, {Grotova}, {Gruen}, {Gueguen}, {H{\"a}mmerich}, {Hamaus}, {Hasinger}, {Haubner}, {Homan}, {Ider Chitham}, {Joseph}, {Joyce}, {K{\"o}nig}, {Kaltenbrunner}, {Khokhriakova}, {Kink}, {Kirsch}, {Kluge}, {Knies}, {Krippendorf}, {Krumpe}, {Kurpas}, {Li}, {Liu}, {Locatelli}, {Lorenz}, {M{\"u}ller},
  {Magaudda}, {Mannes}, {McCall}, {Meidinger}, {Michailidis}, {Migkas}, {Mu{\~n}oz-Giraldo}, {Musiimenta}, {Nguyen-Dang}, {Ni}, {Olechowska}, {Ota}, {Pacaud}, {Pasini}, {Perinati}, {Pires}, {Pommranz}, {Ponti}, {Poppenhaeger}, {P{\"u}hlhofer}, {Rau}, {Reh}, {Reiprich}, {Roster}, {Saeedi}, {Santangelo}, {Sasaki}, {Schmitt}, {Schneider}, {Schrabback}, {Schuster}, {Schwope}, {Seppi}, {Serim}, {Shreeram}, {Sokolova-Lapa}, {Starck}, {Stelzer}, {Stierhof}, {Suleimanov}, {Tenzer}, {Traulsen}, {Tr{\"u}mper}, {Tsuge}, {Urrutia}, {Veronica}, {Waddell}, {Willer}, {Wolf}, {Yeung}, {Zainab}, {Zangrandi}, {Zhang}, {Zhang}, \& {Zheng}}]{Merloni_2024}
{Merloni}, A., {Lamer}, G., {Liu}, T., {et~al.} 2024, \aap, 682, A34

\bibitem[{{Mittal} {et~al.}(2009){Mittal}, {Hudson}, {Reiprich}, \& {Clarke}}]{Mittal_2009}
{Mittal}, R., {Hudson}, D.~S., {Reiprich}, T.~H., \& {Clarke}, T. 2009, \aap, 501, 835

\bibitem[{{Mohan} \& {Rafferty}(2015)}]{pybdsf}
{Mohan}, N., \& {Rafferty}, D. 2015, {PyBDSF: Python Blob Detection and Source Finder}, Astrophysics Source Code Library, record ascl:1502.007

\bibitem[{{Norris} {et~al.}(2021){Norris}, {Marvil}, {Collier}, {Kapi{\'n}ska}, {O'Brien}, {Rudnick}, {Andernach}, {Asorey}, {Brown}, {Br{\"u}ggen}, {Crawford}, {English}, {Rahman}, {Filipovi{\'c}}, {Gordon}, {G{\"u}rkan}, {Hale}, {Hopkins}, {Huynh}, {HyeongHan}, {James Jee}, {Koribalski}, {Lenc}, {Luken}, {Parkinson}, {Prandoni}, {Raja}, {Reiprich}, {Riseley}, {Shabala}, {Sheil}, {Vernstrom}, {Whiting}, {Allison}, {Anderson}, {Ball}, {Bell}, {Bunton}, {Galvin}, {Gupta}, {Hotan}, {Jacka}, {Macgregor}, {Mahony}, {Maio}, {Moss}, {Pandey-Pommier}, \& {Voronkov}}]{Norris_2021}
{Norris}, R.~P., {Marvil}, J., {Collier}, J.~D., {et~al.} 2021, \pasa, 38, e046

\bibitem[{{O'Dea} {et~al.}(2009){O'Dea}, {Daly}, {Kharb}, {Freeman}, \& {Baum}}]{ODea_2009}
{O'Dea}, C.~P., {Daly}, R.~A., {Kharb}, P., {Freeman}, K.~A., \& {Baum}, S.~A. 2009, \aap, 494, 471

\bibitem[{O'Dea {et~al.}(2008)O'Dea, Baum, Privon, Noel-Storr, Quillen, Zufelt, Park, Edge, Russell, Fabian, Donahue, Sarazin, McNamara, Bregman, \& Egami}]{ODea_2008}
O'Dea, C.~P., Baum, S.~A., Privon, G., {et~al.} 2008, The Astrophysical Journal, 681, 1035

\bibitem[{{Pasini} {et~al.}(2021{\natexlab{a}}){Pasini}, {Finoguenov}, {Br{\"u}ggen}, {Gaspari}, {de Gasperin}, \& {Gozaliasl}}]{Pasini_2021}
{Pasini}, T., {Finoguenov}, A., {Br{\"u}ggen}, M., {et~al.} 2021{\natexlab{a}}, \mnras, 505, 2628

\bibitem[{{Pasini} {et~al.}(2021{\natexlab{b}}){Pasini}, {Gitti}, {Brighenti}, {O'Sullivan}, {Gastaldello}, {Temi}, \& {Hamer}}]{PasiniA1668_2021}
{Pasini}, T., {Gitti}, M., {Brighenti}, F., {et~al.} 2021{\natexlab{b}}, \apj, 911, 66

\bibitem[{{Pasini} {et~al.}(2019){Pasini}, {Gitti}, {Brighenti}, {Temi}, {Amblard}, {Hamer}, {Ettori}, {O'Sullivan}, \& {Gastaldello}}]{Pasini_2019}
---. 2019, \apj, 885, 111

\bibitem[{{Pasini} {et~al.}(2022){Pasini}, {Br{\"u}ggen}, {Hoang}, {Ghirardini}, {Bulbul}, {Klein}, {Liu}, {Shimwell}, {Hardcastle}, {Williams}, {Botteon}, {Gastaldello}, {van Weeren}, {Merloni}, {de Gasperin}, {Bahar}, {Pacaud}, \& {Ramos-Ceja}}]{Pasini_2022}
{Pasini}, T., {Br{\"u}ggen}, M., {Hoang}, D.~N., {et~al.} 2022, \aap, 661, A13

\bibitem[{{Peterson} {et~al.}(2001){Peterson}, {Paerels}, {Kaastra}, {Arnaud}, {Reiprich}, {Fabian}, {Mushotzky}, {Jernigan}, \& {Sakelliou}}]{Peterson_2001}
{Peterson}, J.~R., {Paerels}, F.~B.~S., {Kaastra}, J.~S., {et~al.} 2001, \aap, 365, L104

\bibitem[{{Predehl} {et~al.}(2021){Predehl}, {Andritschke}, {Arefiev}, {Babyshkin}, {Batanov}, {Becker}, {B{\"o}hringer}, {Bogomolov}, {Boller}, {Borm}, {Bornemann}, {Br{\"a}uninger}, {Br{\"u}ggen}, {Brunner}, {Brusa}, {Bulbul}, {Buntov}, {Burwitz}, {Burkert}, {Clerc}, {Churazov}, {Coutinho}, {Dauser}, {Dennerl}, {Doroshenko}, {Eder}, {Emberger}, {Eraerds}, {Finoguenov}, {Freyberg}, {Friedrich}, {Friedrich}, {F{\"u}rmetz}, {Georgakakis}, {Gilfanov}, {Granato}, {Grossberger}, {Gueguen}, {Gureev}, {Haberl}, {H{\"a}lker}, {Hartner}, {Hasinger}, {Huber}, {Ji}, {Kienlin}, {Kink}, {Korotkov}, {Kreykenbohm}, {Lamer}, {Lomakin}, {Lapshov}, {Liu}, {Maitra}, {Meidinger}, {Menz}, {Merloni}, {Mernik}, {Mican}, {Mohr}, {M{\"u}ller}, {Nandra}, {Nazarov}, {Pacaud}, {Pavlinsky}, {Perinati}, {Pfeffermann}, {Pietschner}, {Ramos-Ceja}, {Rau}, {Reiffers}, {Reiprich}, {Robrade}, {Salvato}, {Sanders}, {Santangelo}, {Sasaki}, {Scheuerle}, {Schmid}, {Schmitt}, {Schwope}, {Shirshakov}, {Steinmetz}, {Stewart}, {Str{\"u}der},
  {Sunyaev}, {Tenzer}, {Tiedemann}, {Tr{\"u}mper}, {Voron}, {Weber}, {Wilms}, \& {Yaroshenko}}]{Predehl_2021}
{Predehl}, P., {Andritschke}, R., {Arefiev}, V., {et~al.} 2021, \aap, 647, A1

\bibitem[{Rafferty {et~al.}(2008)Rafferty, McNamara, \& Nulsen}]{Rafferty_2008}
Rafferty, D.~A., McNamara, B.~R., \& Nulsen, P. E.~J. 2008, The Astrophysical Journal, 687, 899

\bibitem[{Randall {et~al.}(2010)Randall, Forman, Giacintucci, Nulsen, Sun, Jones, Churazov, David, Kraft, Donahue, Blanton, Simionescu, \& Werner}]{Randall_2011}
Randall, S.~W., Forman, W.~R., Giacintucci, S., {et~al.} 2010, The Astrophysical Journal, 726, 86

\bibitem[{{Reiprich} \& {B{\"o}hringer}(2002)}]{Reiprich_2002}
{Reiprich}, T.~H., \& {B{\"o}hringer}, H. 2002, \apj, 567, 716

\bibitem[{{Rykoff} {et~al.}(2014){Rykoff}, {Rozo}, {Busha}, {Cunha}, {Finoguenov}, {Evrard}, {Hao}, {Koester}, {Leauthaud}, {Nord}, {Pierre}, {Reddick}, {Sadibekova}, {Sheldon}, \& {Wechsler}}]{Rykoff_2014}
{Rykoff}, E.~S., {Rozo}, E., {Busha}, M.~T., {et~al.} 2014, \apj, 785, 104

\bibitem[{{Rykoff} {et~al.}(2016){Rykoff}, {Rozo}, {Hollowood}, {Bermeo-Hernandez}, {Jeltema}, {Mayers}, {Romer}, {Rooney}, {Saro}, {Vergara Cervantes}, {Wechsler}, {Wilcox}, {Abbott}, {Abdalla}, {Allam}, {Annis}, {Benoit-L{\'e}vy}, {Bernstein}, {Bertin}, {Brooks}, {Burke}, {Capozzi}, {Carnero Rosell}, {Carrasco Kind}, {Castander}, {Childress}, {Collins}, {Cunha}, {D'Andrea}, {da Costa}, {Davis}, {Desai}, {Diehl}, {Dietrich}, {Doel}, {Evrard}, {Finley}, {Flaugher}, {Fosalba}, {Frieman}, {Glazebrook}, {Goldstein}, {Gruen}, {Gruendl}, {Gutierrez}, {Hilton}, {Honscheid}, {Hoyle}, {James}, {Kay}, {Kuehn}, {Kuropatkin}, {Lahav}, {Lewis}, {Lidman}, {Lima}, {Maia}, {Mann}, {Marshall}, {Martini}, {Melchior}, {Miller}, {Miquel}, {Mohr}, {Nichol}, {Nord}, {Ogando}, {Plazas}, {Reil}, {Sahl{\'e}n}, {Sanchez}, {Santiago}, {Scarpine}, {Schubnell}, {Sevilla-Noarbe}, {Smith}, {Soares-Santos}, {Sobreira}, {Stott}, {Suchyta}, {Swanson}, {Tarle}, {Thomas}, {Tucker}, {Uddin}, {Viana}, {Vikram}, {Walker}, {Zhang}, \& {DES
  Collaboration}}]{Rykoff_2016}
{Rykoff}, E.~S., {Rozo}, E., {Hollowood}, D., {et~al.} 2016, \apjs, 224, 1

\bibitem[{{Sanders} {et~al.}(2008){Sanders}, {Fabian}, {Allen}, {Morris}, {Graham}, \& {Johnstone}}]{Sanders_2008}
{Sanders}, J.~S., {Fabian}, A.~C., {Allen}, S.~W., {et~al.} 2008, \mnras, 385, 1186

\bibitem[{{Sanders} {et~al.}(2018){Sanders}, {Fabian}, {Russell}, \& {Walker}}]{Sanders_2018}
{Sanders}, J.~S., {Fabian}, A.~C., {Russell}, H.~R., \& {Walker}, S.~A. 2018, \mnras, 474, 1065

\bibitem[{{Sanders} {et~al.}(2025){Sanders}, {Bahar}, {Bulbul}, {Ghirardini}, {Liu}, {Clerc}, {Ramos-Ceja}, {Reiprich}, {Balzer}, {Comparat}, {Kluge}, {Pacaud}, \& {Zhang}}]{Sanders_2025}
{Sanders}, J.~S., {Bahar}, Y.~E., {Bulbul}, E., {et~al.} 2025, \aap, 695, A160

\bibitem[{{Santos} {et~al.}(2008){Santos}, {Rosati}, {Tozzi}, {B{\"o}hringer}, {Ettori}, \& {Bignamini}}]{Santos_2008}
{Santos}, J.~S., {Rosati}, P., {Tozzi}, P., {et~al.} 2008, \aap, 483, 35

\bibitem[{{Scheuer}(1974)}]{Scheuer_1974}
{Scheuer}, P.~A.~G. 1974, \mnras, 166, 513

\bibitem[{{Seppi} {et~al.}(2023){Seppi}, {Comparat}, {Nandra}, {Dolag}, {Biffi}, {Bulbul}, {Liu}, {Ghirardini}, \& {Ider-Chitham}}]{Seppi_2023}
{Seppi}, R., {Comparat}, J., {Nandra}, K., {et~al.} 2023, \aap, 671, A57

\bibitem[{{Shabala} {et~al.}(2008){Shabala}, {Ash}, {Alexander}, \& {Riley}}]{Shabala_2008}
{Shabala}, S.~S., {Ash}, S., {Alexander}, P., \& {Riley}, J.~M. 2008, \mnras, 388, 625

\bibitem[{{Shabala} \& {Godfrey}(2013)}]{Shabala_2013}
{Shabala}, S.~S., \& {Godfrey}, L.~E.~H. 2013, \apj, 769, 129

\bibitem[{{Soker}(2003)}]{Soker_2003}
{Soker}, N. 2003, \mnras, 342, 463

\bibitem[{{Stott} {et~al.}(2012){Stott}, {Hickox}, {Edge}, {Collins}, {Hilton}, {Harrison}, {Romer}, {Rooney}, {Kay}, {Miller}, {Sahl{\'e}n}, {Lloyd-Davies}, {Mehrtens}, {Hoyle}, {Liddle}, {Viana}, {McCarthy}, {Schaye}, \& {Booth}}]{Stott_2012}
{Stott}, J.~P., {Hickox}, R.~C., {Edge}, A.~C., {et~al.} 2012, \mnras, 422, 2213

\bibitem[{{Sunyaev} {et~al.}(2021){Sunyaev}, {Arefiev}, {Babyshkin}, {Bogomolov}, {Borisov}, {Buntov}, {Brunner}, {Burenin}, {Churazov}, {Coutinho}, {Eder}, {Eismont}, {Freyberg}, {Gilfanov}, {Gureyev}, {Hasinger}, {Khabibullin}, {Kolmykov}, {Komovkin}, {Krivonos}, {Lapshov}, {Levin}, {Lomakin}, {Lutovinov}, {Medvedev}, {Merloni}, {Mernik}, {Mikhailov}, {Molodtsov}, {Mzhelsky}, {M{\"u}ller}, {Nandra}, {Nazarov}, {Pavlinsky}, {Poghodin}, {Predehl}, {Robrade}, {Sazonov}, {Scheuerle}, {Shirshakov}, {Tkachenko}, \& {Voron}}]{Sunyaev_2021}
{Sunyaev}, R., {Arefiev}, V., {Babyshkin}, V., {et~al.} 2021, \aap, 656, A132

\bibitem[{{Tamura} {et~al.}(2001){Tamura}, {Kaastra}, {Peterson}, {Paerels}, {Mittaz}, {Trudolyubov}, {Stewart}, {Fabian}, {Mushotzky}, {Lumb}, \& {Ikebe}}]{Tamura_2001}
{Tamura}, T., {Kaastra}, J.~S., {Peterson}, J.~R., {et~al.} 2001, \aap, 365, L87

\bibitem[{{Turner} \& {Shabala}(2015)}]{Turner_2015}
{Turner}, R.~J., \& {Shabala}, S.~S. 2015, \apj, 806, 59

\bibitem[{{Voigt} \& {Fabian}(2004)}]{Voigt_2004}
{Voigt}, L.~M., \& {Fabian}, A.~C. 2004, \mnras, 347, 1130

\bibitem[{{Voigt} {et~al.}(2002){Voigt}, {Schmidt}, {Fabian}, {Allen}, \& {Johnstone}}]{Voigt_2002}
{Voigt}, L.~M., {Schmidt}, R.~W., {Fabian}, A.~C., {Allen}, S.~W., \& {Johnstone}, R.~M. 2002, \mnras, 335, L7

\bibitem[{{Whiting} \& {Humphreys}(2012)}]{Whiting_2012}
{Whiting}, M., \& {Humphreys}, B. 2012, \pasa, 29, 371

\bibitem[{{Whiting}(2020)}]{Whiting_2020}
{Whiting}, M.~T. 2020, in Astronomical Society of the Pacific Conference Series, Vol. 522, Astronomical Data Analysis Software and Systems XXVII, ed. P.~{Ballester}, J.~{Ibsen}, M.~{Solar}, \& K.~{Shortridge}, 469

\bibitem[{{Willott} {et~al.}(1999){Willott}, {Rawlings}, {Blundell}, \& {Lacy}}]{Willott_1999}
{Willott}, C.~J., {Rawlings}, S., {Blundell}, K.~M., \& {Lacy}, M. 1999, \mnras, 309, 1017

\bibitem[{{Wise} {et~al.}(2004){Wise}, {McNamara}, \& {Murray}}]{Wise_2004}
{Wise}, M.~W., {McNamara}, B.~R., \& {Murray}, S.~S. 2004, \apj, 601, 184

\bibitem[{{XRISM Collaboration} {et~al.}(2025){XRISM Collaboration}, {Audard}, {Awaki}, {Ballhausen}, {Bamba}, {Behar}, {Boissay-Malaquin}, {Brenneman}, {Brown}, {Corrales}, {Costantini}, {Cumbee}, {Done}, {Dotani}, {Ebisawa}, {Eckart}, {Eckert}, {Enoto}, {Eguchi}, {Ezoe}, {Foster}, {Fujimoto}, {Fujita}, {Fukazawa}, {Fukushima}, {Furuzawa}, {Gallo}, {Garc{\'\i}a}, {Gu}, {Guainazzi}, {Hagino}, {Hamaguchi}, {Hatsukade}, {Hayashi}, {Hayashi}, {Hell}, {Hodges-Kluck}, {Hornschemeier}, {Ichinohe}, {Ishida}, {Ishikawa}, {Ishisaki}, {Kaastra}, {Kallman}, {Kara}, {Katsuda}, {Kanemaru}, {Kelley}, {Kilbourne}, {Kitamoto}, {Kobayashi}, {Kohmura}, {Kubota}, {Leutenegger}, {Loewenstein}, {Maeda}, {Markevitch}, {Matsumoto}, {Matsushita}, {McCammon}, {McNamara}, {Mernier}, {Miller}, {Miller}, {Mitsuishi}, {Mizumoto}, {Mizuno}, {Mori}, {Mukai}, {Murakami}, {Mushotzky}, {Nakajima}, {Nakazawa}, {Ness}, {Nobukawa}, {Nobukawa}, {Noda}, {Odaka}, {Ogawa}, {Ogorzalek}, {Okajima}, {Ota}, {Paltani}, {Petre}, {Plucinsky}, {Porter},
  {Pottschmidt}, {Sato}, {Sato}, {Sawada}, {Seta}, {Shidatsu}, {Simionescu}, {Smith}, {Suzuki}, {Szymkowiak}, {Takahashi}, {Takeo}, {Tamagawa}, {Tamura}, {Tanaka}, {Tanimoto}, {Tashiro}, {Terada}, {Terashima}, {Trigo}, {Tsuboi}, {Tsujimoto}, {Tsunemi}, {Tsuru}, {Uchida}, {Uchida}, {Uchida}, {Uchiyama}, {Ueda}, {Uno}, {Vink}, {Watanabe}, {Williams}, {Yamada}, {Yamada}, {Yamaguchi}, {Yamaoka}, {Yamasaki}, {Yamauchi}, {Yamauchi}, {Yaqoob}, {Yoneyama}, {Yoshida}, {Yukita}, {Zhuravleva}, {Kondo}, {Werner}, {Pl{\v{s}}ek}, {Sun}, {Hosogi}, \& {Majumder}}]{Xrism_2025}
{XRISM Collaboration}, {Audard}, M., {Awaki}, H., {et~al.} 2025, \nat, 638, 365

\bibitem[{{Yates-Jones} {et~al.}(2021){Yates-Jones}, {Shabala}, \& {Krause}}]{Yates_2021}
{Yates-Jones}, P.~M., {Shabala}, S.~S., \& {Krause}, M. G.~H. 2021, \mnras, 508, 5239

\bibitem[{{Zakamska} \& {Narayan}(2003)}]{Zakamska_2003}
{Zakamska}, N.~L., \& {Narayan}, R. 2003, \apj, 582, 162

\end{thebibliography}

\begin{appendix}
\onecolumn
\section{X-ray and radio properties of the eRASS1/ASKAP clusters}\label{App:A}
\begin{figure}[!ht]
\centering
\includegraphics[width=0.95\columnwidth]{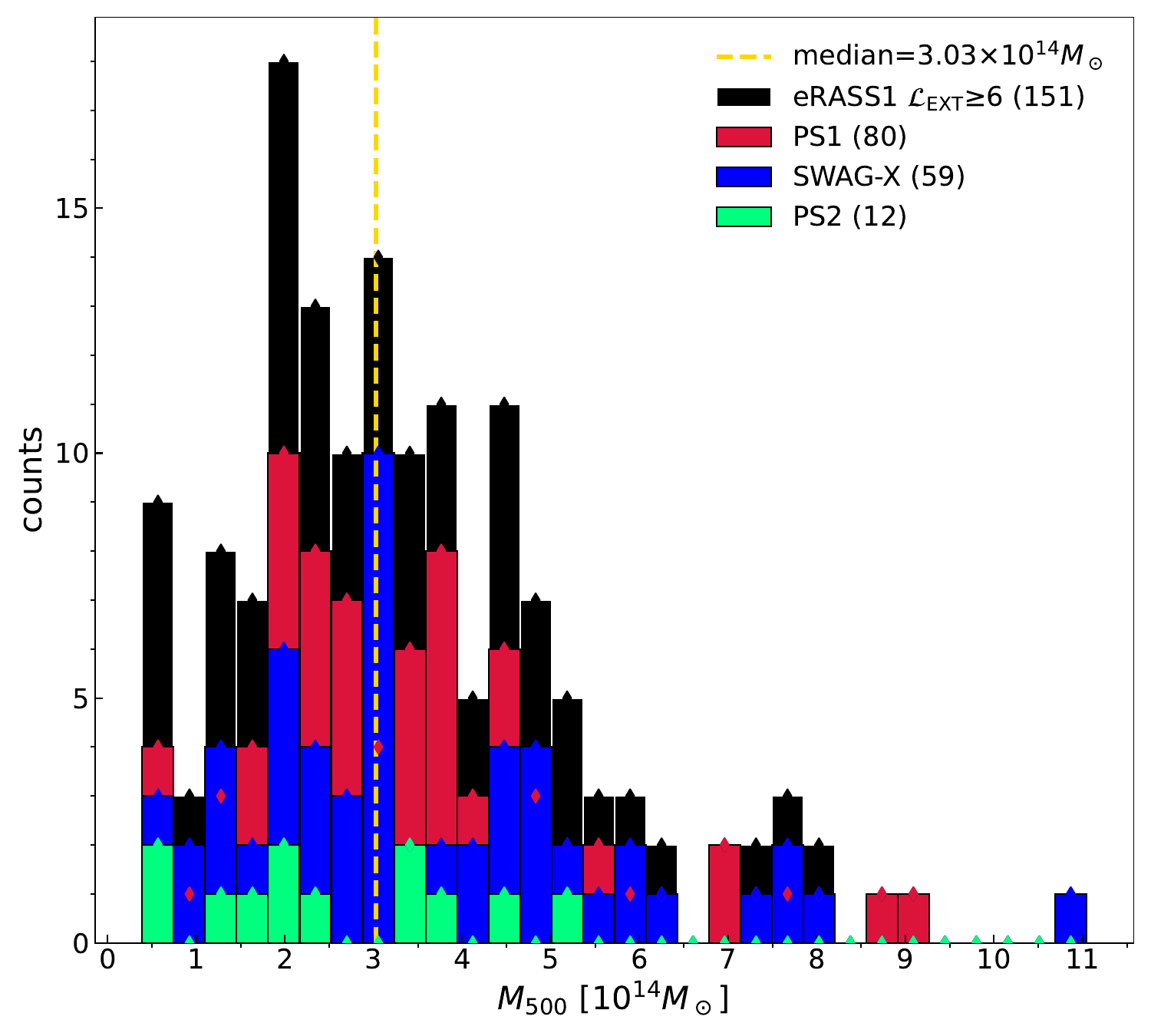}
\includegraphics[width=0.95\columnwidth]{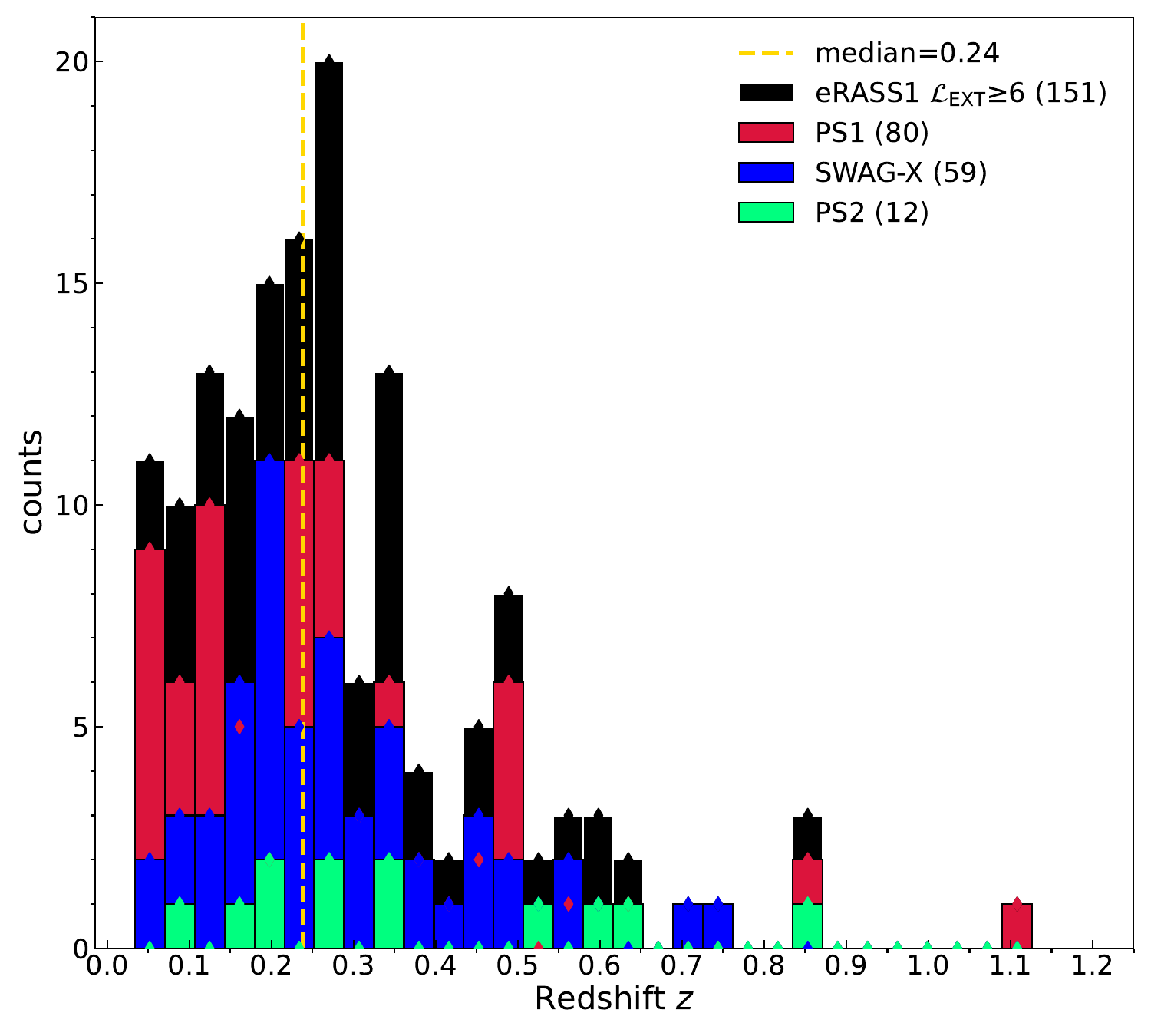}
\includegraphics[width=0.95\columnwidth]{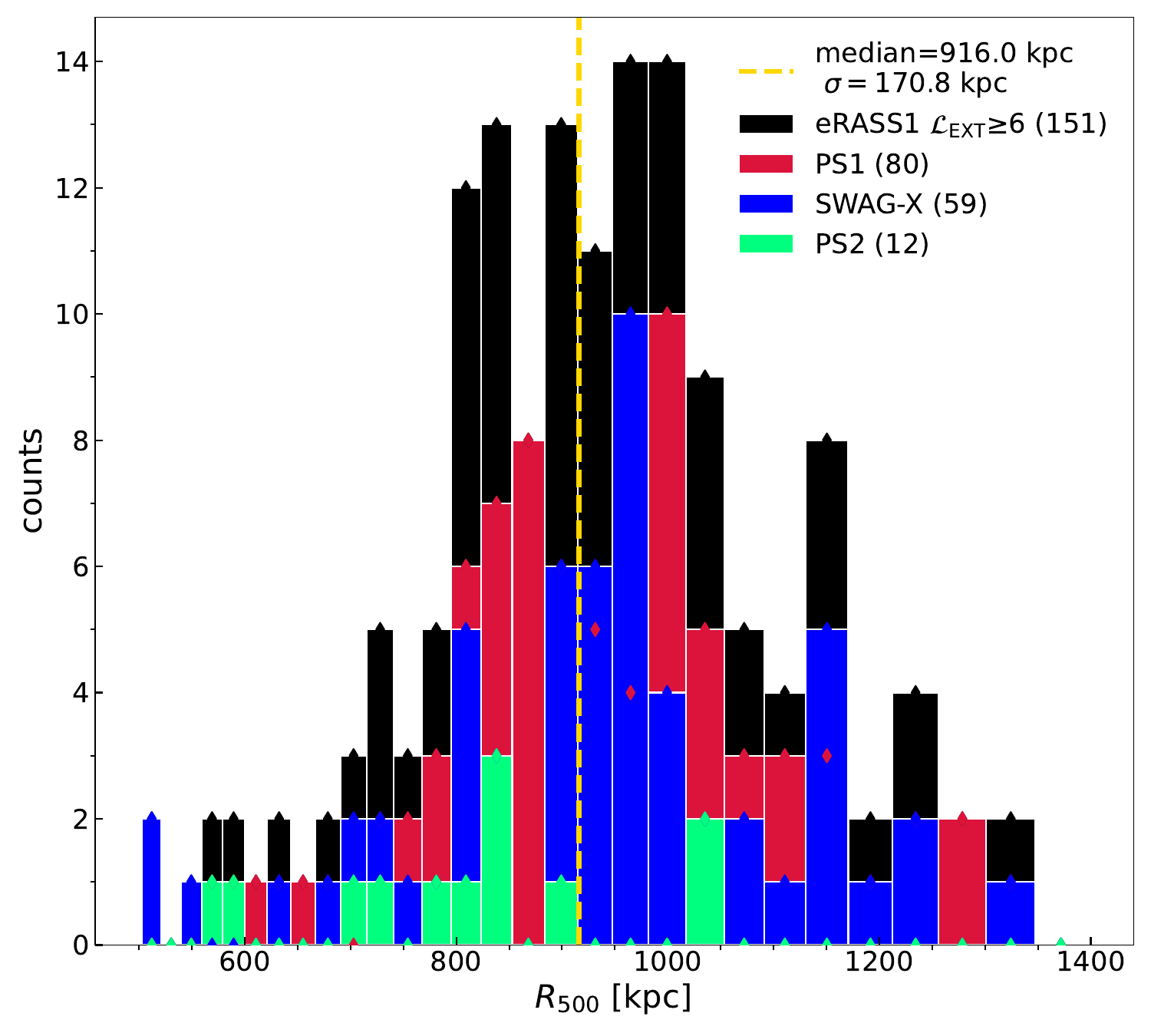}
\caption{The mass (\textit{top}), redshift (\textit{middle}), and $R_{500}$ (\textit{bottom}) distributions of the eRASS1/ASKAP cluster sample. In each plot, PS1, PS2, and SWAG-X subsample is shown in red, green, and blue, respectively, and the whole sample is shown in black. The yellow dashed line denotes the median of the entire sample.}
\label{fig:z-M}
\end{figure}

\begin{figure}
\centering
\includegraphics[width=\columnwidth]{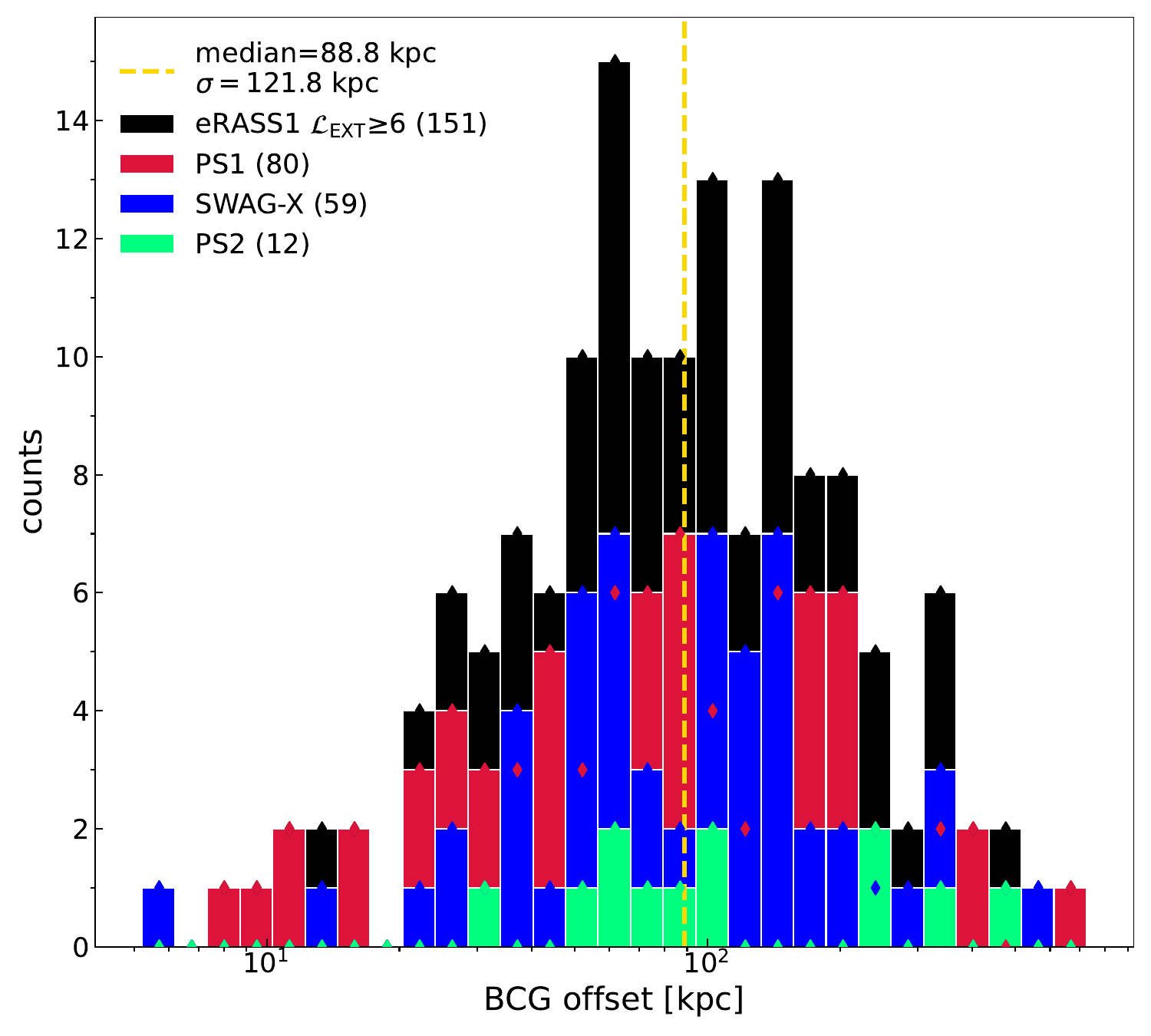}
\caption{Distribution of the BCG offset. The black bars show the whole sample distribution, while PS1, PS2, and SWAG-X are shown in red, green, and blue, respectively. The yellow vertical dashed line indicates the median of the whole sample.}
\label{fig:bcg_offset_hist}
\end{figure}

\begin{figure}
\centering
\includegraphics[width=\columnwidth]{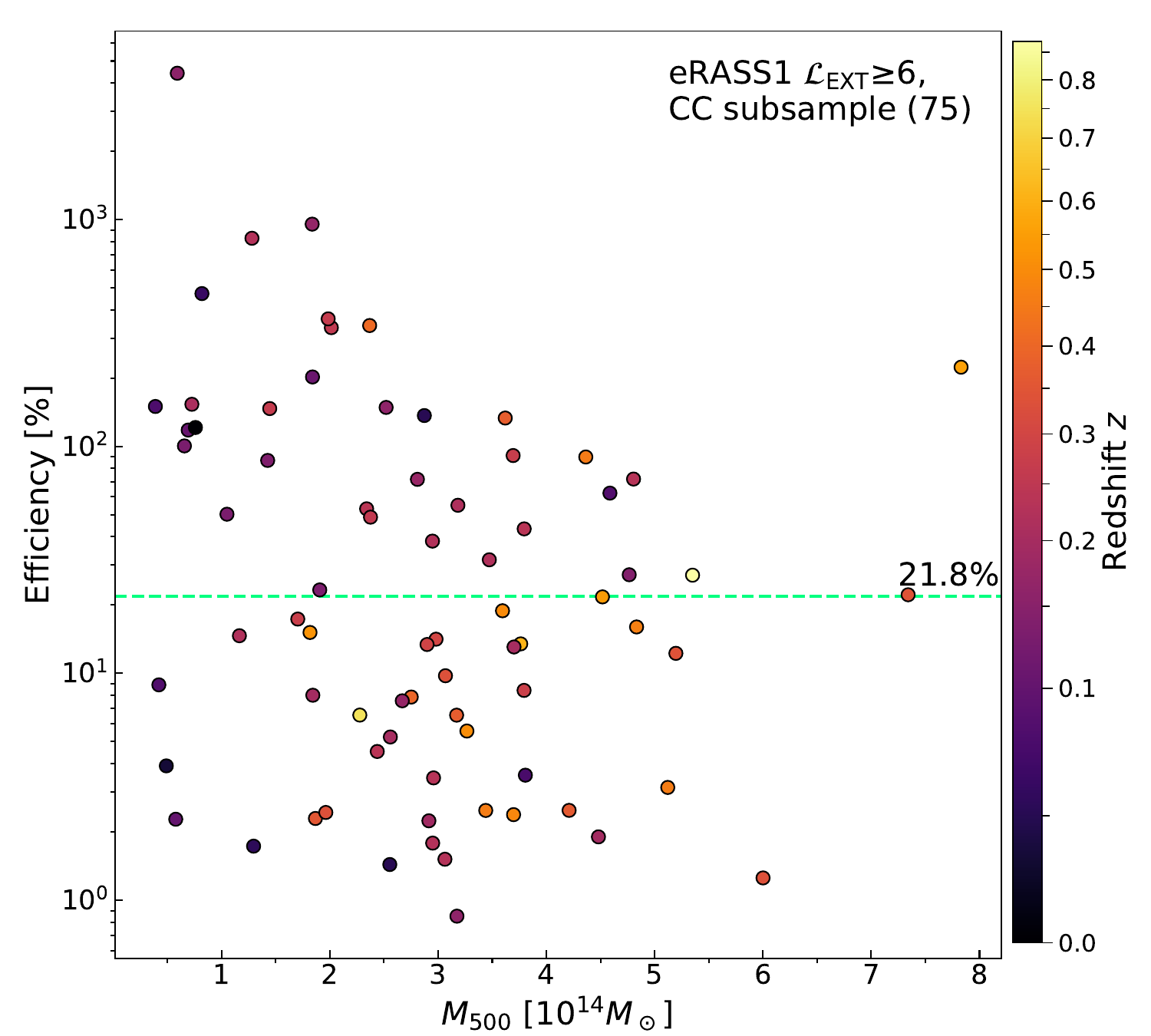}
\caption{Efficiency (logarithmic bias based; see Section~\ref{sect:eff}) calculated using $L_\mathrm{mech,HB+14}$ and $L_{\mathrm{X}~r<R_\mathrm{cool}}$ as a function of cluster mass and redshift for the CC subsample. Green dashed line marks the average efficiency value.}
\label{fig:eff}
\end{figure}

\begin{figure*}
\centering
\includegraphics[width=0.49\textwidth]{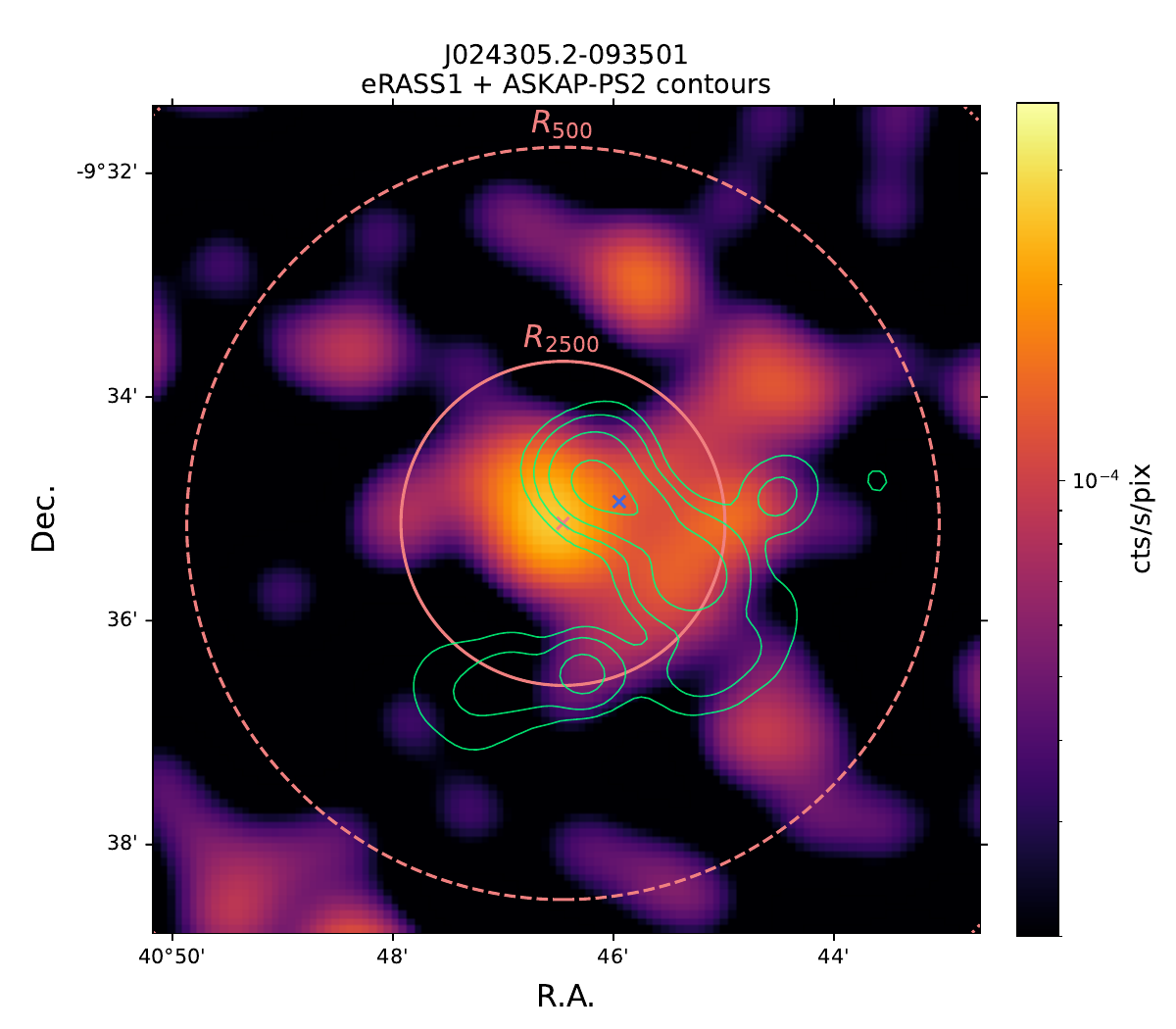}
\includegraphics[width=0.49\textwidth]{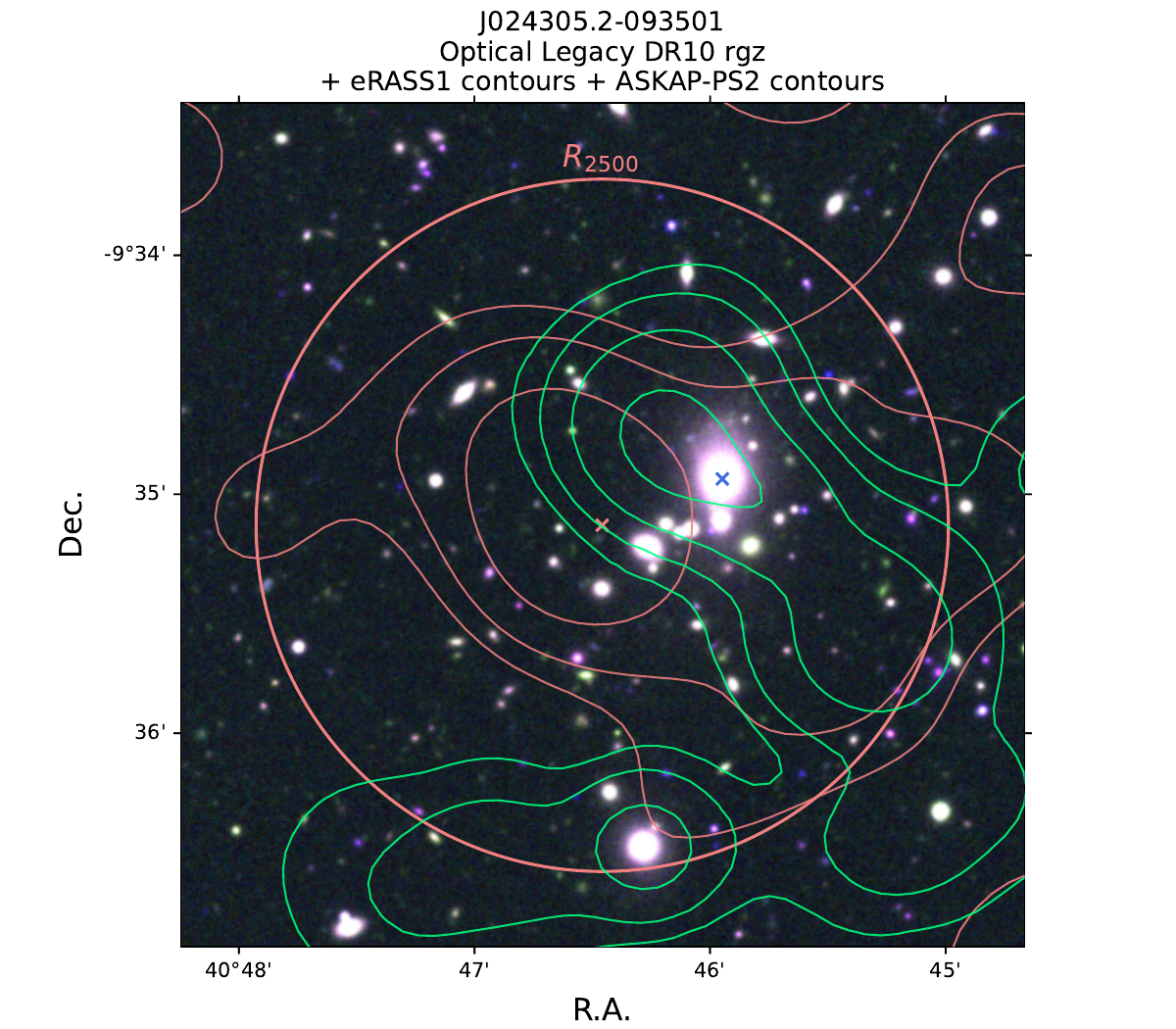}
\caption{Multiwavelength images of J024305.2-093501 ($z=0.1558$ and $R_{500}=3.36'$). The cluster has the highest AGN feedback efficiency among the eRASS1/ASKAP subsample with $L_{\mathrm{X},r<R_\mathrm{cool}}=10^{42.85}\,\mathrm{erg\,s^{-1}}$ and $L_\mathrm{mech,HB+14}=10^{44.49}\,\mathrm{erg\,s^{-1}}$. \textit{Left:} Gaussian smoothed eRASS1 particle-induced background subtracted, exposure-corrected image in the $0.2-2.3\,\mathrm{keV}$ band. \textit{Right:} DESI Legacy Survey DR10 RGB (g,r,z) image. The eRASS1 contours are shown in orange. In both image the X-ray center and BCG position are denoted by the orange and blue crosses, respectively. The ASKAP radio contours are overplotted in green and the $R_{2500}$ ($R_{500}$) radius is plotted as solid (dashed) orange circle.}
\label{fig:example_cl1}
\end{figure*}

\begin{figure*}
\centering
\includegraphics[width=0.49\textwidth]{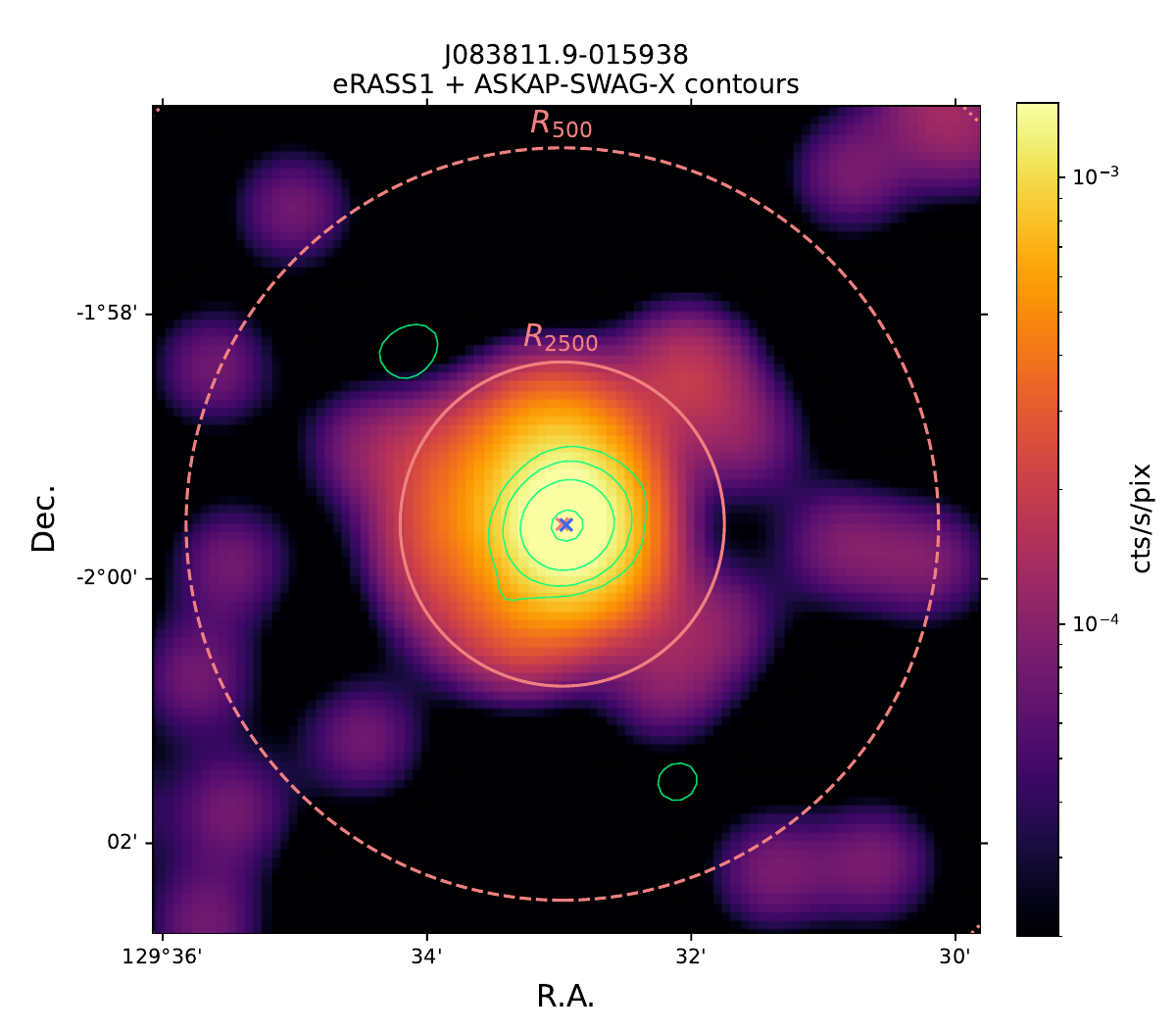}
\includegraphics[width=0.49\textwidth]{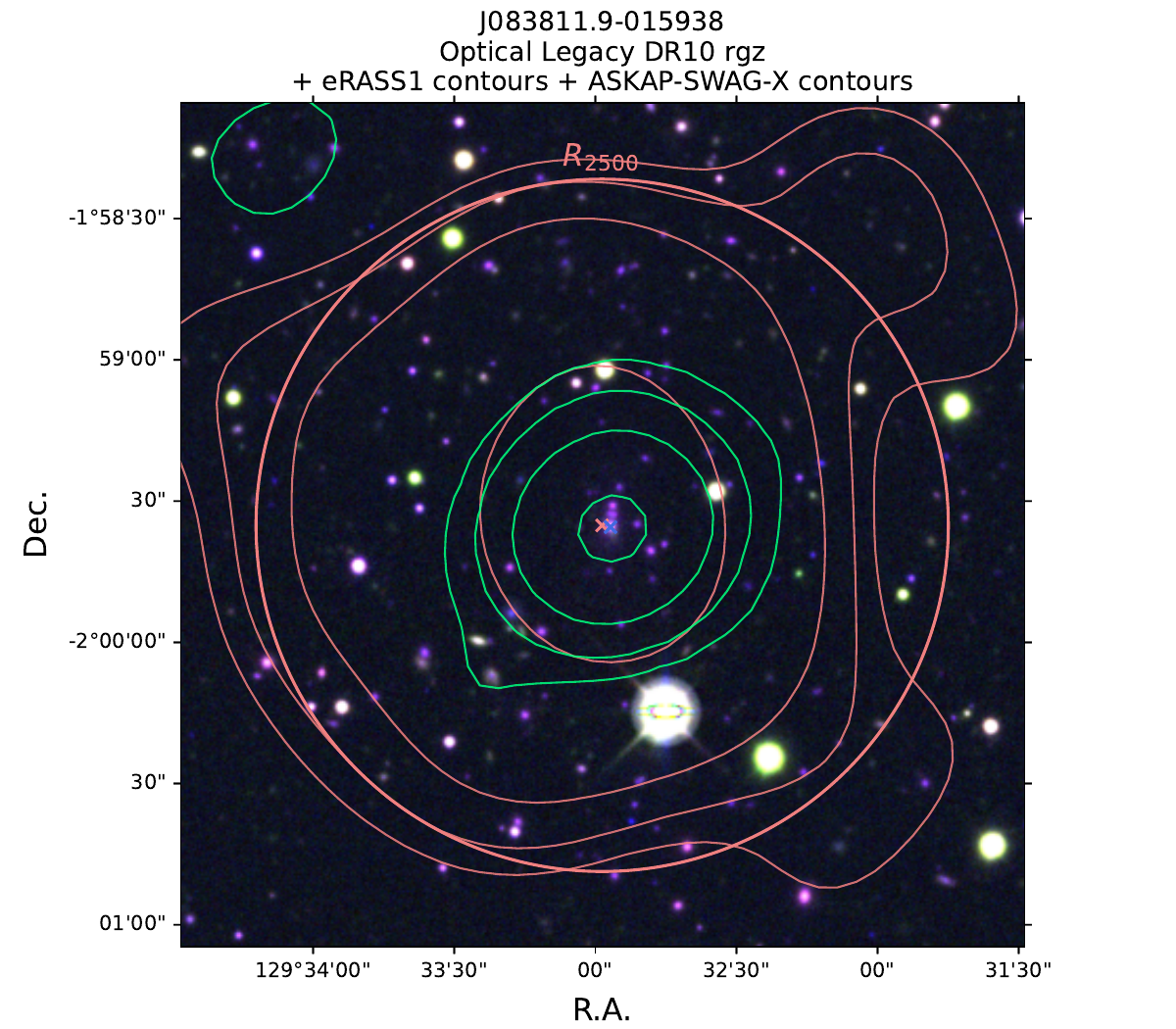}
\caption{Multiwavelength images of J083811.9-015938 ($z=0.5590$ and $R_{500}=2.85'$). The cluster has the highest X-ray cooling luminosity of $L_{\mathrm{X},r<R_\mathrm{cool}}=10^{44.74}\,\mathrm{erg\,s^{-1}}$ with $L_\mathrm{mech,HB+14}=10^{45.09}\,\mathrm{erg\,s^{-1}}$. \textit{Left:} Gaussian smoothed eRASS1 particle-induced background subtracted, exposure-corrected image in the $0.2-2.3\,\mathrm{keV}$ band. \textit{Right:} DESI Legacy Survey DR10 RGB (g,r,z) image. The eRASS1 contours are shown in orange. In both image the X-ray center and BCG position are denoted by the orange and blue crosses, respectively. The ASKAP radio contours are overplotted in green and the $R_{2500}$ ($R_{500}$) radius is plotted as solid (dashed) orange circle.}
\label{fig:example_cl2}
\end{figure*}

We list the BCG properties of the eRASS1 clusters found in the ASKAP fields in Table~\ref{tab:long_tabl}. Columns $S_\mathrm{R,944\,MHz}$ and LLS list the radio flux at $944\,\mathrm{MHz}$ and the largest linear size of the central radio sources, respectively. The LLS is converted from the image-beam-size-convolved largest angular size. Values without errors in these columns indicate non-detected and unresolved sources (see Section~\ref{sec:radio_prop}) In Figure~\ref{fig:z-M} we display the distribution of the properties of the eRASS1/ASKAP clusters: the mass ($M_{500}$), redshift ($z$), and characteristic radius ($R_{500}$) are shown in the top, middle, and bottom panel, respectively. We present the BCG offset distribution in Figure~\ref{fig:bcg_offset_hist} and the AGN feedback efficiency distribution as a function of cluster mass and redshift in Figure~\ref{fig:eff}. Furthermore, multiwavelength images of two clusters are shown in Figures~\ref{fig:example_cl1} and \ref{fig:example_cl2}.

\begin{longtable}{c c c c c c c}
\caption{BCG properties of the eRASS1 clusters found in the ASKAP fields.}
\label{tab:long_tabl} \\
  \hline
  \textbf{1eRASS Name} & \textbf{RA$_\mathrm{BCG}$} & \textbf{DEC$_\mathrm{BCG}$} & \textbf{$z$} & \textbf{$S_\mathrm{R,944\,MHz}$ [mJy]} & \textbf{LLS [kpc]} & \textbf{ASKAP field}\\
  \hline
  \endfirsthead

  \hline
  \textbf{1eRASS Name} & \textbf{RA$_\mathrm{BCG}$} & \textbf{DEC$_\mathrm{BCG}$} & \textbf{$z$} & \textbf{$S_\mathrm{R,944\,MHz}$ [mJy]} & \textbf{LLS [kpc]} & \textbf{ASKAP field}\\
  \hline
  \endhead

  \hline
  \endfoot

  \hline
  \hline
  \endlastfoot

J201601.3-495436$^{\mathrm{CC}}$ & 304.0066 & -49.9066 & 0.2711 & $28.65\pm0.72$ & $240.0\pm9.2$ & PS1 \\
J202154.3-525715 & 305.4725 & -52.9510 & 0.1391 & $142.58\pm2.99$ & $176.0\pm8.1$ & PS1 \\
J202726.9-522215 & 306.8682 & -52.3694 & 0.0644 & $5.98\pm0.26$ & $44.0\pm5.8$ & PS1 \\
J203043.7-563749 & 307.6886 & -56.6322 & 0.3939 & $27.15\pm0.69$ & $286.0\pm10.0$ & PS1 \\
J204008.3-503254 & 310.0406 & -50.5480 & 0.1494 & $54.06\pm1.22$ & $242.0\pm9.2$ & PS1 \\
J204612.1-575544 & 311.5488 & -57.9306 & 0.2251 & $308.33\pm6.31$ & $266.0\pm9.7$ & PS1 \\
J205146.0-604621$^{\mathrm{CC}}$ & 312.9407 & -60.7729 & 0.3372 & $3.04\pm0.20$ & $178.0\pm8.1$ & PS1 \\
J205156.7-523751 & 312.9872 & -52.6297 & 0.0448 & $4.95\pm0.24$ & $32.0\pm5.6$ & PS1 \\
J205316.0-620912$^{\mathrm{CC}}$ & 313.3130 & -62.1517 & 0.3953 & $0.31\pm0.15$ & $218.0\pm8.8$ & PS1 \\
J205555.4-545538$^{\mathrm{CC}}$ & 313.9830 & -54.9302 & 0.1395 & $21.86\pm0.58$ & $374.0\pm11.6$ & PS1 \\
J211250.8-531753$^{\mathrm{CC}}$ & 318.2121 & -53.2983 & 0.2231 & $8.62\pm0.31$ & $130.0\pm7.3$ & PS1 \\
J211652.8-593039 & 319.2222 & -59.5088 & 0.0578 & $60.74\pm1.36$ & $64.0\pm6.1$ & PS1 \\
J212023.5-542845$^{\mathrm{CC}}$ & 320.1055 & -54.4780 & 0.2410 & $5.83\pm0.26$ & $298.0\pm10.2$ & PS1 \\
J213305.7-594531$^{\mathrm{CC}}$ & 323.2733 & -59.7603 & 0.5051 & $0.26\pm0.15$ & $504.0\pm13.8$ & PS1 \\
J213505.3-625454 & 323.7732 & -62.9159 & 0.2228 & $4.26\pm0.23$ & $206.0\pm8.6$ & PS1 \\
J213536.8-572622 & 323.9060 & -57.4419 & 0.4268 & $4.63\pm0.24$ & $432.0\pm12.6$ & PS1 \\
J213800.9-600758 & 324.5035 & -60.1317 & 0.3188 & $5.88\pm0.26$ & $170.0\pm8.0$ & PS1 \\
J214359.4-563717$^{\mathrm{CC}}$ & 325.9968 & -56.6225 & 0.0820 & $168.40\pm3.51$ & $96.0\pm6.7$ & PS1 \\
J214553.0-564448 & 326.4664 & -56.7483 & 0.4810 & $2.76\pm0.20$ & $364.0\pm11.4$ & PS1 \\
J214622.3-571714$^{\mathrm{CC}}$ & 326.5942 & -57.2868 & 0.0733 & $4.69\pm0.24$ & $82.0\pm6.5$ & PS1 \\
J214647.5-573648$^{\mathrm{CC}}$ & 326.7027 & -57.6150 & 0.6109 & $1.02\pm0.16$ & $328.0\pm10.8$ & PS1 \\
J214844.5-611650 & 327.1785 & -61.2795 & 0.5722 & $2.54\pm0.19$ & $316.0\pm10.6$ & PS1 \\
J215129.7-552019 & 327.8746 & -55.3369 & 0.0409 & $2623.45\pm52.61$ & $160.0\pm7.8$ & PS1 \\
J215625.0-513110$^{\mathrm{CC}}$ & 329.1034 & -51.5220 & 0.4942 & $1.94\pm0.18$ & $232.0\pm9.1$ & PS1 \\
J220112.5-614737$^{\mathrm{CC}}$ & 330.3022 & -61.7947 & 0.2382 & $10.27\pm0.35$ & $170.0\pm8.0$ & PS1 \\
J220153.8-595644 & 330.4722 & -59.9454 & 0.1066 & $2.13\pm0.19$ & $96.0\pm6.7$ & PS1 \\
J220920.7-514811 & 332.3402 & -51.8072 & 0.1169 & $0.79\pm0.16$ & $80.0\pm6.4$ & PS1 \\
J221033.0-570945$^{\mathrm{CC}}$ & 332.6328 & -57.1649 & 0.3003 & $0.93\pm0.16$ & $166.0\pm7.9$ & PS1 \\
J221504.0-520501 & 333.7622 & -52.0809 & 0.5002 & $2.12\pm0.18$ & $296.0\pm10.2$ & PS1 \\
J221631.8-532501 & 334.1310 & -53.4125 & 0.1801 & $83.86\pm1.82$ & $160.0\pm7.8$ & PS1 \\
J222251.8-483423 & 335.7113 & -48.5765 & 0.6517 & $0.48\pm0.15$ & $238.0\pm9.2$ & PS1 \\
J203328.8-593552 & 308.3800 & -59.5971 & 0.2004 & $353.54\pm7.21$ & $252.0\pm9.4$ & PS1 \\
J201902.9-564238$^{\mathrm{CC}}$ & 304.7588 & -56.7111 & 0.2313 & $23.98\pm0.60$ & $286.0\pm7.6$ & PS1 \\
J202321.6-553524 & 305.8370 & -55.5972 & 0.2305 & $8.55\pm0.74$ & $175.2\pm7.6$ & PS1 \\
J203153.0-535148$^{\mathrm{CC}}$ & 307.9693 & -53.8626 & 0.1711 & $1.11\pm0.44$ & $108.5\pm6.4$ & PS1 \\
J203220.5-562738 & 308.0586 & -56.4368 & 0.2844 & $48.45\pm0.82$ & $267.2\pm8.9$ & PS1 \\
J203601.2-513931 & 308.9892 & -51.6527 & 0.2719 & $111.78\pm1.64$ & $366.3\pm8.6$ & PS1 \\
J203732.2-540710$^{\mathrm{CC}}$ & 309.3828 & -54.1204 & 0.1314 & $4.57\pm0.44$ & $112.3\pm4.9$ & PS1 \\
J203919.4-624401$^{\mathrm{CC}}$ & 309.7576 & -62.7652 & 0.1014 & $0.10$ & $34.7$ & PS1 \\
J204125.2-551440$^{\mathrm{CC}}$ & 310.3542 & -55.2459 & 0.2186 & $0.72\pm0.08$ & $131.7\pm7.4$ & PS1 \\
J204408.5-603931 & 311.0434 & -60.6559 & 0.1213 & $1.79\pm0.61$ & $87.0\pm4.7$ & PS1 \\
J204822.8-613113 & 312.0900 & -61.5172 & 0.1076 & $0.30\pm0.14$ & $59.0\pm4.4$ & PS1 \\
J205750.2-524548$^{\mathrm{CC}}$ & 314.4707 & -52.7733 & 0.2559 & $6.52\pm0.31$ & $186.7\pm8.2$ & PS1 \\
J205943.1-501908 & 314.9220 & -50.3029 & 0.3307 & $0.59\pm0.11$ & $184.8\pm10.0$ & PS1 \\
J210114.4-554134 & 315.3039 & -55.6940 & 0.2599 & $0.10$ & $74.9$ & PS1 \\
J210452.0-514939$^{\mathrm{CC}}$ & 316.2138 & -51.8235 & 0.0506 & $1.64\pm0.24$ & $55.9\pm2.1$ & PS1 \\
J210604.1-584425 & 316.5194 & -58.7412 & 1.1263 & $7.73\pm0.45$ & $337.9\pm17.1$ & PS1 \\
J210655.8-603441$^{\mathrm{CC}}$ & 316.7334 & -60.5756 & 0.1596 & $40.98\pm0.73$ & $280.6\pm5.7$ & PS1 \\
J211032.5-584227$^{\mathrm{CC}}$ & 317.6634 & -58.7150 & 0.3555 & $0.10$ & $93.0$ & PS1 \\
J212251.3-582948$^{\mathrm{CC}}$ & 320.7058 & -58.4950 & 0.2932 & $0.93\pm0.16$ & $206.0\pm9.2$ & PS1 \\
J212433.5-612500 & 321.1449 & -61.4124 & 0.4365 & $17.47\pm0.57$ & $281.4\pm11.7$ & PS1 \\
J212623.2-553221 & 321.6349 & -55.5222 & 0.8570 & $0.70\pm0.02$ & $287.0\pm15.9$ & PS1 \\
J213151.8-500345$^{\mathrm{CC}}$ & 322.9636 & -50.0624 & 0.4570 & $0.12\pm0.09$ & $144.5\pm14.5$ & PS1 \\
J214528.9-513623$^{\mathrm{CC}}$ & 326.3730 & -51.6071 & 0.0538 & $0.42\pm0.16$ & $39.1\pm2.3$ & PS1 \\
J215406.4-575136 & 328.5175 & -57.8676 & 0.0753 & $1.67\pm0.59$ & $84.2\pm3.1$ & PS1 \\
J215453.8-611541$^{\mathrm{CC}}$ & 328.7207 & -61.2607 & 0.1033 & $11.44\pm0.92$ & $117.3\pm3.9$ & PS1 \\
J215815.7-502328$^{\mathrm{CC}}$ & 329.5744 & -50.3977 & 0.4885 & $0.11\pm0.05$ & $124.9\pm13.5$ & PS1 \\
J215826.3-602359 & 329.6091 & -60.4261 & 0.0849 & $2.88\pm0.21$ & $119.1\pm3.3$ & PS1 \\
J215918.0-564510$^{\mathrm{CC}}$ & 329.8259 & -56.7500 & 0.2794 & $1.03\pm0.09$ & $197.8\pm8.8$ & PS1 \\
J220504.3-592716 & 331.2512 & -59.4547 & 0.3504 & $2.14\pm0.21$ & $204.8\pm10.3$ & PS1 \\
J220609.1-494126$^{\mathrm{CC}}$ & 331.5336 & -49.7188 & 0.1284 & $5.42\pm0.42$ & $105.3\pm4.8$ & PS1 \\
J220731.3-492529 & 331.8876 & -49.4122 & 0.2541 & $13.18\pm0.64$ & $221.7\pm8.2$ & PS1 \\
J220756.6-522333 & 331.9916 & -52.3887 & 0.1106 & $0.24\pm0.12$ & $50.0\pm4.6$ & PS1 \\
J220952.8-553521$^{\mathrm{CC}}$ & 332.4480 & -55.5817 & 0.1680 & $135.89\pm2.37$ & $312.8\pm5.9$ & PS1 \\
J221117.3-483401 & 332.8222 & -48.5727 & 0.2565 & $0.29\pm0.20$ & $148.4\pm9.9$ & PS1 \\
J221454.0-532017 & 333.7287 & -53.3502 & 0.3381 & $145.74\pm1.46$ & $508.9\pm10.0$ & PS1 \\
J221959.1-581546 & 334.9924 & -58.2725 & 0.2814 & $7.00\pm0.79$ & $214.1\pm8.9$ & PS1 \\
J221959.2-482901 & 335.0168 & -48.4767 & 0.8711 & $2.07\pm0.31$ & $378.2\pm16.2$ & PS1 \\
J222000.0-522730 & 335.0352 & -52.4640 & 0.1065 & $0.55\pm0.12$ & $96.5\pm4.1$ & PS1 \\
J222000.3-620520 & 335.0116 & -62.0862 & 0.0634 & $0.77\pm0.29$ & $46.6\pm2.7$ & PS1 \\
J222120.9-501707 & 335.3608 & -50.2793 & 0.1791 & $605.73\pm3.79$ & $349.1\pm6.2$ & PS1 \\
J222327.0-522740 & 335.8409 & -52.4637 & 0.2734 & $0.46\pm0.21$ & $129.6\pm9.4$ & PS1 \\
J222420.4-503901$^{\mathrm{CC}}$ & 336.0797 & -50.6577 & 0.3354 & $0.10$ & $89.5$ & PS1 \\
J211622.6-595403$^{\mathrm{CC}}$ & 319.0925 & -59.9032 & 0.0602 & $101.26\pm6.48$ & $351.2\pm2.4$ & PS1 \\
J213917.7-630845$^{\mathrm{CC}}$ & 324.8008 & -63.1455 & 0.2373 & $0.34\pm0.20$ & $105.1\pm8.9$ & PS1 \\
J222319.6-501552 & 335.8332 & -50.2658 & 0.2659 & $145.86\pm4.40$ & $552.6\pm8.5$ & PS1 \\
J201847.9-524238$^{\mathrm{CC}}$ & 304.6946 & -52.6910 & 0.0502 & $309.06\pm5.06$ & $181.6\pm2.0$ & PS1 \\
J202555.3-511709 & 306.4823 & -51.2745 & 0.2286 & $25.59\pm3.80$ & $438.3\pm7.6$ & PS1 \\
J220052.4-515502$^{\mathrm{CC}}$ & 330.2247 & -51.9221 & 0.2180 & $6.32\pm0.36$ & $209.5\pm7.3$ & PS1 \\
J215919.0-521025 & 329.8301 & -52.1648 & 0.4982 & $3.77\pm0.59$ & $275.3\pm12.8$ & PS1 \\
J082604.4+041910 & 126.5118 & 4.3250 & 0.4742 & $4.52\pm0.29$ & $302.3\pm5.1$ & SWAG-X \\
J083133.9-035152$^{\mathrm{CC}}$ & 127.8926 & -3.8652 & 0.1746 & $12.64\pm0.16$ & $115.4\pm0.5$ & SWAG-X \\
J083205.0+041853 & 128.0177 & 4.3099 & 0.2012 & $152.11\pm0.68$ & $271.3\pm0.2$ & SWAG-X \\
J083652.6+030000 & 129.2144 & 3.0004 & 0.1910 & $5.12\pm0.09$ & $116.8\pm0.7$ & SWAG-X \\
J083714.1-032144$^{\mathrm{CC}}$ & 129.3078 & -3.3617 & 0.0410 & $0.40\pm0.09$ & $30.3\pm2.1$ & SWAG-X \\
J083811.9-015938$^{\mathrm{CC}}$ & 129.5491 & -1.9932 & 0.5590 & $75.68\pm0.31$ & $242.3\pm0.1$ & SWAG-X \\
J084306.7+002834$^{\mathrm{CC}}$ & 130.7832 & 0.4831 & 0.2636 & $10.24\pm0.10$ & $154.6\pm0.5$ & SWAG-X \\
J084342.8+040323$^{\mathrm{CC}}$ & 130.9295 & 4.0596 & 0.2021 & $0.69\pm0.07$ & $123.3\pm4.2$ & SWAG-X \\
J084531.1+022840$^{\mathrm{CC}}$ & 131.3825 & 2.4759 & 0.0768 & $10.93\pm0.14$ & $68.3\pm0.2$ & SWAG-X \\
J085412.1-022058$^{\mathrm{CC}}$ & 133.5539 & -2.3495 & 0.3684 & $0.63\pm0.13$ & $258.0\pm21.1$ & SWAG-X \\
J085751.0+031014 & 134.4750 & 3.1766 & 0.2025 & $8.24\pm0.25$ & $267.9\pm3.3$ & SWAG-X \\
J090131.4+030055$^{\mathrm{CC}}$ & 135.3795 & 3.0157 & 0.1936 & $0.36\pm0.09$ & $115.4\pm9.1$ & SWAG-X \\
J091235.0-042929 & 138.1475 & -4.4918 & 0.2060 & $16.72\pm0.17$ & $124.8\pm0.2$ & SWAG-X \\
J091414.7+001922 & 138.5613 & 0.3252 & 0.1665 & $10.53\pm0.18$ & $138.1\pm0.9$ & SWAG-X \\
J091453.6+041611 & 138.7190 & 4.2654 & 0.1477 & $84.60\pm0.24$ & $115.3\pm0.1$ & SWAG-X \\
J092023.1+013433 & 140.0980 & 1.5789 & 0.6939 & $5.45\pm0.16$ & $337.9\pm3.8$ & SWAG-X \\
J092546.2-014321$^{\mathrm{CC}}$ & 141.4387 & -1.7288 & 0.2265 & $95.17\pm0.56$ & $300.1\pm0.3$ & SWAG-X \\
J092647.6+050038$^{\mathrm{CC}}$ & 141.6970 & 5.0011 & 0.4659 & $2.09\pm0.18$ & $567.4\pm50.9$ & SWAG-X \\
J093150.9-002220$^{\mathrm{CC}}$ & 142.9627 & -0.3705 & 0.3353 & $0.59\pm0.15$ & $280.8\pm27.9$ & SWAG-X \\
J093346.9-025503$^{\mathrm{CC}}$ & 143.4550 & -2.9139 & 0.1290 & $4.07\pm0.22$ & $143.7\pm2.9$ & SWAG-X \\
J093347.9+071429 & 143.4528 & 7.2459 & 0.2527 & $0.81\pm0.09$ & $137.1\pm4.6$ & SWAG-X \\
J093527.1+073020$^{\mathrm{CC}}$ & 143.8644 & 7.5073 & 0.2150 & $8.18\pm0.22$ & $153.1\pm0.9$ & SWAG-X \\
J093640.3+064147 & 144.1691 & 6.7115 & 0.2816 & $73.06\pm0.37$ & $345.2\pm0.3$ & SWAG-X \\
J093831.3-012524$^{\mathrm{CC}}$ & 144.6255 & -1.4258 & 0.4094 & $27.17\pm0.18$ & $252.1\pm0.4$ & SWAG-X \\
J094023.3+022824 & 145.1024 & 2.4776 & 0.1509 & $10.12\pm0.37$ & $211.5\pm2.4$ & SWAG-X \\
J083102.8-041145$^{\mathrm{CC}}$ & 127.7634 & -4.1959 & 0.0334 & $79.36\pm3.56$ & $121.0\pm3.3$ & SWAG-X \\
J083257.5-022735$^{\mathrm{CC}}$ & 128.2403 & -2.4601 & 0.7484 & $0.23\pm0.04$ & $159.8\pm33.6$ & SWAG-X \\
J083934.2-014035 & 129.8798 & -1.6729 & 0.2698 & $0.16$ & $74.0$ & SWAG-X \\
J083948.5+072232 & 129.9572 & 7.4120 & 0.2006 & $0.16$ & $59.2$ & SWAG-X \\
J084015.0-032654$^{\mathrm{CC}}$ & 130.0625 & -3.4478 & 0.2261 & $0.16$ & $64.9$ & SWAG-X \\
J084147.8-031154 & 130.4492 & -3.1946 & 0.2218 & $0.16$ & $64.0$ & SWAG-X \\
J084527.7+032736$^{\mathrm{CC}}$ & 131.3657 & 3.4608 & 0.3336 & $0.16$ & $85.7$ & SWAG-X \\
J084641.8-034334$^{\mathrm{CC}}$ & 131.6886 & -3.7410 & 0.1892 & $0.16$ & $56.5$ & SWAG-X \\
J085217.5-010126$^{\mathrm{CC}}$ & 133.0695 & -1.0266 & 0.4589 & $0.16$ & $104.4$ & SWAG-X \\
J085642.7-025957$^{\mathrm{CC}}$ & 134.1780 & -2.9934 & 0.2313 & $0.31\pm0.07$ & $91.6\pm20.4$ & SWAG-X \\
J090128.8-013837 & 135.3778 & -1.6548 & 0.3152 & $0.16$ & $82.5$ & SWAG-X \\
J090539.7+043433$^{\mathrm{CC}}$ & 136.4161 & 4.5775 & 0.2277 & $0.16$ & $65.3$ & SWAG-X \\
J091214.5-021729$^{\mathrm{CC}}$ & 138.0637 & -2.2944 & 0.1592 & $0.16$ & $49.1$ & SWAG-X \\
J091756.4+050849 & 139.4822 & 5.1524 & 0.5890 & $0.39\pm0.13$ & $237.8\pm69.7$ & SWAG-X \\
J092050.7+024512 & 140.2071 & 2.7539 & 0.2819 & $1.86\pm0.15$ & $152.7\pm16.4$ & SWAG-X \\
J092121.0+031735$^{\mathrm{CC}}$ & 140.3380 & 3.2870 & 0.3439 & $5.70\pm0.27$ & $223.5\pm17.1$ & SWAG-X \\
J092210.0+034626 & 140.5319 & 3.7664 & 0.2679 & $0.16$ & $73.6$ & SWAG-X \\
J092211.2-002812 & 140.5456 & -0.4603 & 0.3198 & $4.85\pm0.35$ & $192.7\pm18.8$ & SWAG-X \\
J092339.7+052647$^{\mathrm{CC}}$ & 140.9125 & 5.4467 & 0.3760 & $20.55\pm0.48$ & $246.4\pm16.2$ & SWAG-X \\
J093025.3+021707$^{\mathrm{CC}}$ & 142.6027 & 2.2903 & 0.5399 & $1.20\pm0.17$ & $227.9\pm33.9$ & SWAG-X \\
J093358.7+071512$^{\mathrm{CC}}$ & 143.4932 & 7.2531 & 0.2526 & $37.96\pm1.08$ & $276.6\pm13.1$ & SWAG-X \\
J093459.9+005438$^{\mathrm{CC}}$ & 143.7528 & 0.9041 & 0.3610 & $0.16$ & $90.3$ & SWAG-X \\
J093521.3+023222 & 143.8376 & 2.5432 & 0.4993 & $21.94\pm0.89$ & $471.6\pm23.8$ & SWAG-X \\
J085435.8+003858$^{\mathrm{CC}}$ & 133.6525 & 0.6426 & 0.1073 & $61.60\pm1.91$ & $152.8\pm7.2$ & SWAG-X \\
J093512.7+004735 & 143.8012 & 0.8256 & 0.3570 & $15.21\pm2.31$ & $692.2\pm71.4$ & SWAG-X \\
J093828.0-041933 & 144.6500 & -4.3351 & 0.0807 & $900.81\pm24.12$ & $576.1\pm7.0$ & SWAG-X \\
J093439.4+054141 & 143.6624 & 5.6957 & 0.5590 & $0.24\pm0.07$ & $160.8\pm40.5$ & SWAG-X \\
J091315.0+034847$^{\mathrm{CC}}$ & 138.3116 & 3.8136 & 0.4561 & $14.39\pm0.42$ & $282.6\pm18.7$ & SWAG-X \\
J091510.8+051440$^{\mathrm{CC}}$ & 138.7951 & 5.2418 & 0.1361 & $13.72\pm0.45$ & $153.4\pm8.0$ & SWAG-X \\
J085932.4+030832$^{\mathrm{CC}}$ & 134.8847 & 3.1447 & 0.1955 & $0.58\pm0.12$ & $116.0\pm20.3$ & SWAG-X \\
J091608.3-002355 & 139.0385 & -0.4045 & 0.3234 & $25.05\pm0.79$ & $269.0\pm15.5$ & SWAG-X \\
J085029.4+001453 & 132.6159 & 0.2502 & 0.1955 & $81.03\pm3.04$ & $487.3\pm14.6$ & SWAG-X \\
J082809.6-000958$^{\mathrm{CC}}$ & 127.0383 & -0.1665 & 0.0805 & $0.34\pm0.07$ & $36.1\pm7.7$ & SWAG-X \\
J091406.0-034208 & 138.5319 & -3.7364 & 0.1622 & $37.98\pm0.98$ & $218.8\pm9.1$ & SWAG-X \\
J022201.8-113233 & 35.5030 & -11.5399 & 0.3256 & $1.48\pm0.54$ & $353.5\pm10.3$ & PS2 \\
J022813.4-115703$^{\mathrm{CC}}$ & 37.0569 & -11.9541 & 0.2003 & $1.86\pm0.62$ & $250.4\pm7.2$ & PS2 \\
J022940.1-121406$^{\mathrm{CC}}$ & 37.4136 & -12.2356 & 0.2760 & $0.98\pm0.54$ & $521.7\pm9.8$ & PS2 \\
J023043.3-105615 & 37.6769 & -10.9365 & 0.6498 & $3.03\pm0.51$ & $390.4\pm14.6$ & PS2 \\
J023337.6-094238$^{\mathrm{CC}}$ & 38.4123 & -9.7154 & 0.2629 & $20.45\pm1.22$ & $269.9\pm8.4$ & PS2 \\
J023419.1-095138 & 38.5830 & -9.8656 & 0.6047 & $1.25\pm0.28$ & $418.9\pm14.2$ & PS2 \\
J023419.6-094007$^{\mathrm{CC}}$ & 38.5798 & -9.6676 & 0.8699 & $2.13\pm0.54$ & $437.4\pm16.5$ & PS2 \\
J023720.6-080141$^{\mathrm{CC}}$ & 39.3375 & -8.0473 & 0.2106 & $3.82\pm0.51$ & $232.1\pm7.2$ & PS2 \\
J024037.9-120201$^{\mathrm{CC}}$ & 40.1725 & -12.0638 & 0.5204 & $0.22$ & $116.3$ & PS2 \\
J024106.3-111241 & 40.2841 & -11.2125 & 0.0961 & $5.72\pm0.38$ & $88.7\pm3.7$ & PS2 \\
J024305.2-093501$^{\mathrm{CC}}$ & 40.7658 & -9.5823 & 0.1558 & $295.23\pm0.55$ & $398.2\pm5.6$ & PS2 \\
J024424.7-073636 & 41.1028 & -7.6059 & 0.3474 & $4.96\pm0.92$ & $261.4\pm10.3$ & PS2 \\
\hline
\multicolumn{7}{l}{$^\mathrm{\textbf{CC}}$Cool-core cluster, classified based on the concentration parameter cut, $c_{\mathrm{R_{500}}}>0.26$ (see Section~\ref{sec:dynamic_state}).}
\label{tab:list_clusters}
\end{longtable}

\clearpage
\twocolumn

\clearpage
\section{AGN mechanical feedback considering source age}\label{app:Lmech_SG}
\begin{figure}[h!]
\centering
\includegraphics[width=\columnwidth]{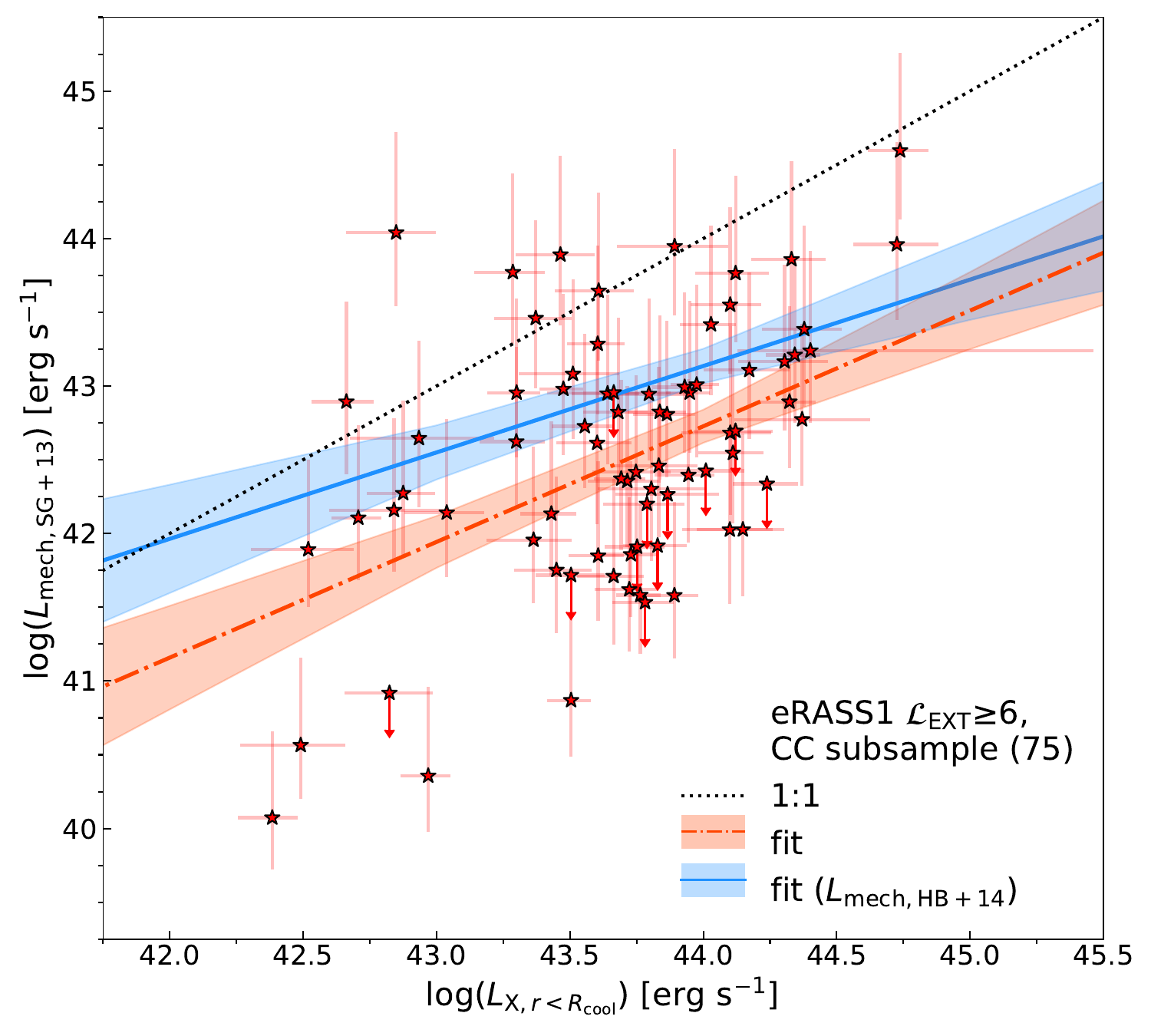}
\caption{Central AGN mechanical luminosity scaled from the monochromatic radio luminosity using Equation~\ref{eq:Lmech_SG} from \cite{Shabala_2013} against X-ray luminosity within the cooling radius for the CC subsample. The orange dashed-dotted line and shaded area are the linear fit and the $1\sigma$ band constrained from the CC subsample. The solid blue and its shaded area is the fit from the dataset where $L_\mathrm{mech}$ was calculated using Equation~\ref{eq:Lkin} from \cite{Heckman_2014}, identical to those plotted in Figure~\ref{fig:lkin} (Section~\ref{sect:lkin-lx_cc}). The dotted line marks the 1-to-1 line.}
\label{fig:lkin_SG}
\end{figure}

The $L_\mathrm{mech,SG+13}$ values range from $1.18\times10^{40}\,\mathrm{erg\,s^{-1}}$ to $3.95\times10^{44}\,\mathrm{erg\,s^{-1}}$. The $\log L_\mathrm{mech,SG+13}$–$\log L_{\mathrm{X},~r<R_\mathrm{cool}}$ correlation estimated from the CC subsample is shown in Figure~\ref{fig:lkin_SG} as the orange dash-dotted line and shaded area. The slope obtained from this dataset is $A= 0.78\pm0.19$ and the normalization is $B=-0.09\pm0.10$. Although the generalised Kendall's $\tau$ $p$ values is more significant with 0.0032, the scatters are still large, $\sigma=0.85$.
\par
The efficiency of the AGN feedback as estimated with the first method (see Section~\ref{sect:eff}) at $L_{\mathrm{X},~r<R_\mathrm{cool}}=10^{42}\,\mathrm{erg\,s^{-1}}$, $10^{43.74}\,\mathrm{erg\,s^{-1}}$, and $10^{45}\,\mathrm{erg\,s^{-1}}$ are $14.4\pm11.7\%$, $6.1\pm1.4\%$, and $3.2\pm2.0\%$, respectively.

\end{appendix}

\end{document}